\documentclass[journal,dvipsnames]{IEEEtran}
\ifCLASSINFOpdf
\else
\fi
\usepackage{times}
\usepackage{epsfig}
\usepackage{cite}
\usepackage{amsmath,amssymb,amsfonts}
\usepackage{graphicx}
\usepackage[table]{xcolor}
\usepackage{amsthm}
\usepackage{caption, subfig}
\usepackage{url}
\usepackage{lineno}
\usepackage{float}
\usepackage{array} 
\usepackage{listings}
\usepackage{algorithm}
\usepackage[noend]{algpseudocode}
\usepackage[breaklinks=true,bookmarks=false,pagebackref=false,colorlinks]{hyperref}
\usepackage{cancel}
\usepackage{soul}

\definecolor{deepgreen}{rgb}{0,0.5,0}
\definecolor{deepblue}{rgb}{0,0,0.5}

\theoremstyle{definition}

\DeclareMathOperator*{\argmin}{arg\,min}

\DeclareFixedFont{\ttb}{T1}{txtt}{bx}{n}{10} 
\DeclareFixedFont{\ttm}{T1}{txtt}{m}{n}{10}  

\graphicspath{{plots/}}

\begin{document}
%
\title{Benchmarking optimization algorithms for auto-tuning GPU kernels}
%
%

\author{Richard Schoonhoven\textsuperscript{\rm 1,3}\qquad~Ben~van~Werkhoven\textsuperscript{\rm 1,2}\qquad~K.~Joost~Batenburg\textsuperscript{\rm 1,3}\\
		{\textsuperscript{\rm 1}Computational Imaging Group, Centrum Wiskunde \& Informatica, Amsterdam, Netherlands}\\
		{\textsuperscript{\rm 2}Netherlands eScience Center, Amsterdam, Netherlands}\\
		{\textsuperscript{\rm 3}Leiden Institute of Advanced Computer Science, Leiden, Netherlands}\\
		\small{\texttt{\{richard.schoonhoven, ben.van.werkhoven, k.j.batenburg\}@cwi.nl}}
}

%
%

\markboth{IEEE Transactions on Evolutionary Computation}%
{Shell \MakeLowercase{\textit{et al.}}: Bare Demo of IEEEtran.cls for IEEE Journals}
%



\maketitle

\begin{abstract}
Recent years have witnessed phenomenal growth in the application, and capabilities of Graphical Processing Units (GPUs) due to their high parallel computation power at relatively low cost. However, writing a computationally efficient GPU program (kernel) is challenging, and generally only certain specific kernel configurations lead to significant increases in performance.
%
%
Auto-tuning is the process of automatically optimizing software for highly-efficient execution on a target hardware platform.
%
%
Auto-tuning is particularly useful for GPU programming, as a single kernel requires re-tuning after code changes, for different input data, and for different architectures.
However, the discrete, and non-convex nature of the search space creates a challenging optimization problem. In this work, we investigate which algorithm produces the fastest kernels if the time-budget for the tuning task is varied. We conduct a survey by performing experiments on 26 different kernel spaces, from 9 different GPUs, for 16 different evolutionary black-box optimization algorithms. We then analyze these results and introduce a novel metric based on the PageRank centrality concept as a tool for gaining insight into the difficulty of the optimization problem. We demonstrate that our metric correlates strongly with observed tuning performance.
\end{abstract}


%
\IEEEpeerreviewmaketitle

\section{Introduction}

\IEEEPARstart{G}{raphics} Processing Units (GPUs) have revolutionized the HPC landscape in the past decade~\cite{heldens2020landscape}, and are seen as one of enabling factors in recent breakthroughs in Artificial Intelligence (AI)~\cite{lecun2015deep}. 
GPUs originated as processors for gaming and then adapted to more general workloads as co-processors in many HPC systems. Over the past decade, GPUs have started to again penetrate new markets such as IoT devices~\cite{mittal2019jetson} and autonomous vehicles~\cite{liu2017computer}. The range of applications of GPUs as such continues to expand. Because of their relatively low cost with respect to their parallel processing power, more and more supercomputers come equipped with GPUs, and in 2020, the majority of modern supercomputers use GPUs~\cite{Top5002020} as the major source of compute power.

The sections of code that run on a GPU, called \emph{kernels}, can be challenging to configure such that they run efficiently for a varying combinations of datasets and GPU architectures~\cite{werkhoven2020lessons}. The kernel parameters can be split into those defined by the program, and those that are a consequence of the underlying architecture and models behind the GPU. The hardware-specific parameters define how the thousands of threads in a GPU are grouped. An ineffective layout can cause underutilization of GPU resources. In general, the computational efficiency can drop by an order of magnitude depending on certain implementation choices. Typically, only a small subset of the possible configurations lead to a large increase in performance~\cite{sclocco2014auto}. Therefore, it is vital to be able to select an efficient kernel configuration. 


The search space for this problem is formed by all feasible combinations of GPU kernel parameters. This space is discrete and non-convex~\cite{petrovivc2020benchmark}, making it hard to carry out the optimization. For most GPU kernels used in practice, the size of this search space is such that traversing the options by hand or brute-force is infeasible. An additional complication in optimizing kernel parameters is that evaluating the performance of each configuration requires costly recompilation and test runs. Furthermore, the same GPU kernel often requires re-tuning for different input data, hardware, or after changes to the code~\cite{li2009note, kamil2010auto, nukada2009auto, nugteren2015cltune}. Large throughput pipelines often rely on computationally expensive GPU kernels that consume large amount of resources \cite{SCLOCCO2020100549, 7418239}, and cannot be tuned exhaustively due to the aforementioned reasons.

Automatic performance tuning (auto-tuning) techniques rely on empirical results and feedback to optimize the kernel parameters with respect to desired performance metrics. These techniques aim to be widely applicable across architectures. For this reason, auto-tuning can be used to find configurations with increased performance for GPU programs. As the search space for the auto-tuning task depends on various aspects (kernel source, code layout, input data, GPU-architecture), the optimization framework must deal with a broad variety of search spaces and constraints. We, therefore, treat the problem as a black-box optimization task. This raises the question of which optimization algorithm is best suited to find highly efficient settings for GPU kernels, and how these optimization algorithms need to be configured to tune GPU kernels.

The main contribution of this work is to determine which optimization algorithms produce the fastest GPU kernels for different tuning-time ranges. To do so, we conduct a survey of 16 evolutionary optimization algorithms for 9 different NVidia and AMD GPUs, and run 3 real-world applicable benchmark kernels. We select our benchmark problems such that we are able (given ample time) to compute the entire search space, and make these spaces publicly available. To benchmark the GPU kernels we use the Kernel Tuner package \cite{kerneltuner}. We use the wide range of optimization algorithms present in Kernel Tuner for a large-scale comparison, and provide favourable default hyperparameters for GPU tuning for each algorithm. In addition, we extend Kernel Tuner with several highly-efficient optimization algorithms, including iterative local search (ILS) and dual annealing that cannot be found in any other generic auto-tuning framework.

Secondly, we aim to quantify tuning difficulty for these seemingly challenging and capricious search spaces. To do so, we introduce the \emph{fitness flow graph} (FFG), which is a network of the points in the search space, with directed edges between neighbours with a better fitness.
By computing the likelihood of local search walks terminating in good local minima, we use FFGs to better understand the discrepancies between optimization algorithms, and subsequently tailor them to better suit GPU tuning. In addition, FFGs can help explain the differences across GPU manufactures, architectures, and kernel programs, and such knowledge can help steer future development. To quantify tuning difficulty per kernel, we introduce a novel metric based on Google's PageRank algorithm~\cite{brin1998anatomy, page1999pagerank}.

This work is structured as follows. In section \ref{sec:relatedwork} we discuss existing GPU kernel tuning approaches. In section \ref{sec:method} we introduce the preliminaries on GPU kernels, and describe the optimization algorithms that are considered in this survey. In section \ref{sec:implementation}, we describe certain implementation details of Kernel Tuner and our Python optimization package BlooPy. We discuss the setup of our experiments in section \ref{sec:experimentsetup}. In section \ref{sec:algoresults} we tune the hyperparameters of the algorithms, and present our findings on optimization algorithm performance. In section \ref{sec:spaceresults} we introduce fitness flow graphs (FFGs) and quantify tuning difficulty for kernel search spaces. Finally, we present our conclusions in section \ref{sec:conclusion}.

\section{Related work}
\label{sec:relatedwork}

\subsection{Automated performance tuning}
\label{sec:autotuningwork}
It is well-known that GPU tuning can yield considerable gains in computational efficiency and utilization for large-scale, high-throughput pipelines that run on compute clusters. As an example, we mention the AMBER pipeline~\cite{SCLOCCO2020100549, 7418239}, which is used to detect Fast Radio Bursts (FRBs) and other single pulse radio transients in astronomy. The pipeline has a throughput of 2 TB/s, and uses a large amount of resources. Benchmarking a single configuration is expensive, and the search space consists of millions of configurations, meaning that sophisticated tuning approaches have to be developed.

Research in automated performance tuning (auto-tuning) can be grouped into two main categories: (1) auto-tuning compiler-generated code
optimizations~\cite{tiwari2009scalable,puschel2005spiral, SRTuner, compilePDEtune}, and (2) software auto-tuning~\cite{li2009note,zhang2012auto}. Ashouri et al.~\cite{ashouri2018survey} wrote an excellent survey on machine-learning methods for compiler-based auto-tuning. 
In this paper, we limit our scope to (2), i.e., optimizations methods for software auto-tuning, which is sometimes referred to as automated design space exploration~\cite{nardi2019hypermapper}. Software auto-tuning allows developers to automatically optimize individual functions and allows, for example, to tune for entirely different implementations and parallelizations that solve the same problem.

As such, auto-tuning techniques are often employed to optimize the source code of high-performance libraries and applications for the CPU, e.g.  ATLAS~\cite{whaley1998automatically} or FFTW~\cite{frigo2005design}, as well as for GPUs~\cite{grewe2011automatically,li2009note,tomov2010dense,zhang2012auto,mametjanov2012autotuning,vanWerkhoven2014optimizing,sclocco2014auto}.

A number of generic auto-tuning frameworks have been introduced in recent years. 
%
%
OpenTuner~\cite{ansel_opentuner_2014} was one of the first generic software auto-tuning frameworks, supporting a number of different search optimization algorithms, but with no support for tuning individual GPU kernels. GPTune~\cite{liu2021gptune} and HyperMapper~\cite{nardi2019hypermapper} are recently proposed frameworks that both use Bayesian Optimization for auto-tuning on different platforms, but do not target GPUs.

Grauer-Gray et al.~\cite{grauer2012auto} have applied auto-tuning to the high-level directive-based HMPP framework, which can compile to CUDA or OpenCL code. They demonstrate significant performance improvements using auto-tuning over unoptimized HMPP kernels in the PolyBench benchmark suite.
Wang et al.~\cite{wang2014automatic} take a similar compiler-based approach to automatically convert shared memory OpenMP applications into OpenCL code for GPUs. They use a machine-learning approach, which based on the number of compute operations and memory accesses in the kernels, predicts the best performing hardware platform to execute the kernels either the multi-core CPU using OpenMP, or on the GPU using OpenCL. 
Hou et al.~\cite{hou2017auto} proposed a data-sensitive auto-tuning framework for sparse matrix vector (SpMV) multiplication that automatically finds the best parallelization strategy. They use a two-step machine learning approach in which they first determine the optimal way to group data into bins and then select the most suitable kernel to process the rows in each bin.


CLTune~\cite{nugteren2015cltune} was the first generic auto-tuning framework with specific support for directly tuning GPU kernels written in the OpenCL programming language. CLTune supports several optimization algorithms, including simulated annealing and particle swarm optimization, but these do not outperform random search~\cite{nugteren2015cltune}.
Kernel Tuning Toolkit (KTT)~\cite{filipovivc2017autotuning} is developed specifically to support online auto-tuning and pipeline tuning, which allows for the exploration of combinations of tunable parameters over multiple kernels. An interesting feature of KTT is the support to keep track of hardware performance counters, such as L2 cache utilization, during benchmarking, which can also be used in advanced search strategies~\cite{filipovivc2021using}. 
Auto-Tuning Framework (ATF)~\cite{raschTACO} implements an innovative way to generate auto-tuning search spaces, for efficient storage and fast exploration of constrained search spaces, but does not focus on introducing new optimization algorithms.

In earlier work, we have introduced Kernel Tuner~\cite{kerneltuner}, a generic auto-tuning framework specifically designed to be an easy-to-use and easy to extend tool for researching auto-tuning optimization algorithms.
Kernel Tuner is a state-of-the-art framework that implements the largest range of search optimization strategies of all generic auto-tuning frameworks, and was the first generic framework to implement multiple search strategies that consistently outperformed random search~\cite{kerneltuner}.

\subsection{Analyzing auto-tuning search spaces}

In this paper, we do not only compare the performance of different optimization algorithms on the GPU auto-tuning problem, but we also investigate the properties of the search spaces to understand why certain optimization algorithms outperform others, and gain insight into the difficulty of the optimization problem. 

Ryoo et al.~\cite{ryoo2008program} were one of the first to study the properties of optimization spaces for GPU applications. They defined two performance metrics to model the efficiency and utilization of a CUDA kernel and used these to find kernel configurations on the pareto curve that maximizes the two metrics. A downside of this approach is that the performance metrics have to be constructed for each kernel individually and require manual inspection and counting instructions in assembly code.
Lim et al.~\cite{lim2017autotuning} also look into search space properties to preselect certain parameter values using static code analysis in order to try to limit exploration of the search space by an auto-tuner.

In \cite{ochoa2008study}, Ochoa et al introduce the concept of {\em Local Optima Networks} (LONs). When constructing LONs the search space is partitioned into basins of attraction, i.e., sets of points where a local search algorithm will terminate in the same local minimum.  A LON is a graph with the local optima as vertices, and a directed edge between nodes if a local search step transforms a solution from one basin of attraction to another. We build upon this idea to define {\em fitness flow graphs} (FFGs), which as opposed to LONs contain all the points in the search space. Due to the large number of points with ``failure fitnesses'' (when a configuration fails to compile) in GPU kernel spaces, defining the basin of attraction is difficult. Instead, we simplify the ideas behind LONs to the entire search space, and quantify how likely local search algorithms terminate in good local optima. To do so, we look at PageRank centrality of local optima. In \cite{herrmann2015predicting} the idea of using PageRank centrality for LONs as predictor of performance for local-seach based heuristics was proposed, and in \cite{herrmann2016determining} the PageRank was used to rank space difficulty. We extend this idea to FFGs to determine GPU tuning difficulty.


\section{Method: Optimization problem}
\label{sec:method}

In this section, we define the performance optimization of GPU kernels as a mathematical optimization problem, and we present the optimization algorithms that are part of our experiments.

\subsection{GPU kernels}
\label{sec:kernels}

GPU kernels are executed by millions of threads in parallel to perform data-parallel computations on the GPU. However, the compute performance of a GPU kernel depends on how the software has been optimized for the hardware.

There are various different design choices that have an impact on the performance of GPU kernels, and this impact is challenging to accurately predict. For example, the way that a computation is parallelized and mapped on the thread blocks and individual threads affects the utilization of the GPU cores. Other design choices include what data types and data layouts to use in the various memory spaces available to GPU applications. There may also be entirely different algorithms to choose from to implement certain parts of the computation.

Other tunable parameters are introduced through code optimizations that can be enabled and may in turn introduce new parameters, such as tiling factors, vector data types, or partial loop unrolling factors. GPU kernels also have a number inherent parameters in terms of the number of thread blocks and the number of threads per block that are used to execute the kernel. The multitude of implementation choices for GPU kernels result in sizeable, non-convex, and discontinuous kernel design spaces.

To automate the kernel design space exploration process, GPU code can be parameterized, either using a kernel template or a code generator. An auto-tuner can take such a kernel template or code generator and empirically benchmark different kernel configurations, until it has found an efficient implementation. The search performed by the auto-tuner can be treated as a mathematical optimization problem of the form:
\begin{equation}
    x^{*} = \argmin\limits_{x \in X} f(x)
\end{equation}
where $f(x)$ is the performance metric to be minimized, for kernel configuration $x$ for a combination of kernel, GPU device, and input settings. In this work the performance metric to be minimized will be the runtime of the kernel.

\subsection{GPU kernel search spaces}
\label{sec:gpuspace}
 The search space of possible settings for a GPU kernel can be characterized as a finite subset $X\subset\mathbb{Z}^n$ for $n$ different parameters. Specifically, for each dimension $1\leq i\leq n$, every entry $x_i$ of a point $x\in X$ takes values from a finite set $S_i\subset\mathbb{Z}$. For example, the block dimension might allow values in $\{16,32,64,128\}$. The total search space is the Cartesian product of these finite sets
\[X=S_1\times S_2\times\cdots\times S_n.\]

The local structure of a search space depends on the definition of \emph{neighbouring points}. A common definition of the neighbours of a point are the points which differ only in one dimension, and are equal for all other dimensions. Mathematically, according to this definition the set of neighbours $N(x)$ of a point $x$ is
\begin{equation}
\label{eq:hammingnbour}
    N(x):=\bigcup_{i=1}^n\left\{y\in X\setminus\{x\}\;\big|\; y_j=x_j,\;\forall j\neq i\right\}.
\end{equation}

Here, we will consider a more restrictive type of neighbourhood concept where we place the additional requirement that the parameter that differs from $x_i$ should have a value adjacent to $x_i$ in the list $S_i$. For example, if the block dimension is allowed to be $[16,32,64,128]$, then neighbours of $x_i=64$ would be $32$ and $128$, and the neighbour of $128$ would only be $64$. We consider this restriction because this definition gives information on whether closely related parameter values are related in performance.

Points of special interest in the search space are local minima. A point is a local minimum if all neighbouring points have a worse fitness. In other words, there are no improvements to be found in the local neighbourhood. Algorithms that scan local neighbourhoods can get stuck in local minima as there are no close points with better fitness.

\subsection{Black-box optimization algorithms}
\label{sec:algos}
As the search space is typically too large to iterate over all feasible points, a range of more sophisticated optimization algorithms are used in practice. In this section we describe the optimization algorithms that are considered in the experiments. As a categorization of these algorithms, we distinguish continuous, and discrete algorithms, and algorithms that learn about the stochasticity of the problem.

\vspace{1.0em}
\subsubsection{Optimization - discrete algorithms}
\label{sec:discretealgos}
Since tuning GPU kernels involves choosing the best from a finite set of possibilities, it makes sense to consider \emph{discrete optimization algorithms}. These algorithms are considered in this work:

\begin{itemize}
\item \textbf{Random sampling} randomly generates solutions and records the highest scoring one. This strategy serves as a baseline comparison to determine if optimization algorithms offer significant benefits.
\end{itemize}

We consider several local search (or hill climb) algorithms which iteratively check for a neighbouring solution of lower fitness to visit, until a local minimum is reached. For all local search algorithms we distinguish between \emph{best-improvement} search, where we move to the best neighbour next, and \emph{first-improvement}, where we examine neighbours in a random order and move to one when we encounter an improvement.

Local search algorithms can vary the neighbourhood function that they use to generate new candidates. The algorithms can use both the version outlined in equation \ref{eq:hammingnbour} (called \emph{Hamming}), and the more restrictive neighbourhood definition in section \ref{sec:gpuspace} (called \emph{adjacent}). First-improvement variants can decide whether they continue checking the remaining variables first after finding an improvement, or if they restart the search (hyperparameter \emph{restart search}).

\begin{itemize}
\item \textbf{Multi-start local search (MLS)} repeatedly generates random starting solutions, and hill climbs them until a local minimum is reached.

\emph{Hyperparameters: neighbourhood, restart.}

\item \textbf{Iterative local search (ILS)}~\cite{lourencco2003iterated} is similar to MLS but inherits part of the original local minimum when generating a new starting solution. After reaching a minimum, ILS performs several random permutations to generate a new starting solution. This \emph{perturbation size} is a tunable parameter. In addition, a tunable \emph{exit after no improve} hyperparameter randomly restarts if no improvement is found after a that many iterations, which helps to escape basins of attraction for small perturbation sizes.

\emph{Hyperparameters: perturbation size, exit after no improve, neighbourhood, restart.}

\item \textbf{Tabu search}~\cite{glover1989tabu} maintains a queue of previously visited solutions which the algorithm is not allowed to visit. The tunable hyperparameter \emph{tabu size} defines the queue size, and ensures that the new solution has not been visited for \emph{tabu size} iterations. Tabu search always picks a new solution, whether it is an improvement or not.

\emph{Hyperparameters: tabu size, neighbourhood.}

\item \textbf{Simulated annealing (SA)}~\cite{kirkpatrick1983optimization} maintains a temperature parameter that, together with the fitness values, determines the probability that we move to a (potentially worse) neighbouring solution. The temperature parameter is decreased each iteration to mimic the behaviour of cooling processes in material physics. A tunable \emph{exploration} parameter determines the size of the mutation of the current solution that is performed at each iteration. A \emph{hill climber} subsequently optimizes the new solution, which is accepted with a certain probability.

\emph{Hyperparameters: exploration, hill climber, neighbourhood}
\end{itemize}

We also consider two discrete population-based algorithms, which require a \emph{population size} parameter to determine the number of solutions they maintain. Population-based methods iteratively create a next generation of solutions by mixing solutions from the previous generation. A \emph{reproduction} operator creates new initial solutions from existing ones, e.g., with two-point crossover a section of the solution vector is swapped between two solutions. After new solutions have been created, and their fitnesses determined, a \emph{selection} mechanism determines which solutions are kept for this generation. For example, in tournament selection a number of randomly picked solutions compete for a spot in the next generation.

\begin{itemize}
\item \textbf{Genetic local search (GLS)}~\cite{gls} (or memetic algorithm) is a population-based method where every solution in a generation is subsequently \emph{hill climbed}. The initial population is made up of randomly generated solutions, which after hill climbing are all local minima. Next, a number of children is created by reproduction, and a new initial starting population is selected from the batch. These solutions are subsequently hill climbed and the procedure is repeated.

\emph{Hyperparameters: hill climber, population size, reproduction, selection.}

\item \textbf{Genetic algorithm (GA)}~\cite{mitchell1998introduction} is similar to GLS, but instead of hill climbing each solution, it performs a single mutation only, e.g., permuting a single parameter. Instead of a hill climbing algorithm, GA has a tunable hyperparamter \emph{mutation} that determines the fraction of variables of a solution that are mutated each generation.

\emph{Hyperparameters: mutation, population size, reproduction, selection.}
\end{itemize}

\begin{table*}
    \centering
    \small
    \begin{tabular}{l|cccccc|}
       & \multicolumn{6}{c|}{\textbf{GPU Specifications}} \\
       & & CUDA cores/ & Device & & Bandwidth & Peak compute \\
       \textbf{GPU model} & Year & Stream processors & memory & Boost clock & (GB/s) & (GFLOP/s) \\
       \hline
        NVidia Tesla K20            & 2012  & 2496 & 5~GB & 0.76 GHz & 208 & 3524 \\
        NVidia GTX Titan X          & 2015  & 3072 & 12~GB & 1.08 GHz & 336 & 6605 \\
        NVidia Tesla P100 PCIe      & 2016  & 3584 & 12~GB & 1.30 GHz & 549 & 9340 \\
        NVidia GTX 1080 Ti          & 2017  & 3584 & 11~GB & 1.58 GHz & 484 & 11340 \\
        NVidia Tesla V100 PCIe      & 2017  & 5120 & 32~GB & 1.37 GHz & 900 & 14899 \\
        AMD Radeon Instinct MI50    & 2018  & 3840 & 16~GB & 1.73 GHz & 1024 & 13300 \\
        NVidia Titan RTX            & 2018  & 4608 & 24~GB & 1.77 GHz & 672 & 16312 \\
        NVidia RTX 2070 Super       & 2019  & 2560 & 8~GB & 1.77 GHz & 448 & 9060 \\
        NVidia A100 PCIe            & 2020  & 6912 & 40GB & 1.41 GHz & 1555 & 19500 \\
    \end{tabular}
    \caption{\scriptsize Specifications of graphical processing unints (GPUs) used to create experimental data.}
    \label{tab:gpuspecs}
\end{table*}

\vspace{1.0em}
\subsubsection{Optimization - continuous algorithms}
\label{sec:continuousalgos}
As an alternative to discrete algorithms, we can consider \emph{continuous optimization algorithms} which operate on real-valued solutions. In order to apply algorithms which assume continuous variables to a discrete problem such as GPU kernel tuning, we need to define a mapping between a real-valued vector, and the discrete values in the search space. Suppose the search space allows values $x_1,x_2,\ldots,x_n$ for a particular variable, then a continuous variable $y\in[0,1]$ gets mapped to the closest grid point $\bar{y}$:
\begin{align*}
B&=\{\frac{1}{2n},\frac{3}{2n},\ldots,\frac{2n-1}{2n}\}\\
j^*&=\argmin_{%
        \substack{%
         i=1,\ldots,n}
     }
\{|B_i-y|\}\\
\bar{y}&=x_{j^*}.
\end{align*}
Effectively, this ensures that all possible discrete values are equally spaced across the interval $[0,1]$, and the continuous variable is mapped to the closest one. The continuous optimization algorithms operate on real-valued vectors with dimensions equal to the number of parameters that are to be optimized. Each entry is bounded to the unit interval.

The mapping ensures that points close together in real-valued space can get mapped to the same point in the GPU tuning space. While this can negatively impact the performance of continuous algorithms, it does not automatically lead to poor performance, as illustrated by the strong performance of continuous algorithms in Kernel Tuner \cite{kerneltuner}. Furthermore, this mapping allows us to explore a new class of algorithms. Here, we consider two local search algorithms.

\begin{itemize}
\item \textbf{Basin hopping}~\cite{wales1997global} is a global stepping algorithm that chooses new starting positions for local minimization. It requires the local minimizer \emph{method} and a \emph{temperature} parameter to be chosen. The temperature parameter determines the accept--reject criterion. Currently supported minimization methods are the nonlinear conjugate gradient (CG) \cite{nocedal2006conjugate}, simplex (Nelder-Mead) \cite{nelder1965simplex}, conjugate direction (Powell) \cite{powell1964efficient}, L-BFGS-B \cite{byrd1995limited}, Constrained Optimization BY Linear Approximation (COBYLA) \cite{powell1994direct}, and Sequential Least Squares Programming (SLSQP) \cite{kraft1988software} methods.

\emph{Hyperparameters: minimizer method, temperature.}

\item \textbf{Dual annealing}~\cite{tsallis1996generalized} is an extension of generalized simulated annealing, paired with a local minimization method. It combines global and local search procedures, and it requires users to choose a local minimization \emph{method}.

\emph{Hyperparameters: minimizer method.}
\end{itemize}

Lastly, we consider two population-based algorithms.
\begin{itemize}
%
\item \textbf{Particle swarm optimization (PSO)}~\cite{kennedy1995particle} initializes a \emph{number of particles} at random in the search space. Each iteration, these particles update their position and velocity. Particles transmit information to a certain \emph{number of neighbours}, thereby influencing the movement of the other particles. 

\emph{Hyperparameters: \#particles, neighbours evaluated.}

\item \textbf{Differential evolution}~\cite{storn1997differential} is similar to a genetic algorithm, but mixing strategies are based on real-valued solutions. Typically they involve mixing the best solution with a random candidate, and accepting the result with a certain probability.

\emph{Hyperparameters: mixing method, population size, mutation size, recombination probability.}
\end{itemize}

\vspace{1.0em}
\subsubsection{Optimization - tuning algorithms for stochastic optimization}
\label{sec:stochasticalgos}
In this survey we consider two state-of-the-art parameter tuning algorithms;
\begin{itemize}
\item \textbf{Sequential Model Algorithm Configuration (SMAC)} \cite{lindauer2021smac3} is a random forest-based Bayesian optimization method that is designed for optimization of stochastic problems. However, it can also be used to optimize deterministic problems. SMAC requires the \emph{model type} of its Bayesian optimizer to be chosen. The \emph{gp-mcmc} model was significantly slower and worse than \emph{gp} in the preliminary experiments. We therefore use the \emph{gp} model type in this work. The \emph{acquisition function} of the BO is another tunable hyperparameter.

\emph{Hyperparameters: acquisition function.}

\item \textbf{Iterated racing (irace)} \cite{lopez2016irace} is a statistical approach for selecting the best configuration out of a set of candidates for stochastic optimization problems. After consulting the authors \cite{lopez2016irace}, we set \emph{firstTest} and \emph{nbConfigurations} as tunable hyperparameters for irace.

\emph{Hyperparameters: firstTest, nbConfigurations.}
\end{itemize}
%

\section{Implementation}
\label{sec:implementation}
In this section we comment on certain implementation details for the software developed for this work.
The algorithms and analysis tools are implemented in the BlooPy Python package, and the tuning of the GPU kernels is performed by Kernel Tuner.

\subsection{BlooPy and SOTA packages}
\label{sec:bloopy}
The algorithms evaluated in this work are implemented in the discrete optimization package \textbf{BlooPy} (BLackbOx Optimization Python) \cite{bloopy}. The package implements the algorithms by encoding solutions as bitstrings. BlooPy implements several functions for converting discrete solutions that are encoded as lists or arrays to bitstrings. Similarly, continuous solutions are mapped to discrete solution vectors using the mapping outlined in section \ref{sec:continuousalgos}. BlooPy requires the search space to be finite. Using the bitstring encodings, BlooPy's algorithms can make use of the computationally efficient Python module \verb|bitarray| which implements fast low-level bitstrings in C. In addition, the algorithms are automatically applicable to benchmark bitstring-based optimization problems such as randomized $Nk$-landscapes \cite{randomizednk}.

Optimization algorithms in BlooPy maintain a cache of previously visited solutions. This means that a solution that has been visited before does not count as a function evaluation, and instead the cached value is returned. In addition to a variety of optimization algorithms, BlooPy implements several search space analysis tools. For example, it implements functions to determine the type of points in the search space, e.g., local minima and saddle points. Furthermore, BlooPy implements functions to compute the fitness flow graphs outlined in section \ref{sec:critgraphs}. BlooPy can be installed from the GitHub source repository \cite{bloopy}, or by package manager. To perform experiments with SMAC we used the Python package \cite{smacgit}, and for irace we used the R package \cite{iracegit}.

\subsection{Kernel Tuner}
\label{sec:kerneltuner}
Kernel Tuner~\cite{kerneltuner} implements a wide range of optimization algorithms, and builds on top of various backends (e.g. PyOpenCL, PyCUDA, Cupy, GCC) that take care of the compilation process.

Kernel Tuner runs Python code, provided by the user, which calls the tuner function. In addition, the user needs to provide a code generator or parameterized template for the kernel they wish to optimize. An optimization algorithm then selects different kernel configurations for benchmarking.

\begin{table}
    \centering
    \small
    \begin{tabular}{l|cccc|}
        & & \#local & \#variables & \#failed\\
       \textbf{Kernel} & \#points & minima & to tune & points\\
       \hline
        Convolution$^*$     & 18432  & 89 & 6 & 12656$^{*}$ \\
        GEMM              & 82944  & 64 & 5 & 64988 \\
        Point-in-polygon  & 8184  & 220 & 10 & 335 \\
    \end{tabular}
    \caption{\footnotesize Statistics (averaged across GPU models) of kernel spaces. Number of failed points refers to the average number of configurations in the kernel space that failed to compile.\\ $^*$Convolution has 864 points for AMD MI50, and 450 fail points.}
    \label{tab:searchspace}
\end{table}

\begin{table*}
    \scriptsize
    \newcommand{\hght}{0.5em}
    \centering
    \begin{tabular}{c|c|c|c|c|c|c|c|c|}
       & \multicolumn{7}{c|}{\small Maximum number of function evaluations (budget)} \\
       \hline
       \hline
       \textbf{Basin hopping} & \color{Maroon}{25} & \color{Maroon}{50} & \color{Maroon}{100} & \color{Maroon}{200} & \color{Maroon}{400} & \color{Maroon}{800} & \color{Maroon}{1600} \\
       \hline
        method & Powell & \multicolumn{4}{c|}{COBYLA} & \multicolumn{2}{c|}{SLSQP} \\
        \hline
        temperature & \multicolumn{5}{c|}{0.1} & \multicolumn{2}{c|}{1.0} \\
        \hline
        \hline
       \textbf{Dual annealing} & \color{Maroon}{25} & \color{Maroon}{50} & \color{Maroon}{100} & \color{Maroon}{200} & \color{Maroon}{400} & \color{Maroon}{800} & \color{Maroon}{1600} \\
       \hline
        \multicolumn{1}{c|}{method} & COBYLA & \multicolumn{6}{c|}{Powell} \\
        \hline
        \hline 
       \textbf{Differential evolution} & \color{Maroon}{25} & \color{Maroon}{50} & \color{Maroon}{100} & \color{Maroon}{200} & \color{Maroon}{400} & \color{Maroon}{800} & \color{Maroon}{1600} \\
       \hline
        population size & 1 & \multicolumn{2}{c|}{2} & 4 & 8 & 16 & 32 \\
        \hline
        method & \multicolumn{4}{c|}{best1bin} & best2bin & \multicolumn{2}{c|}{best1exp} \\
        \hline
        recombination & 0.5 & \multicolumn{6}{c|}{0.7} \\
        \hline
        mutation & \multicolumn{7}{c|}{(0.2, 0.7)} \\ 
        \hline
        \hline
       \textbf{Particle swarm optimization} & \color{Maroon}{25} & \color{Maroon}{50} & \color{Maroon}{100} & \color{Maroon}{200} & \color{Maroon}{400} & \color{Maroon}{800} & \color{Maroon}{1600} \\
       \hline
        Number of particles & \multicolumn{2}{c|}{25} & 10 & 20 & 40 & 80 & 160 \\
        \hline
        neighbours evaluated & \multicolumn{3}{c|}{5} & 10 & 20 & 26 & 32 \\
        \hline
        \hline
       \textbf{FirstILS} & \color{Maroon}{25} & \color{Maroon}{50} & \color{Maroon}{100} & \color{Maroon}{200} & \color{Maroon}{400} & \color{Maroon}{800} & \color{Maroon}{1600} \\
       \hline
        perturbation size & \multicolumn{5}{c|}{1.0} & \multicolumn{2}{c|}{0.05} \\
       \hline
        Exit after no improve & \multicolumn{5}{c|}{25} & \multicolumn{2}{c|}{10} \\
       \hline
        neighbour method & \multicolumn{5}{c|}{Hamming} & \multicolumn{2}{c|}{adjacent}\\
       \hline
        restart search & \multicolumn{4}{c|}{False}  & \multicolumn{3}{c|}{True} \\
        \hline
        \hline
       \textbf{BestILS} & \color{Maroon}{25} & \color{Maroon}{50} & \color{Maroon}{100} & \color{Maroon}{200} & \color{Maroon}{400} & \color{Maroon}{800} & \color{Maroon}{1600} \\
       \hline
        perturbation size & \multicolumn{5}{c|}{1.0} & \multicolumn{2}{c|}{0.05} \\
       \hline
        Exit after no improve & \multicolumn{7}{c|}{25} \\
       \hline
        neighbour method & \multicolumn{2}{c|}{adjacent} & \multicolumn{3}{c|}{Hamming} & \multicolumn{2}{c|}{adjacent} \\
        \hline
        \hline
       \textbf{FirstTabu} & \color{Maroon}{25} & \color{Maroon}{50} & \color{Maroon}{100} & \color{Maroon}{200} & \color{Maroon}{400} & \color{Maroon}{800} & \color{Maroon}{1600} \\
       \hline
        tabu size & \multicolumn{2}{c|}{4} & \multicolumn{5}{c|}{2000} \\
       \hline
        neighbour method & \multicolumn{7}{c|}{Hamming} \\
        \hline
        \hline
       \textbf{BestTabu} & \color{Maroon}{25} & \color{Maroon}{50} & \color{Maroon}{100} & \color{Maroon}{200} & \color{Maroon}{400} & \color{Maroon}{800} & \color{Maroon}{1600} \\
       \hline
        tabu size & \multicolumn{7}{c|}{2000} \\
       \hline
        neighbour method & \multicolumn{7}{c|}{Hamming} \\
        \hline
        \hline
       \textbf{FirstMLS} & \color{Maroon}{25} & \color{Maroon}{50} & \color{Maroon}{100} & \color{Maroon}{200} & \color{Maroon}{400} & \color{Maroon}{800} & \color{Maroon}{1600} \\
       \hline
        restart search & \multicolumn{2}{c|}{True} & \multicolumn{2}{c|}{False} & \multicolumn{3}{c|}{True}\\
       \hline
        neighbour method & \multicolumn{7}{c|}{Hamming} \\
        \hline
        \hline
       \textbf{BestMLS} & \color{Maroon}{25} & \color{Maroon}{50} & \color{Maroon}{100} & \color{Maroon}{200} & \color{Maroon}{400} & \color{Maroon}{800} & \color{Maroon}{1600} \\
       \hline
        neighbour method & \multicolumn{3}{c|}{adjacent} & \multicolumn{4}{c|}{Hamming}\\
        \hline
        \hline
       \textbf{Simulated annealing} & \color{Maroon}{25} & \color{Maroon}{50} & \color{Maroon}{100} & \color{Maroon}{200} & \color{Maroon}{400} & \color{Maroon}{800} & \color{Maroon}{1600} \\
       \hline
        explore ($p$) & \multicolumn{4}{c|}{1.0} & 0.7 & \multicolumn{2}{c|}{0.1} \\
        \hline
        hill climber & None & \multicolumn{6}{c|}{RandomFirst}\\
        \hline
        neighbour method & \multicolumn{7}{c|}{Hamming} \\
        \hline
        \hline
       \textbf{Genetic local search} & \color{Maroon}{25} & \color{Maroon}{50} & \color{Maroon}{100} & \color{Maroon}{200} & \color{Maroon}{400} & \color{Maroon}{800} & \color{Maroon}{1600} \\
       \hline
        hill climber & \multicolumn{7}{c|}{RandomFirst} \\
       \hline
        population size & 2 & 2 & 4 & 16 & 20 & 40 & 80 \\
       \hline
        reproductor & \multicolumn{2}{c|}{uniform} & \multicolumn{2}{c|}{2point} & \multicolumn{3}{c|}{uniform} \\
       \hline
        selector & \multicolumn{7}{c|}{RTS} \\
       \hline
        neighbour method & \multicolumn{7}{c|}{Hamming}\\
        \hline
        \hline
       \textbf{Genetic algorithm} & \color{Maroon}{25} & \color{Maroon}{50} & \color{Maroon}{100} & \color{Maroon}{200} & \color{Maroon}{400} & \color{Maroon}{800} & \color{Maroon}{1600} \\
       \hline
        mutation & \multicolumn{3}{c|}{0.02} & \multicolumn{2}{c|}{0.05} & \multicolumn{2}{c|}{0.02} \\
       \hline
        population size & 8 & 10 & 20 & 40 & 80 & 128 & 320 \\
       \hline
        reproductor & \multicolumn{5}{c|}{1point} & \multicolumn{2}{c|}{2point} \\
       \hline
        selector & \multicolumn{5}{c|}{tour8} & \multicolumn{2}{c|}{tour4} \\
        \hline
        \hline
       \textbf{SMAC} & \color{Maroon}{25} & \color{Maroon}{50} & \color{Maroon}{100} & \color{Maroon}{200} & \color{Maroon}{400} & \color{Maroon}{800} & \color{Maroon}{1600} \\
       \hline
        model type & \multicolumn{5}{c|}{gp} & \multicolumn{2}{c|}{NA} \\
       \hline
        acquisition function & \multicolumn{5}{c|}{LCB} & \multicolumn{2}{c|}{NA} \\
        \hline
        \hline
       \textbf{irace} & \color{Maroon}{25} & \color{Maroon}{50} & \color{Maroon}{100} & \color{Maroon}{200} & \color{Maroon}{400} & \color{Maroon}{800} & \color{Maroon}{1600} \\
       \hline
        firstTest & \multicolumn{3}{c|}{NA} & \multicolumn{4}{c|}{2} \\
       \hline
        nbConfigurations & \multicolumn{3}{c|}{NA} & \multicolumn{4}{c|}{0} \\
        \hline
    \end{tabular}
    \caption{\scriptsize Selected hyperparameters across different budgets (50 to 2000) for the optimization algorithms used in this work. The budgets columns are given in red. These hyperparameters were optimized using the convolution, GEMM, and PnPoly kernels on the NVidia P100 GPU.}
    \label{tab:hyperpars}
\end{table*}

\section{Experimental Setup}
\label{sec:experimentsetup}
To analyze the structure of different kernel spaces, and find the optimization algorithm best suited to finding strong kernel settings, we run experiments on 9 different GPUs, for 3 real-world applicable kernel programs (26 kernels in total).

\begin{itemize}
    \item {\bf Convolution}~\cite{vanWerkhoven2014optimizing} operations are an essential tool in image processing, and are often used for tasks such as edge detection, blurring, or sharpening. They also feature prominently in deep learning methods for image processing as they form the backbone of the convolutional neural network (CNN).
    \item {\bf GEMM} (Generalized dense matrix–matrix multiplication)~\cite{clblast} is one of the most widely-used kernels across many application domains, including neural networks. Here we perform the calculation $C=\alpha A\cdot B+\beta C$ for $4096\times 4096$ matrices $A,B,C$, and constants $\alpha$ and $\beta$.
    \item {\bf PnPoly} (Point-in-Polygon) kernel is used by Goncalves et al.~\cite{goncalves2016spatial} as part of a geospatial database management system to, for example, return all objects within the outline of a specific area.
\end{itemize}

Some statistics on the kernel spaces is given in Table \ref{tab:searchspace}. For convolution and GEMM the majority of the points in the kernel space fail to compile, 68\% and 78\% respectively. In the case of a failed compilation we attribute a ``fail'' fitness of $10^{10}$ to this point. The exact kernel spaces can be found in the table of tunable parameters (Table \ref{tab:tuneparams}) in Appendix \ref{app:tuneparams}.

We selected these kernels programs since they are tunable common subroutines in real-world applications, but also have a compact parameter space which can be fully explored, given ample computation time. Note that this is not feasible for many other kernels used in practice (see section \ref{sec:autotuningwork}). We have generated cache files of the entire search space for each kernel by brute-force calculation. This allows us to know the optimal settings for each problem, and therefore score solutions returned by algorithms. It also allows us to develop analysis metrics on the entire search space, which could at a later stage be adapted to work %
when sampling only small parts of the space. 
We supply our cache files for benchmarking optimization algorithms \cite{gpurepo}, similar to other computationally expensive applications such as neural architecture search~\cite{natsbench}.


The 9 GPUs that are used for testing are given in Table~\ref{tab:gpuspecs}. The convolution kernel is implemented in CUDA, GEMM in OpenCL, and PnPoly is a heterogeneous kernel that runs partly on the CPU, and partly on the GPU using CUDA. The PnPoly kernel uses CUDA-specific features that are not available on AMD GPUs. We have used CUDA Version 11.2, OpenCL 1.2, Python 3.8.5, PyCUDA v2021.1, PyOpenCL v2020.3.1, and BlooPy version 0.4.2.

\vspace{1.0em}
\textbf{Experimental setup:} The GPU kernels are tuned with respect to runtime (ms). The runtime of a GPU kernel is stochastic, and can vary slightly per execution. Kernel Tuner automatically benchmarks a given configuration 32 times to acquire a mean runtime per configuration. In most cases, the compilation time for a given kernel configuration significantly exceeds the time needed to benchmark 32 runs. Therefore, most of our experiments are performed in a deterministic setting where the fitness of a configuration is the mean runtime. However, we also perform a stochastic experiment where a single kernel runtime is returned for every evaluation. This means that the fitness for the same point in the search space can vary, and algorithms that learn stochastic information, such as irace and SMAC, can potentially benefit.

After discussion with the authors \cite{lopez2016irace} we decided to benchmark irace only for the stochastic experiment. This was decided as it was deemed inappropriate for the deterministic setting since the point of using irace is to dynamically handle stochasticity in expensive problems.

The algorithms are evaluated based on the fraction of the optimal runtime they can find within a limited budget of evaluations. For each algorithm, we run experiments with a \emph{maximum function evaluation} limit (budget) of 25, 50, 100, 200, 400, 800, 1600. The goal of our experiments is to benchmark algorithms when traversing only a fraction of the search space. Therefore, we set the highest budget limit at 1600 since it is already approximately 20\% of the Point-in-polygon search space. Every run is performed 50 times in order to get an indication of the spread. Due to computational demand, SMAC and irace experiments are ran 20 times. SMAC is only run up to a budget of 400 evaluations due to the high tuning time. Data and scripts for the experiments and figures can be found in the GitHub repository \cite{gpurepo}.

\section{Results: Benchmarking optimization algorithms on runtime}
\label{sec:algoresults}
In this section, we first discuss how to initialize each algorithm with favourable hyperparameters. Next, we discuss which algorithms are best suited to tuning GPU kernels.

\subsection{Setting hyperparameters}
\label{sec:hyperparameters}
In order to compare the optimization algorithms fairly for the GPU tuning problem, we need to choose sensible hyperparameters. Which hyperparameters can be varied per algorithm is outlined in italics in section \ref{sec:algos}. We test different combinations of hyperparameters on the P100, RTX 2070 Super, and GTX 1080Ti, for all three kernels (9 out of 26 kernels). This way we select reasonable hyperparameters across various architectures and kernels, which users can use as defaults for new GPU tuning problems. We performed a bruteforce search over all combinations of parameter values, and ran each set 20 times for each algorithm. For PSO we kept the $w, c_1, c_2, p$ parameters constant as these appeared to have little effect on algorithm performance. Note that the time required to tune hyperparameters varies greatly between algorithms due to this exhaustive search. The hyperparameters chosen are in Table \ref{tab:hyperpars}.

To choose hyperparameters, we first group settings which perform similarly statistically, and attempt to find one set of hyperparameters that performs well across all 9 kernels. For a budget $p$, let $f_{p,best}$ be the lowest average fitness achieved for a set of hyperparameters, and $\sigma_{p,best}$ the standard deviation. We perform the following selection approach:

\begin{enumerate}
    \item For every kernel, we create a set of hyperparameter settings whose average found fitness is within $k\cdot\sigma_{p,best}$ of $f_{p,best}$. 
    \item For each budget, intersect the acceptable settings for convolution, GEMM, and PnPoly, across the 3 GPUs.
    \item For each budget, if this intersection is non-empty, reduce $k$ and repeat. If the intersection is empty, increase $k$ and repeat. Repeat until only one set of hyperparameters remains in the intersection.
\end{enumerate}

\begin{figure*}
    \centering
    \newcommand{\wdth}{0.46}
    \newcommand{\bspc}{0.0}
    \newcommand{\hspc}{-0.2}
    \newcommand{\vspc}{-0.91}
    \captionsetup[subfigure]{labelformat=empty}
    \subfloat{\makebox[\bspc\textwidth][c]{
    \subfloat{\includegraphics[width=\wdth\textwidth]{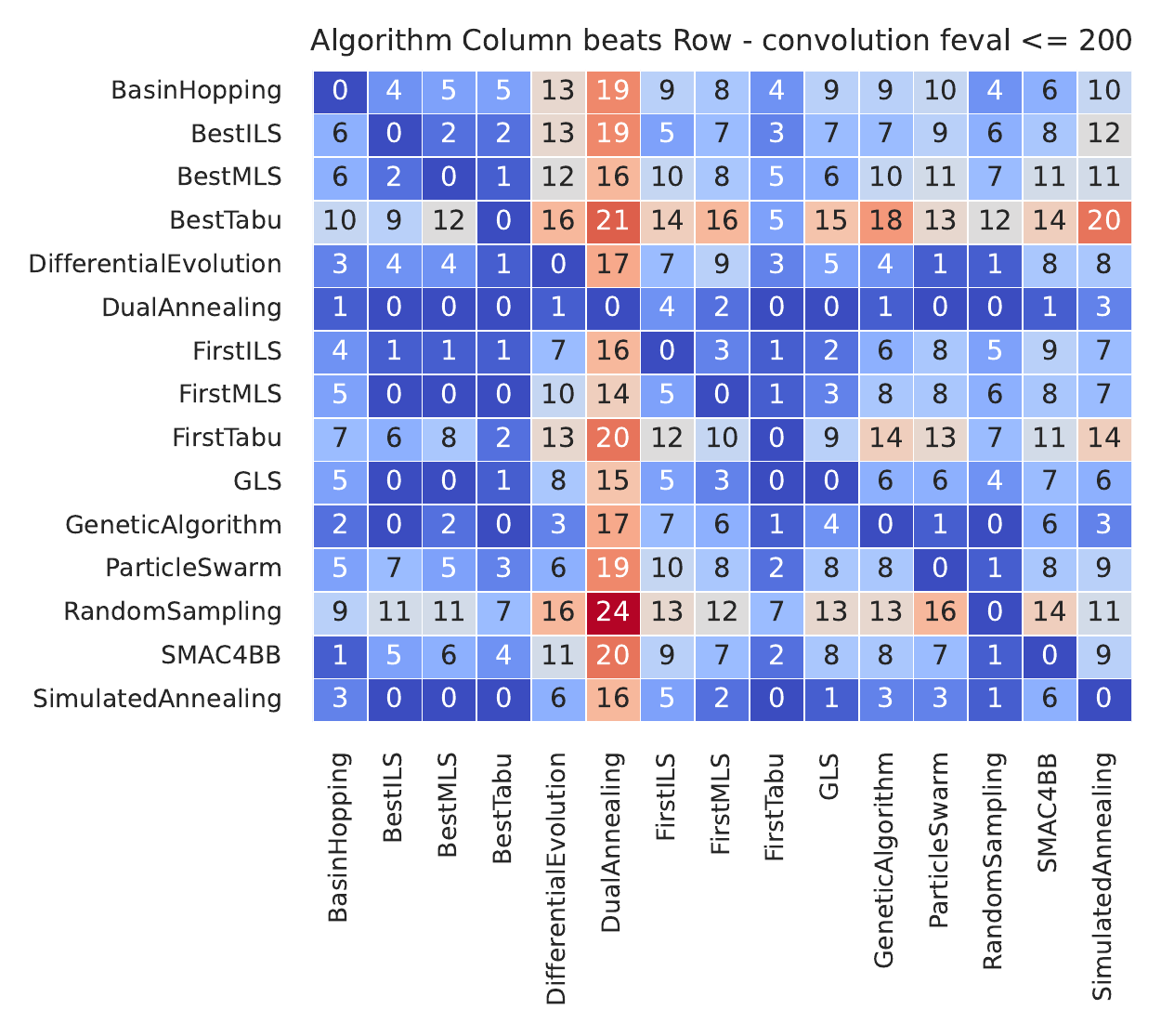}}
    \hspace{\hspc cm}
    \subfloat{\includegraphics[width=\wdth\textwidth]{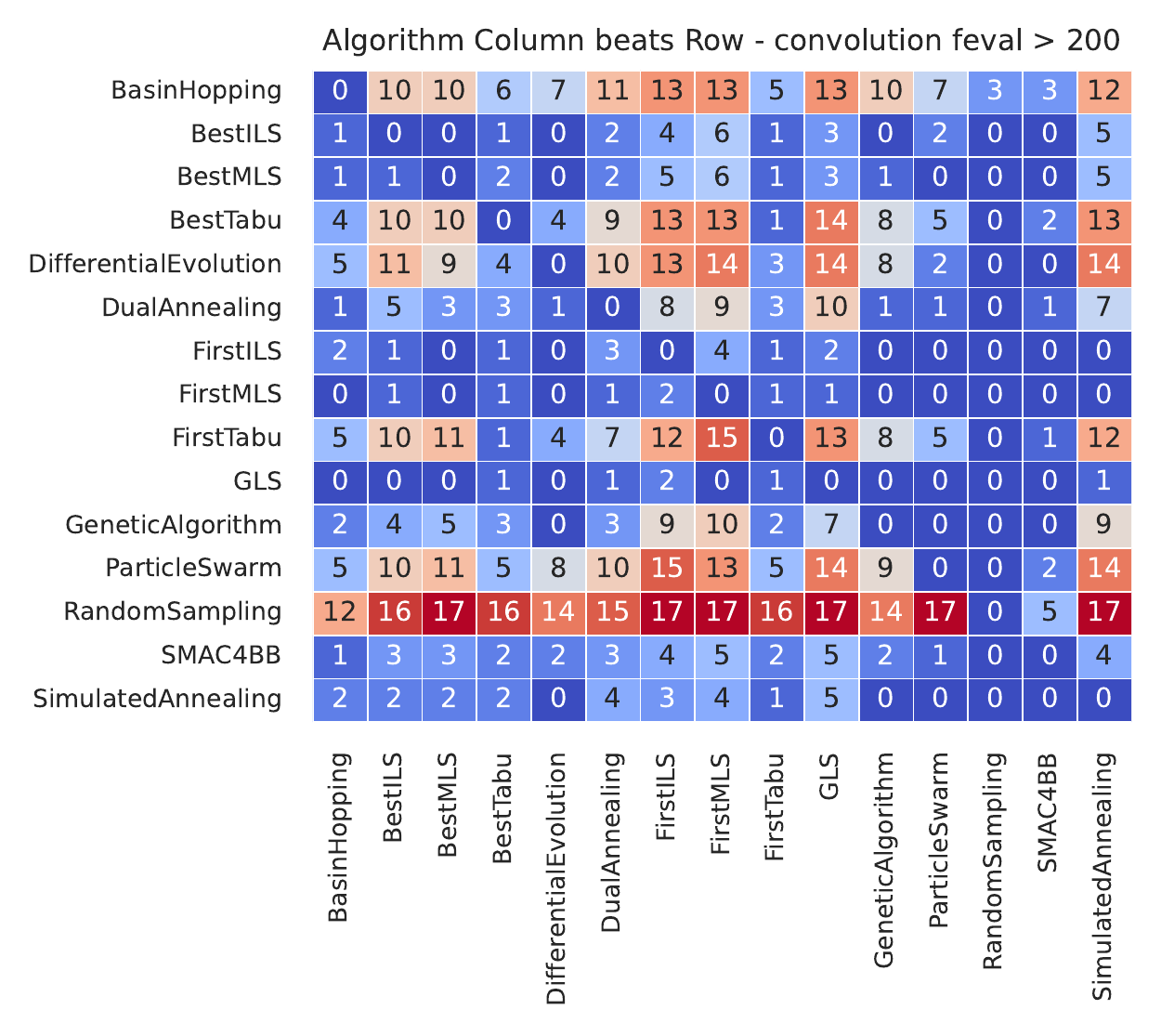}}
    }}\\
    \vspace{\vspc cm}
    \subfloat{\makebox[\bspc\textwidth][c]{
    \subfloat{\includegraphics[width=\wdth\textwidth]{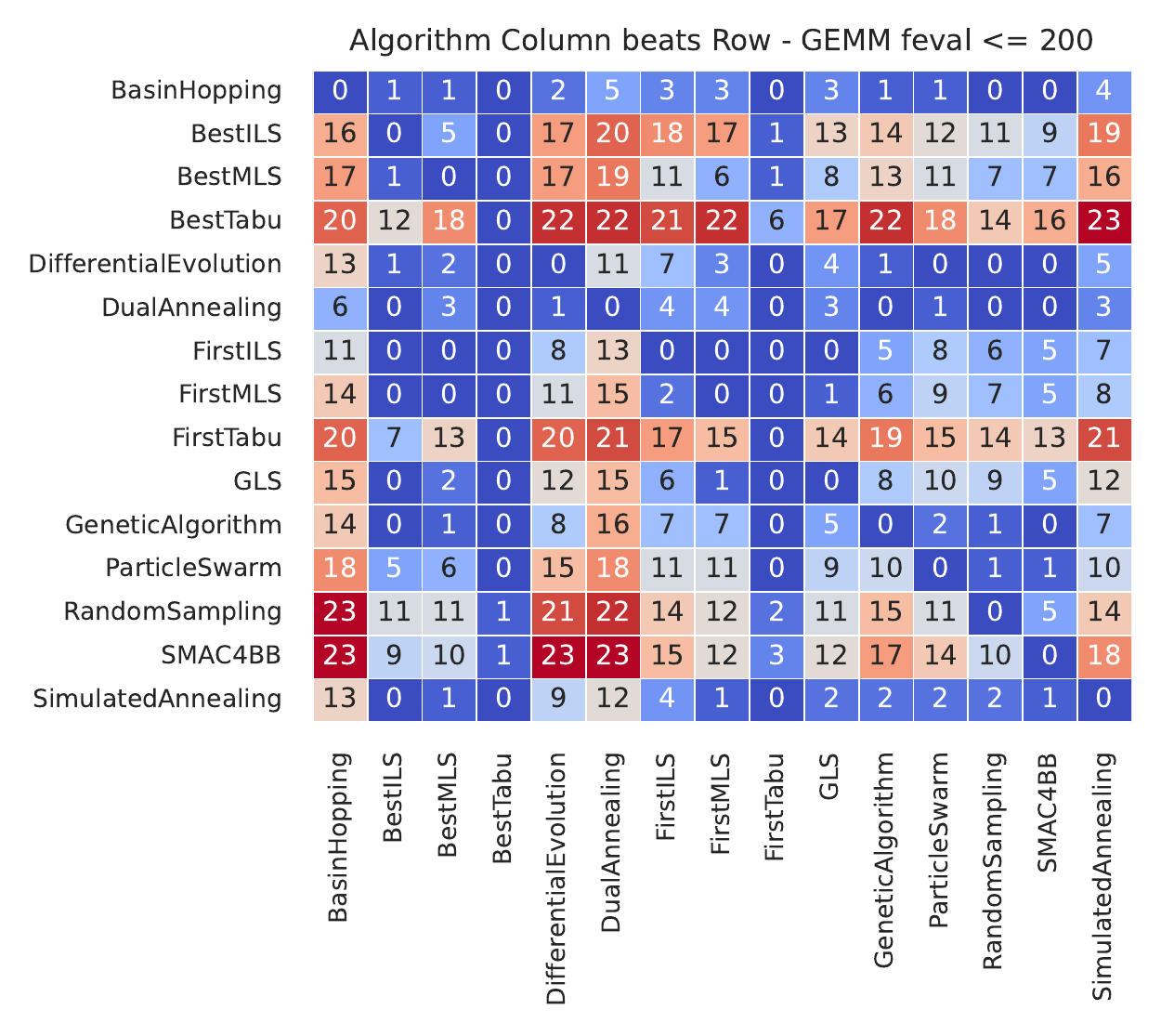}}
    \hspace{\hspc cm}
    \subfloat{\includegraphics[width=\wdth\textwidth]{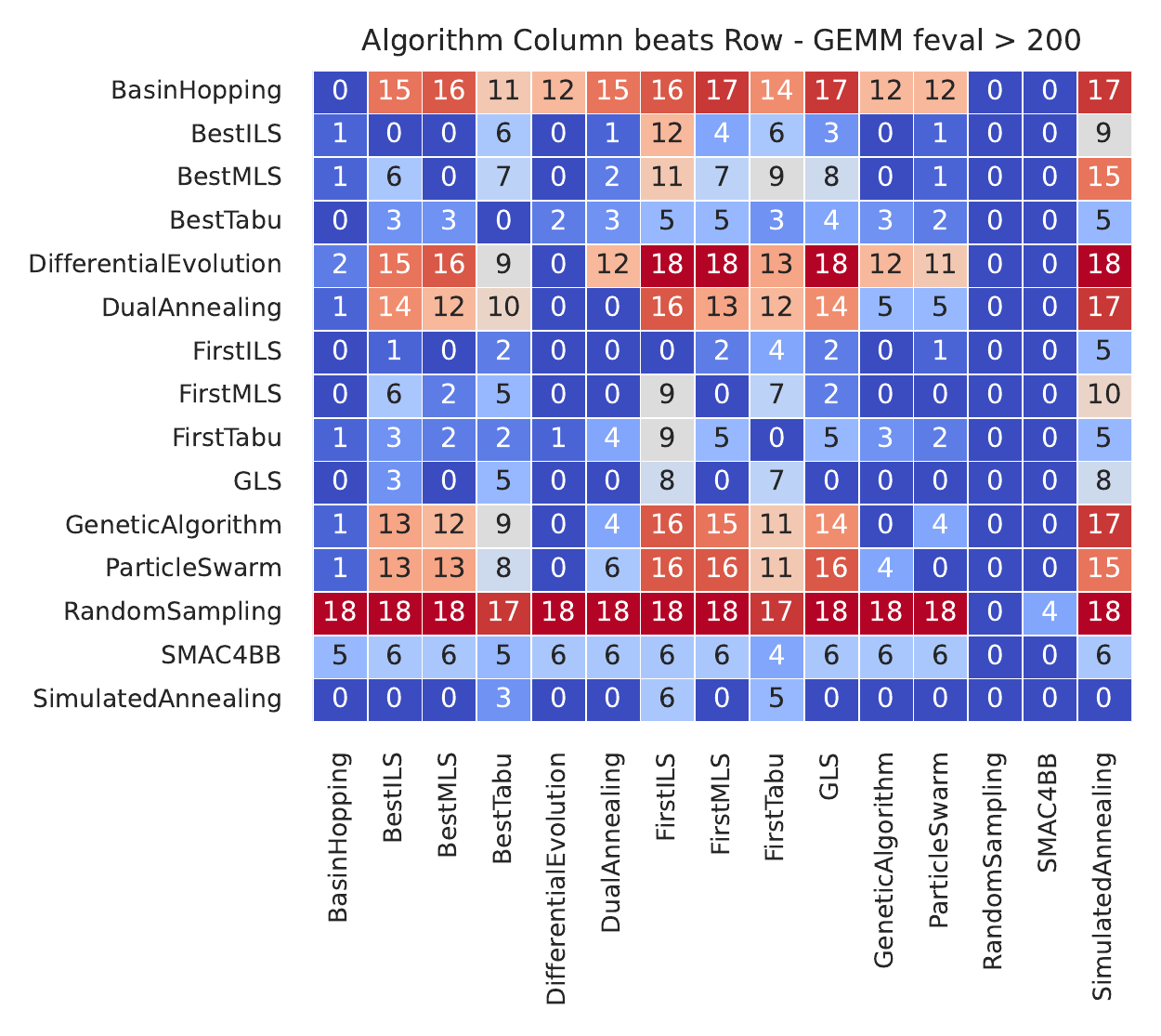}}
    }}\\
    \vspace{\vspc cm}
    \subfloat{\makebox[\bspc\textwidth][c]{
    \subfloat{\includegraphics[width=\wdth\textwidth]{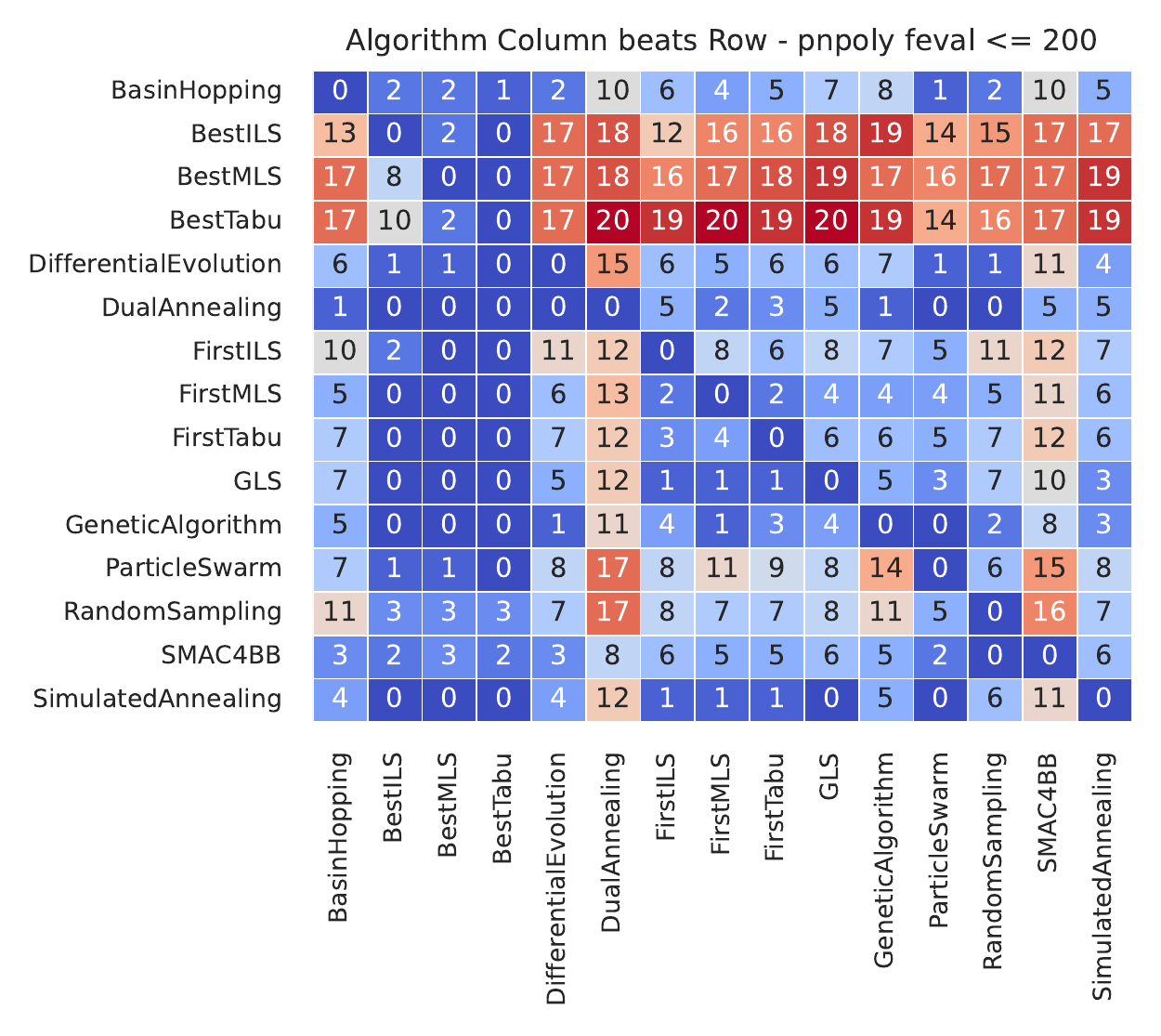}}
    \hspace{\hspc cm}
    \subfloat{\includegraphics[width=\wdth\textwidth]{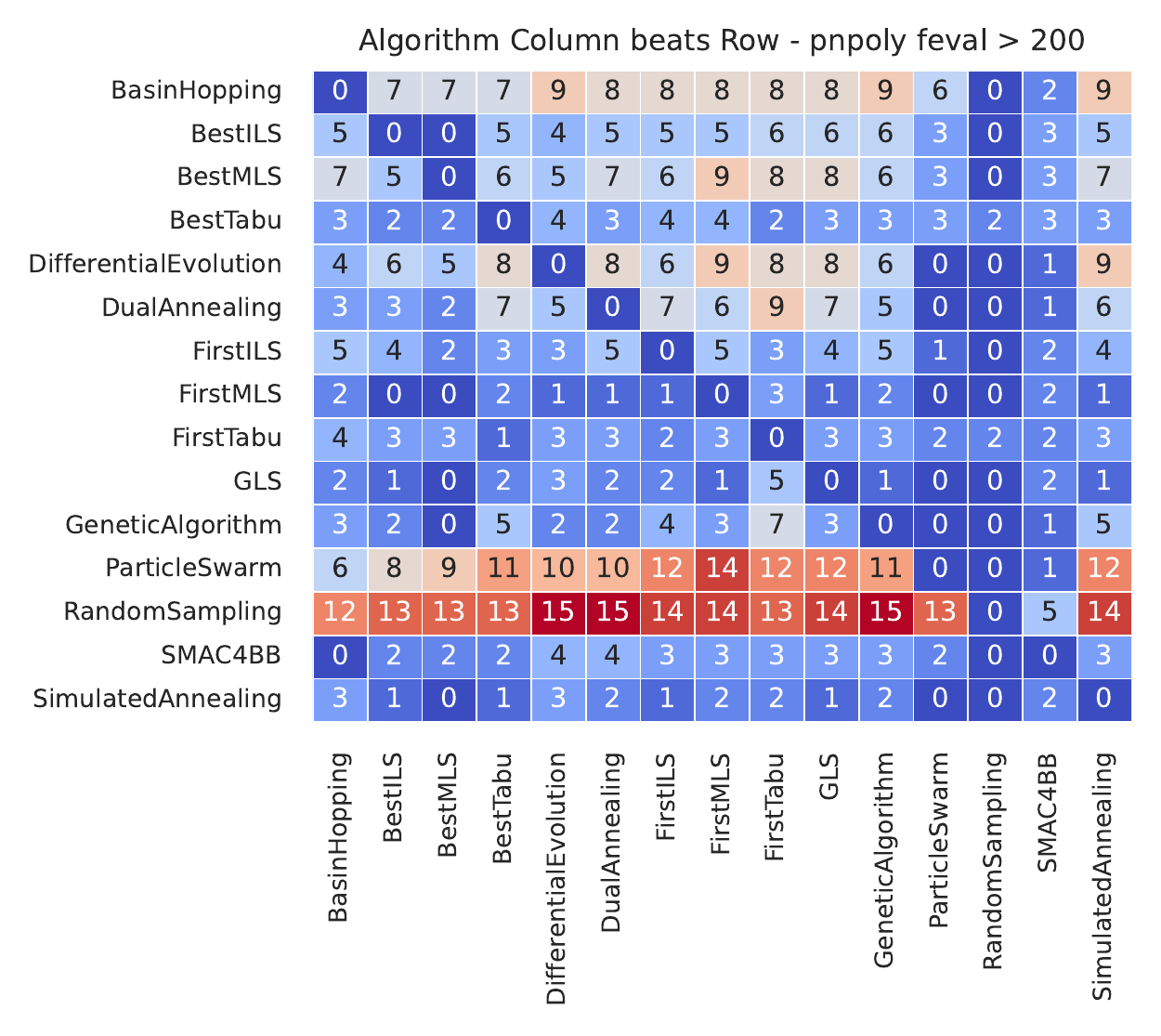}}
    }}
    \vspace{-0.15 cm}
    \caption{\scriptsize Heatmaps counting the occurrences when the column algorithm found statistically better solutions than the row algorithm for the (top) convolution, (middle) GEMM, and (botttom) PnPoly kernels. An occurrence is counted when 50 runs for a budget are statistically significantly better according to a two-sample independent t-test ($\alpha=0.05$). (Left): Heatmap for low $\leq200$ budgets, i.e., 25, 50, 100, and 200. (Right): Heatmap for mid and high $>200$ budgets, i.e., 400, 800, 1600. Algorithms with low values (blue) in their rows were not often beaten for those budgets, and algorithms with high values in their column (red) often beat other algorithms.}
    \label{fig:heatmapconv}
\end{figure*}

\begin{figure*}
    \centering
    \newcommand{\wdth}{0.34}
    \newcommand{\bspc}{0.0}
    \newcommand{\hspc}{-0.35}
    \captionsetup[subfigure]{labelformat=empty}
    \subfloat{\makebox[\bspc\textwidth][c]{
    \subfloat{\includegraphics[width=\wdth\textwidth]{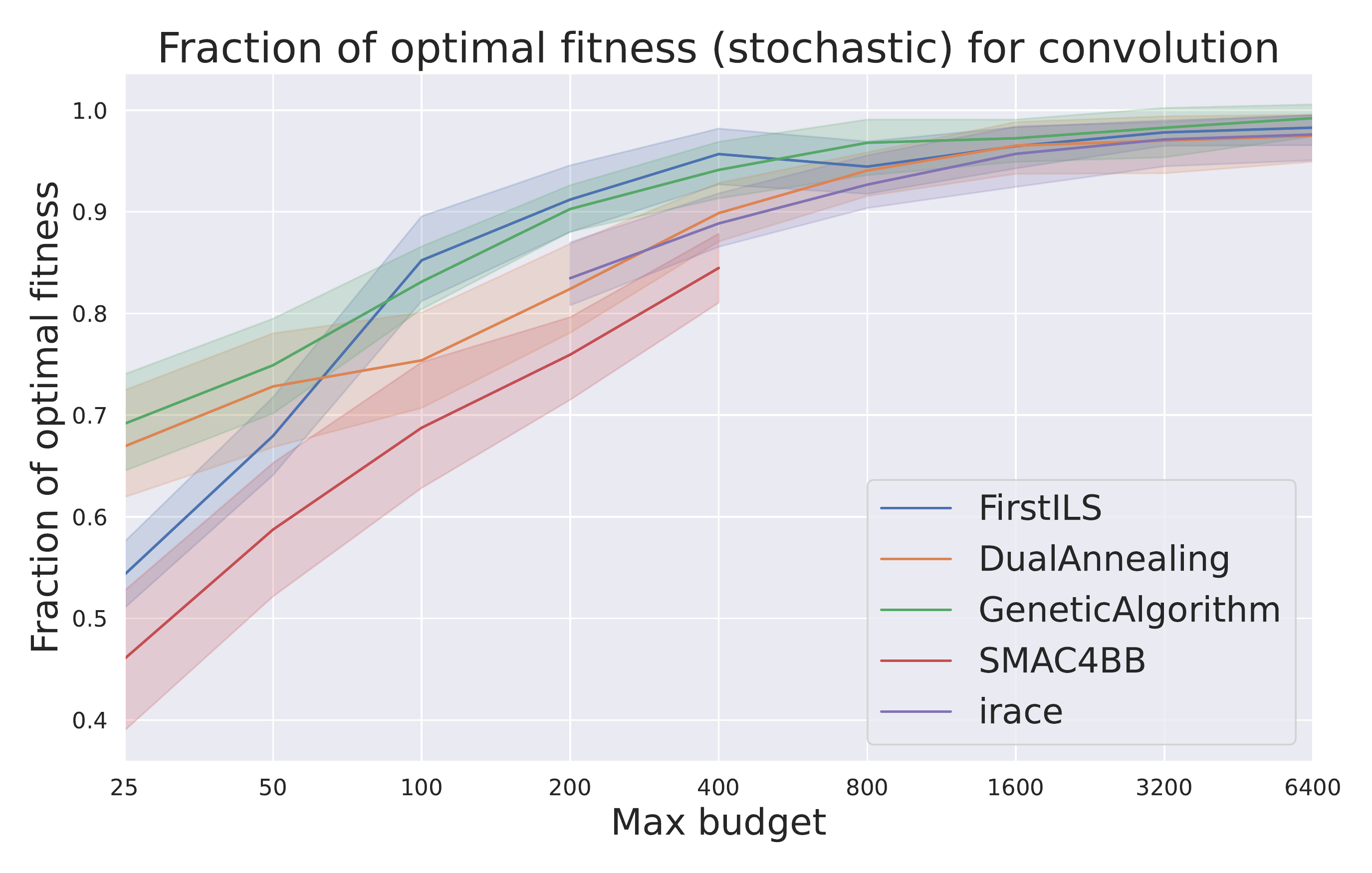}}
    \hspace{\hspc cm}
    \subfloat{\includegraphics[width=\wdth\textwidth]{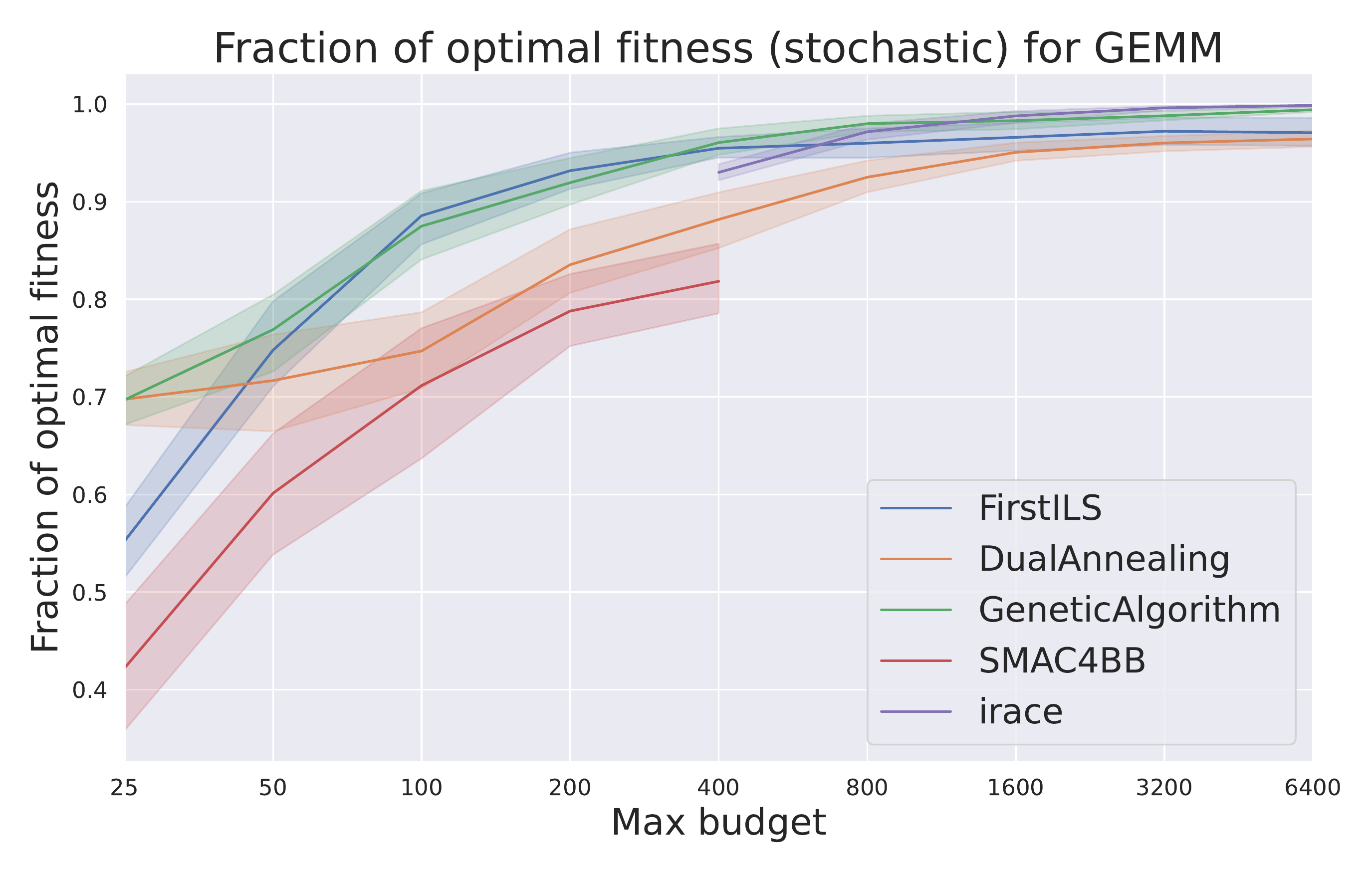}}
    \hspace{\hspc cm}
    \subfloat{\includegraphics[width=\wdth\textwidth]{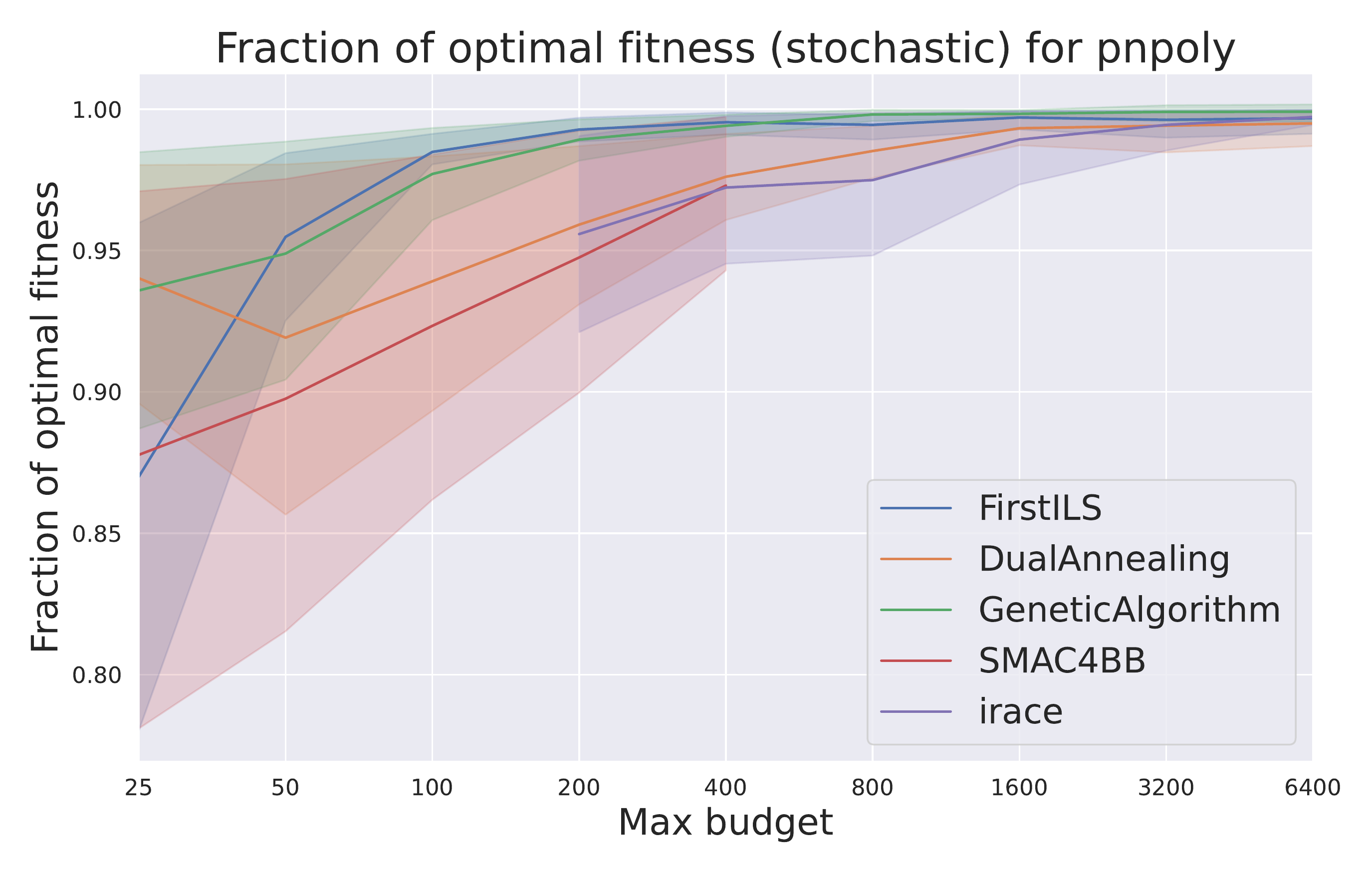}}
    }}
    \caption{\scriptsize Fraction of optimal runtime for max budget supplied over all GPUs. Each point is the mean fraction of optimal runtime found ($y$-axis) for each budget limit ($x$-axis) over all GPUs, with the shaded region indicating 95\% confidence interval. Left: convolution kernel. Middle: GEMM kernel. Right: PnPoly kernel (logarithmic $x$-axis).}
    \label{fig:stochastic_plot}
\end{figure*}

\subsection{Kernel tuning algorithm comparison}
To quantify which kernel tuning algorithms perform best for certain budgets, we can check whether an algorithm provided statistically significantly better results than others for a certain experiment. To do so, we use a \emph{two-sample independent t-test} with $\alpha=0.05$. For each GPU, kernel, and budget combination, we perform the $t$-test to see if an algorithm A performed significantly better than algorithm B. We subsequently combine the total number of ``wins'' for algorithm A across all GPUs (excluding those that were used for tuning).

We split the competitions into low-range (200 evaluations or fewer), and medium-range budgets. The competition tables at different splits can be found in the Appendix \ref{app:compheatmaps}. The full plots per GPU and algorithm can be found in Appendix \ref{app:separate_algos}. The results of these inter-algorithm competitions are given in the heatmaps displayed in Figure \ref{fig:heatmapconv}. The competition heatmaps display how often the column algorithm found a statistically better solution than the row algorithm in that budget range.

\vspace{1.0em}
\subsubsection{Kernel tuning: deterministic fitness}
In this section we present the results of our deterministic experiments, i.e., the algorithms have to minimize the runtime of a GPU kernel where the runtime is fixed at the mean of 32 runs.

\textbf{Low-range budget:} For 200 function evaluations or fewer, for convolution dual annealing was statistically better than all other algorithms for all GPU models (see column with ``DualAnnealing'' for convolution $\leq200$ function evaluations), with simulated annealing as second. For GEMM, basin hopping and dual annealing perform equally with dual annealing beating basin hopping 5 times, and basin hopping beating dual annealing 6 times. For PnPoly dual annealing was again best, followed by SMAC. We show the total number of wins and losses across all kernels and GPUs in Table \ref{tab:totalwins}. Here we see that for low budgets dual annealing has significantly more wins, and fewer losses than all other algorithms.

Interestingly, SMAC was second for PnPoly, in the best half of algorithms for convolution, but did not even beat random sampling for GEMM. We hypothesize this is either due to the number of variables to optimize for each of the three kernels (see Table \ref{tab:searchspace}), or the increasing fraction of fail fitnesses for these kernels. We hypothesize that the Bayesian optimizer could not fit a proper surrogate for GEMM with a low budget, and many failed compilations.

\textbf{Medium budget:} For more than 200 function evaluations, FirstMLS and GLS performed best for convolution, followed by FirstILS. For GEMM, FirstILS and simulated annealing performed best, followed by GLS. For PnPoly, FirstMLS is the strongest algorithm, followed by simulated annealing and GLS. As can be seen from Table \ref{tab:totalwins}, FirstILS and simulated annealing have the most number of total wins, and simulated annealing and genetic local search have the least number of losses.

\begin{table}
    \scriptsize
    \newcommand{\hght}{0.5em}
    \centering
    \begin{tabular}{c|c|c||c|c|}
     & \multicolumn{2}{c|}{\footnotesize Low budget} & \multicolumn{2}{c|}{\footnotesize Medium budget} \\
     & total wins & total losses & total wins & total losses \\
    \hline
    \hline
    \textbf{Basin hopping} & 403 & 204 & 131 & 393 \\
    \hline
    \textbf{Dual annealing} & \cellcolor{LimeGreen!50}\textbf{680} & \cellcolor{LimeGreen!50}\textbf{65} & 227 & 233\\
    \hline
    \textbf{Differential evolution} & \cellcolor{LimeGreen!50}426 & 192 & 150 & 347\\
    \hline
    \textbf{PSO} & 290 & 327 & 136 & 368\\
    \hline
    \textbf{FirstILS} & 352 & 233 & \cellcolor{LimeGreen!50}\textbf{361} & 77\\
    \hline
    \textbf{BestILS} & 125 & 472 & 257 & 126\\
    \hline
    \textbf{FirstTabu} & 148 & 430 & 255 & 183\\
    \hline
    \textbf{BestTabu} & 35 & 677 & 220 & 185\\
    \hline
    \textbf{FirstMLS} & 317 & 205 & \cellcolor{LimeGreen!50}341 & \cellcolor{LimeGreen!50}64\\
    \hline
    \textbf{BestMLS} & 143 & 466 & 226 & 174\\
    \hline
    \textbf{Simulated annealing} & \cellcolor{LimeGreen!50}412 & \cellcolor{LimeGreen!50}150 & \cellcolor{LimeGreen!50}360 & \cellcolor{LimeGreen!50}\textbf{59}\\
    \hline
    \textbf{Genetic local search} & 320 & 216 & 329 & \cellcolor{LimeGreen!50}\textbf{59}\\
    \hline
    \textbf{Genetic algorithm} & 376 & \cellcolor{LimeGreen!50}162 & 201 & 207\\
    \hline
    \textbf{SMAC} & 356 & 344 & 48 & 145\\
    \hline
    \textbf{Random sampling} & 222 & 463 & 7 & 629\\
    \hline
    \end{tabular}
    \caption{\scriptsize Total number of wins: sum of occurrences when the algorithm found statistically better solutions than other algorithms (summed over all kernels). A win (and corresponding loss for the other algorithm) is counted when 50 runs for a budget are statistically significantly better according to a two-sample independent t-test ($\alpha=0.05$). Low budget is for $\leq200$ budgets, i.e., 25, 50, 100, and 200. Medium budget is for $>200$ budgets, i.e., 400, 800, 1600. The top 3 cells are coloured green.}
    \label{tab:totalwins}
\end{table}

\vspace{1.0em}
\textbf{Additional remarks:} In general, best-improvement local search algorithms performed significantly worse than the first-improvement variants. In fact, for the low range, they proved statistically worse than random sampling for PnPoly, and in general have fewer wins and more losses. This can be explained due to the fact that exploring all the neighbours before taking a step costs many evaluations, and leads to only exploring a single neighbourhood for low budgets. For the population-based methods, GLS is the best performing algorithm for medium budgets, but does worse than differential evolution and GA for low budgets. PSO performed significantly worse.

Interestingly, dual annealing, which works on real-valued solution vectors, performs well for low budgets. It seems that the mapping from $[0,1]^n$ to discrete space does not prevent dual annealing from finding strong solutions quickly. One of the main drawbacks of using continuous algorithms is that if a continuous algorithm updates its real-valued solution vector, it could mean that it does not actually update the discrete solution vector it is mapped to. However, since our algorithms cache previously visited solutions (only for the deterministic experiments), such redundant optimization steps do not cost any budget. We think this may negatively impact gradient-based algorithms as the subroutines Powell and COBYLA, which do not require derivatives to be known, are the selected solvers for dual annealing during hyperparameter tuning.

As a final remark, we notice that SMAC performs poorly in the medium budgets. Note that SMAC only has $1/3$ as many data points in the medium budget since we do not perform the 800 and 1600 budget experiments for SMAC. Nevertheless the algorithm performs poorly on the 400 budget compared to other methods. It seems that SMAC is unsuccessful in fitting a meaningful surrogate model for kernel tuning. This could be due to the deterministic setup of this experiment, or due to the high number of fail configurations with ``infinite'' fitness.

\vspace{1.0em}
\subsubsection{Kernel tuning: stochastic fitness}
For the stochastic experiments the algorithms have to minimize the runtime of a GPU kernel where the runtime is a random draw from the 32 timings. In addition to running SMAC and irace, we also run FirstILS, GA and dual annealing which did well for deterministic fitnesses. We remark that irace throws an error if the budget is too small with respect to the number of variables, and therefore starts at a budget of 200 for convolution and PnPoly, and 400 for GEMM.

The experimental results are shown in Figure \ref{fig:stochastic_plot}. Here we aggregate the results per kernel for all GPU models by showing the mean fraction of optimum (and 95\% confidence interval) for a given max budget. We see that GA and dual annealing are best for low budgets in the stochastic experiments. FirstILS does well for budgets $\geq 100$. Irace is the best method for GEMM with budgets $\geq 800$, but for convolution and pnpoly irace is not as good as GA, dual annealing, and FirstILS.

SMAC consistently achieves a lower fraction of optimality than the competing algorithms across kernels and budgets. Again, we hypothesize that this is because of the high number of fail fitnesses in the search spaces (see Table \ref{tab:searchspace}). This makes it hard for the Bayesian optimizer to fit a meaningful surrogate.

\textbf{Stochastic or deterministic:} Overall, we notice that higher budgets are necessary to find good solutions for the stochastic experiments than in the deterministic case. This leads to a higher overall tuning time. We therefore recommend to treat GPU kernel tuning as a deterministic optimization problem, with the mean runtime as fitness. The added stochastic information does not appear to allow SMAC or irace to consistently outperform conventional black-box algorithms. This could be because the runtime does not vary much; the average (normalized) runtime and standard deviation is $1.000\pm 0.011$. Second, the high number of failure configurations could confuse models that try to learn stochastic information.

\begin{figure*}
    \centering
    \newcommand{\wdth}{0.9}
    \includegraphics[width=\wdth\textwidth]{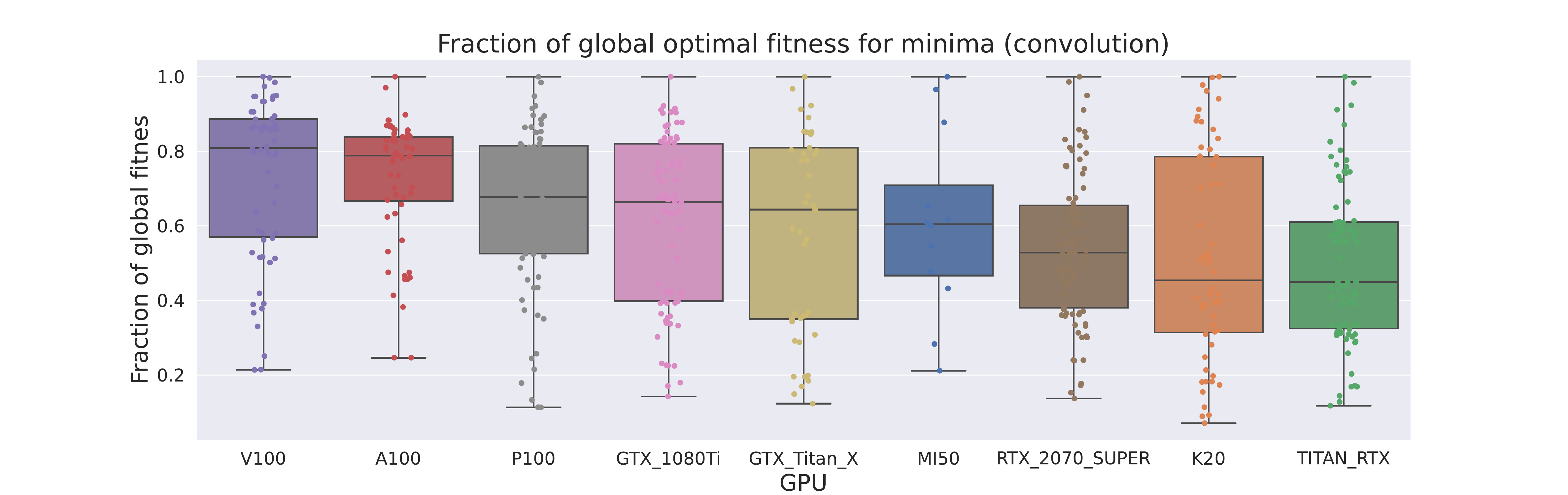}
    \caption{\scriptsize Fraction of optimal fitness of local minima for each GPU model, for the convolution kernel. The box plots shows the median line, the box designates the quartiles, and the whiskers the full extend of the distribution. Additionally, a scatter plot of the fitness for each local minimum is shown. The GPUs are ordered in descending median fraction of optimal fitness from left to right.}
    \label{fig:boxstrip_frac_minima_conv_pergpu}
\end{figure*}

\begin{figure*}
    \centering
    \newcommand{\wdth}{0.33}
    \newcommand{\bspc}{0.0}
    \newcommand{\hspc}{-0.22}
    \captionsetup[subfigure]{labelformat=empty}
    \subfloat{\makebox[\bspc\textwidth][c]{
    \subfloat{\includegraphics[width=\wdth\textwidth]{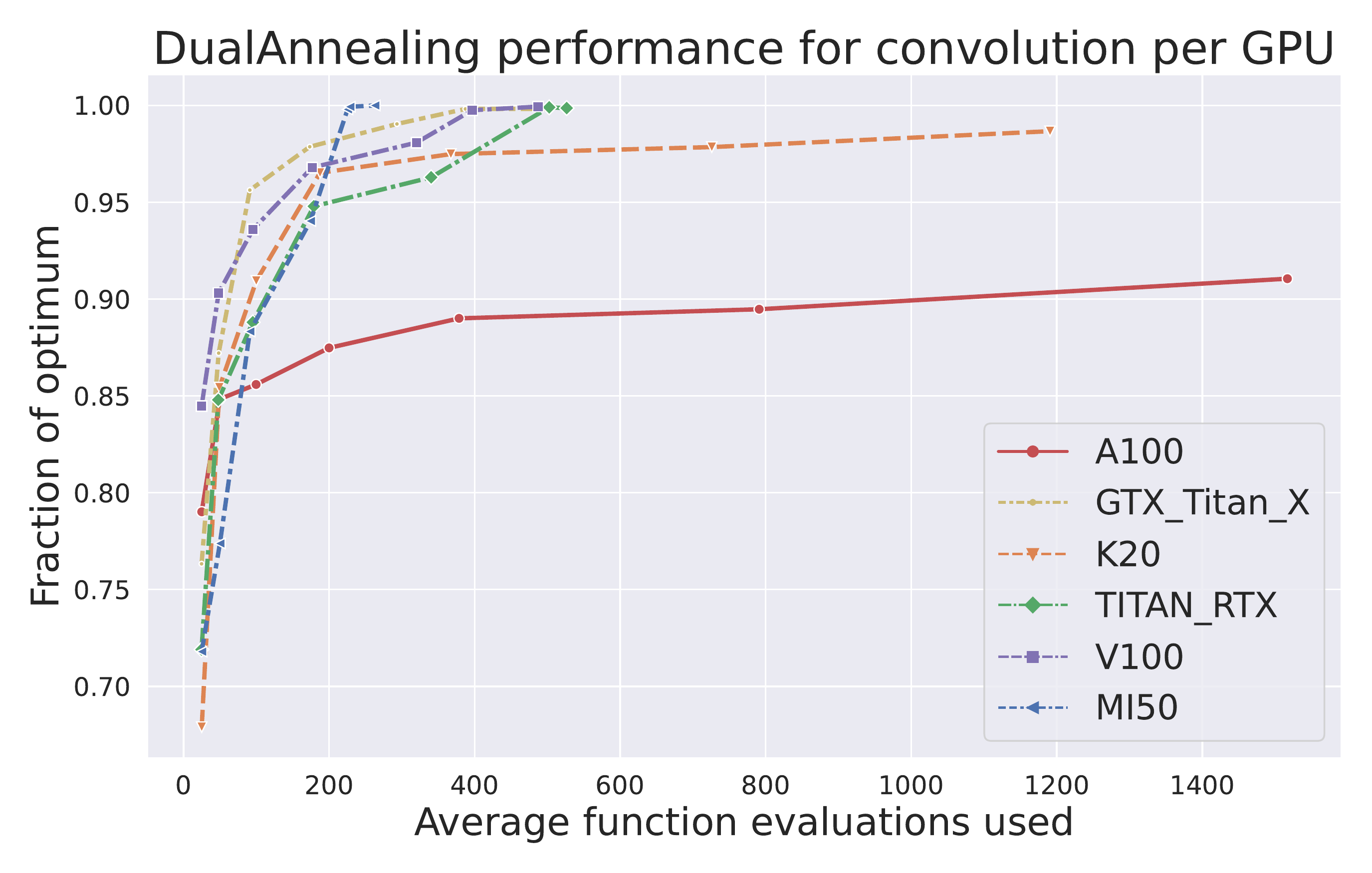}}
    \hspace{\hspc cm}
    \subfloat{\includegraphics[width=\wdth\textwidth]{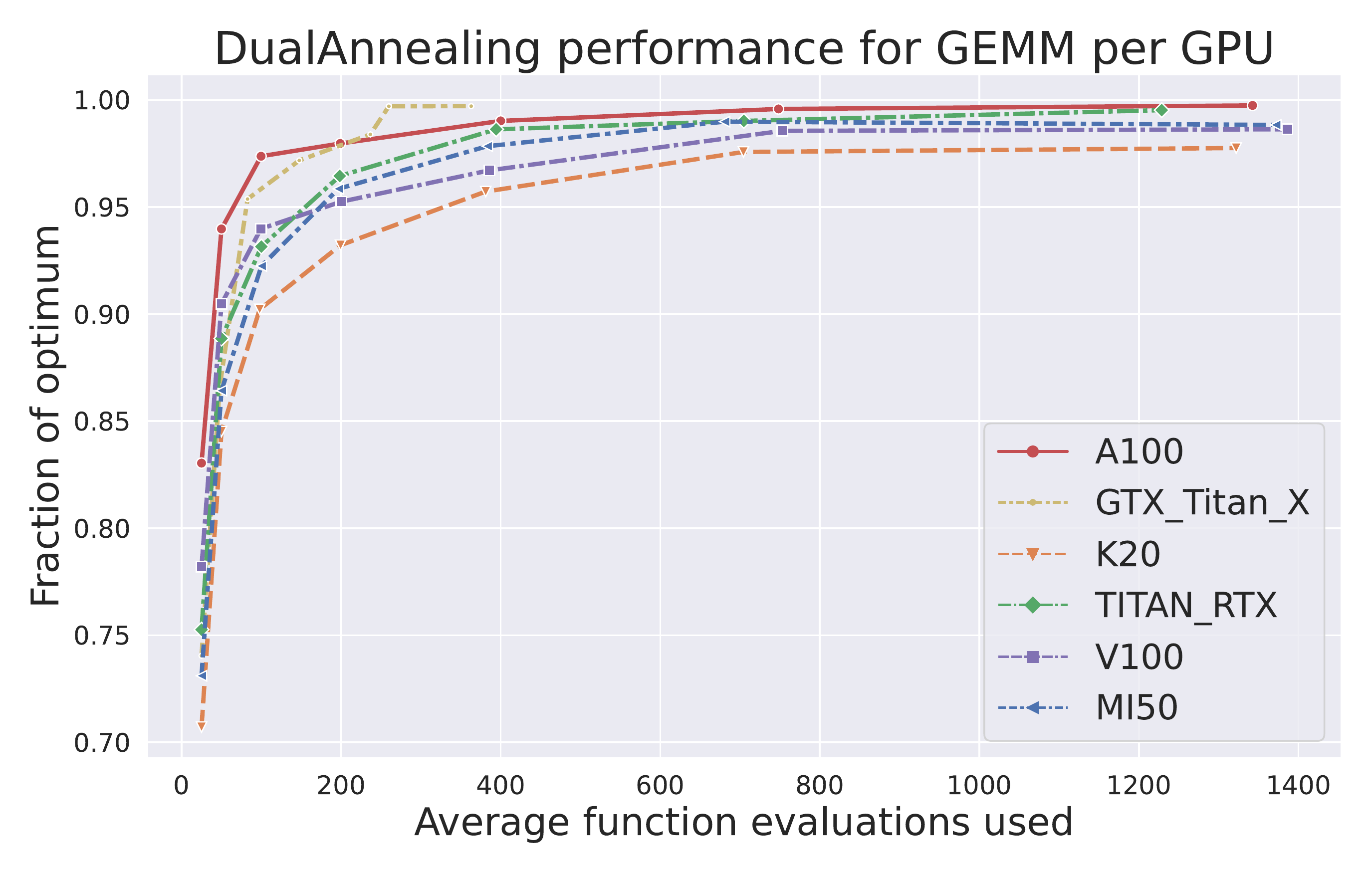}}
    \hspace{\hspc cm}
    \subfloat{\includegraphics[width=\wdth\textwidth]{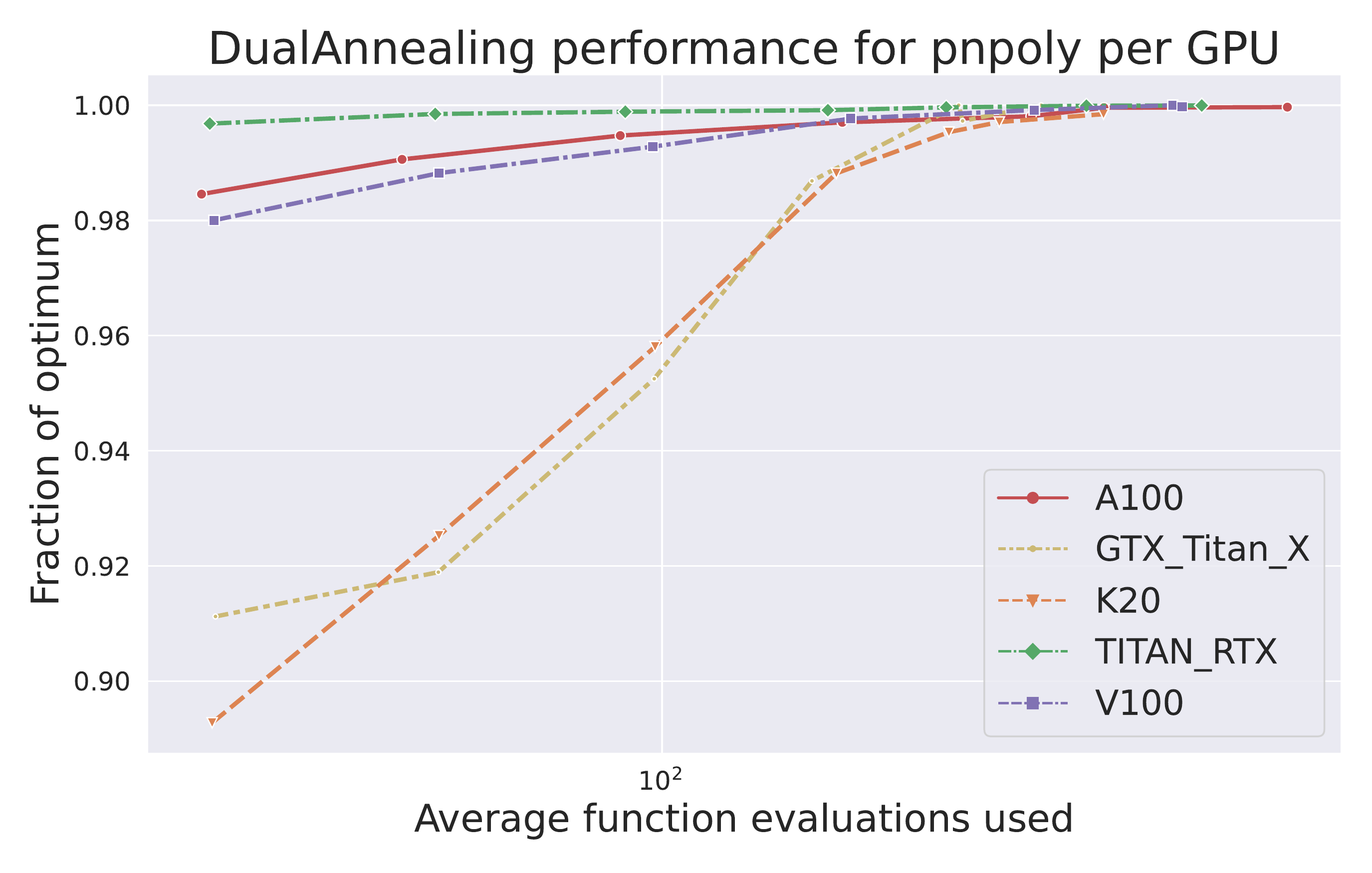}}
    }}
    \caption{\scriptsize \textbf{Dual annealing:} Fraction of optimal runtime for different budgets (per GPU). Each point is the average fraction of optimal runtime found ($y$-axis) for each budget, with respect to the average number of evaluations actually used ($x$-axis) for that budget. Function evaluations used counts only the unique settings that were visited. Left: convolution kernel. Middle: GEMM kernel. Right: PnPoly kernel (logarithmic $x$-axis).}
    \label{fig:dsa_pergpu}
    \vspace{-0.5cm}
\end{figure*}

\begin{figure*}
    \centering
    \newcommand{\wdth}{0.34}
    \newcommand{\bspc}{0.0}
    \newcommand{\hspc}{-0.35}
    \captionsetup[subfigure]{labelformat=empty}
    \subfloat{\makebox[\bspc\textwidth][c]{
    \subfloat{\includegraphics[width=\wdth\textwidth]{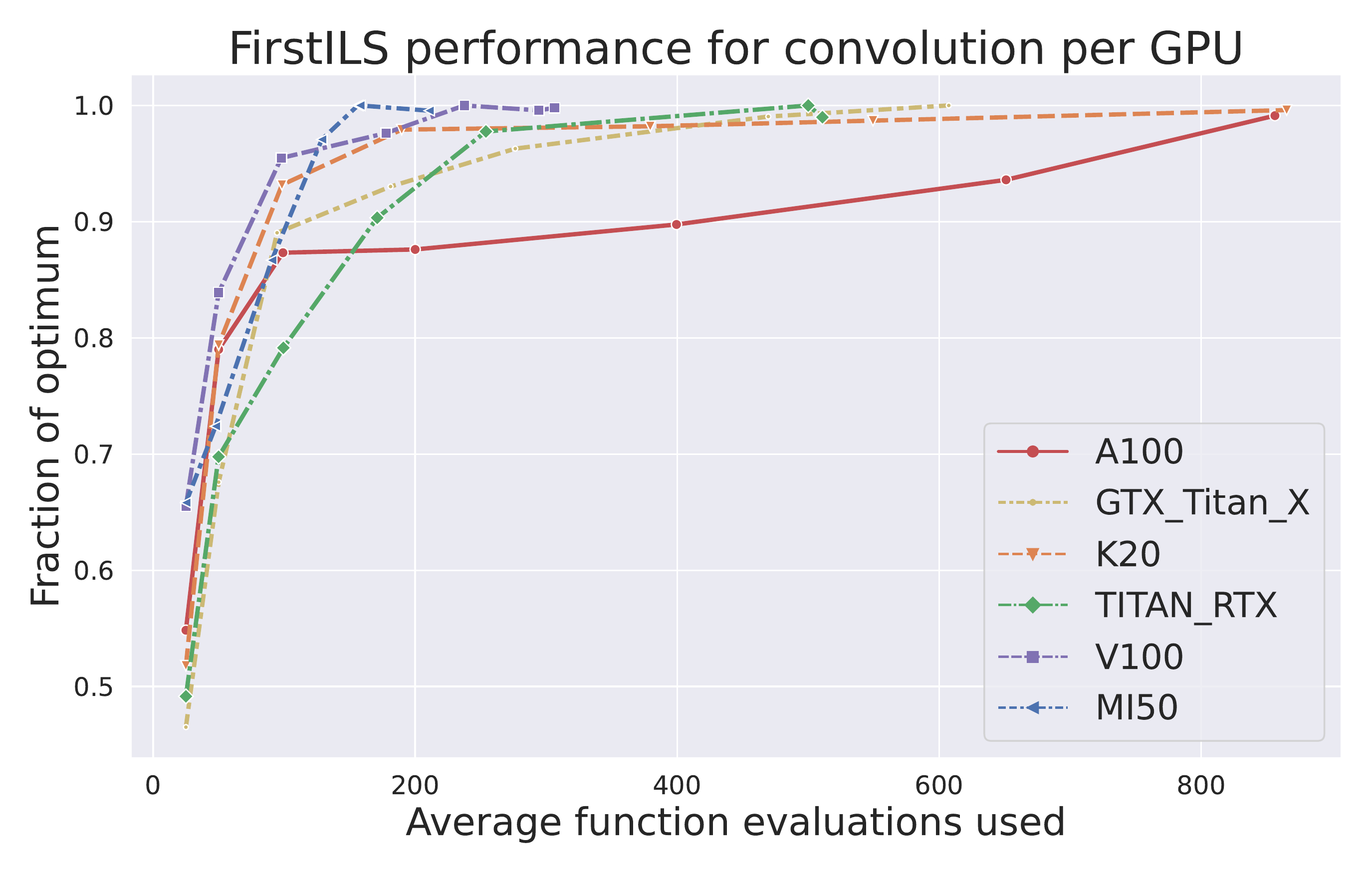}}
    \hspace{\hspc cm}
    \subfloat{\includegraphics[width=\wdth\textwidth]{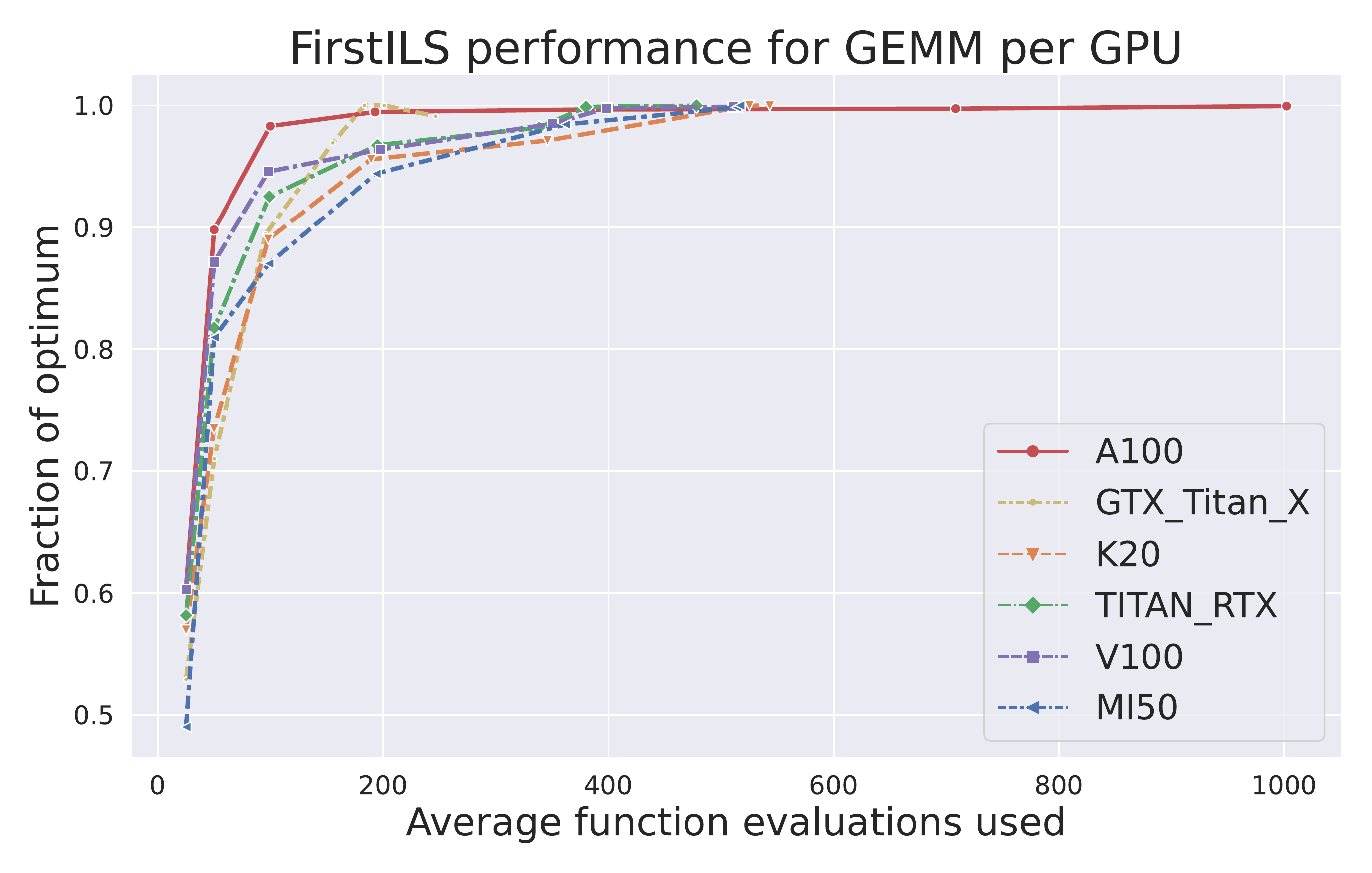}}
    \hspace{\hspc cm}
    \subfloat{\includegraphics[width=\wdth\textwidth]{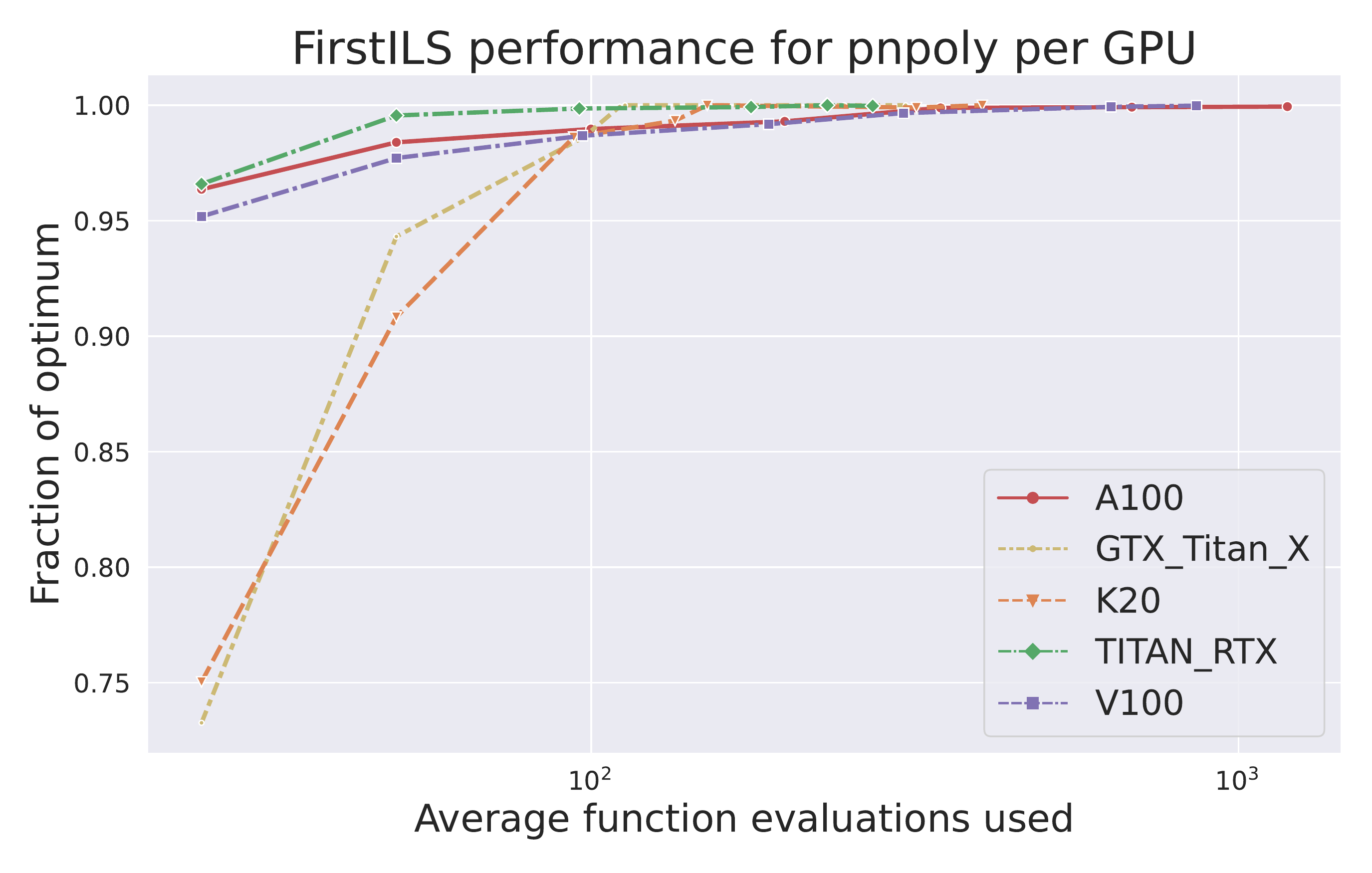}}
    }}
    \caption{\scriptsize \textbf{FirstILS:} Fraction of optimal runtime for different budgets (per GPU). Each point is the average fraction of optimal runtime found ($y$-axis) for each budget, with respect to the average number of evaluations actually used ($x$-axis) for that budget. Function evaluations used counts only the unique settings that were visited. Left: convolution kernel. Middle: GEMM kernel. Right: PnPoly kernel (logarithmic $x$-axis).}
    \vspace{-0.2cm}
    \label{fig:greedyils_pergpu}
\end{figure*}

\section{Quantifying GPU tuning difficulty}
\label{sec:spaceresults}

In this section we want to gain insight into the difficulty of the GPU kernel tuning optimization problem, and quantify kernel spaces according to tuning difficulty. When attempting to understand why certain GPU kernel spaces appear difficult to optimize we found that relatively simple metrics do not coincide with our experimental results. We outline discrepancies between an intuitive simple metric and our experimental results, and introduce a novel refined approach that does correlate with our results.

\begin{figure*}
    \centering
    \newcommand{\wdth}{0.46}
    \includegraphics[width=\wdth\textwidth]{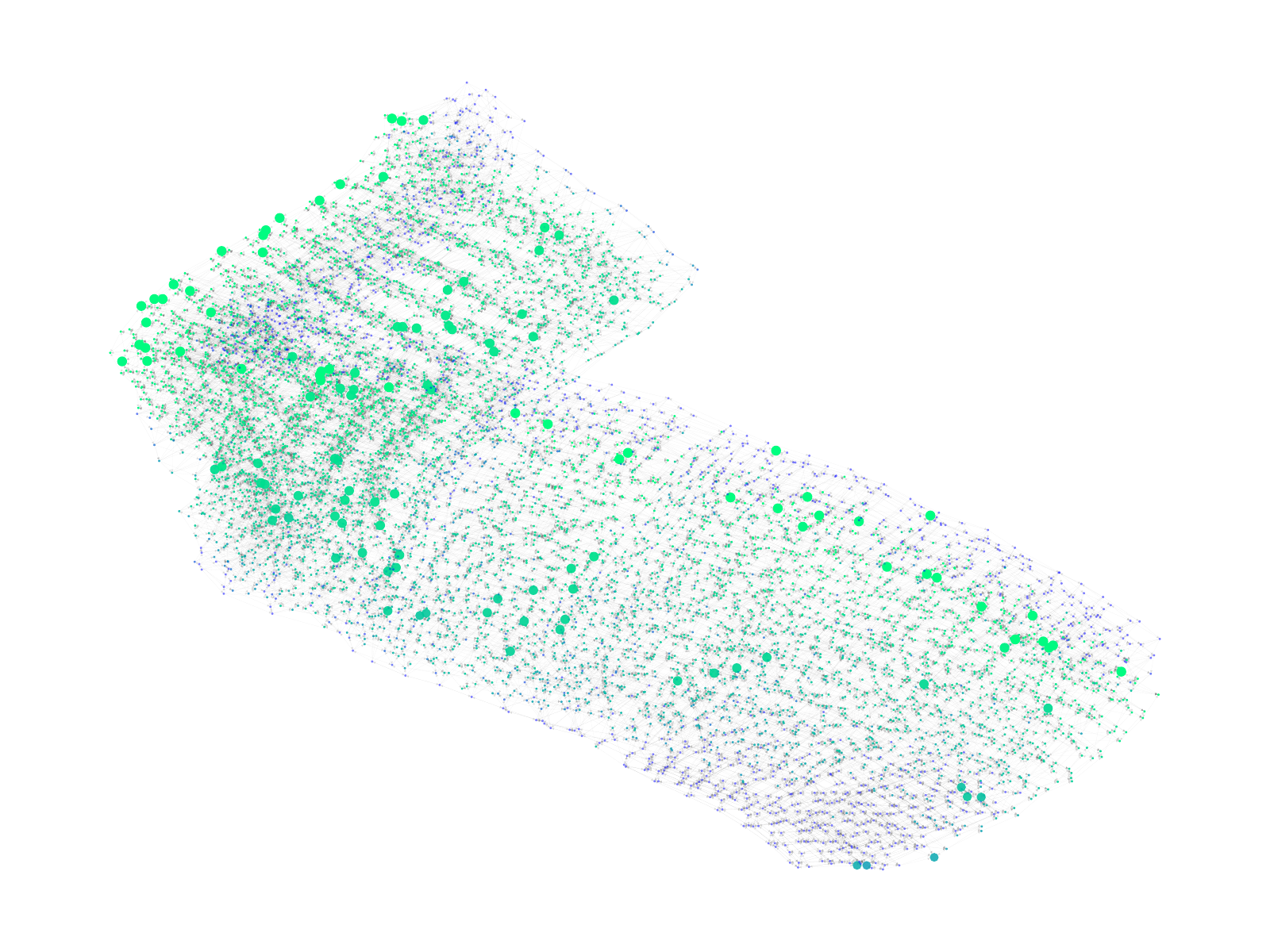}
    \includegraphics[width=\wdth\textwidth]{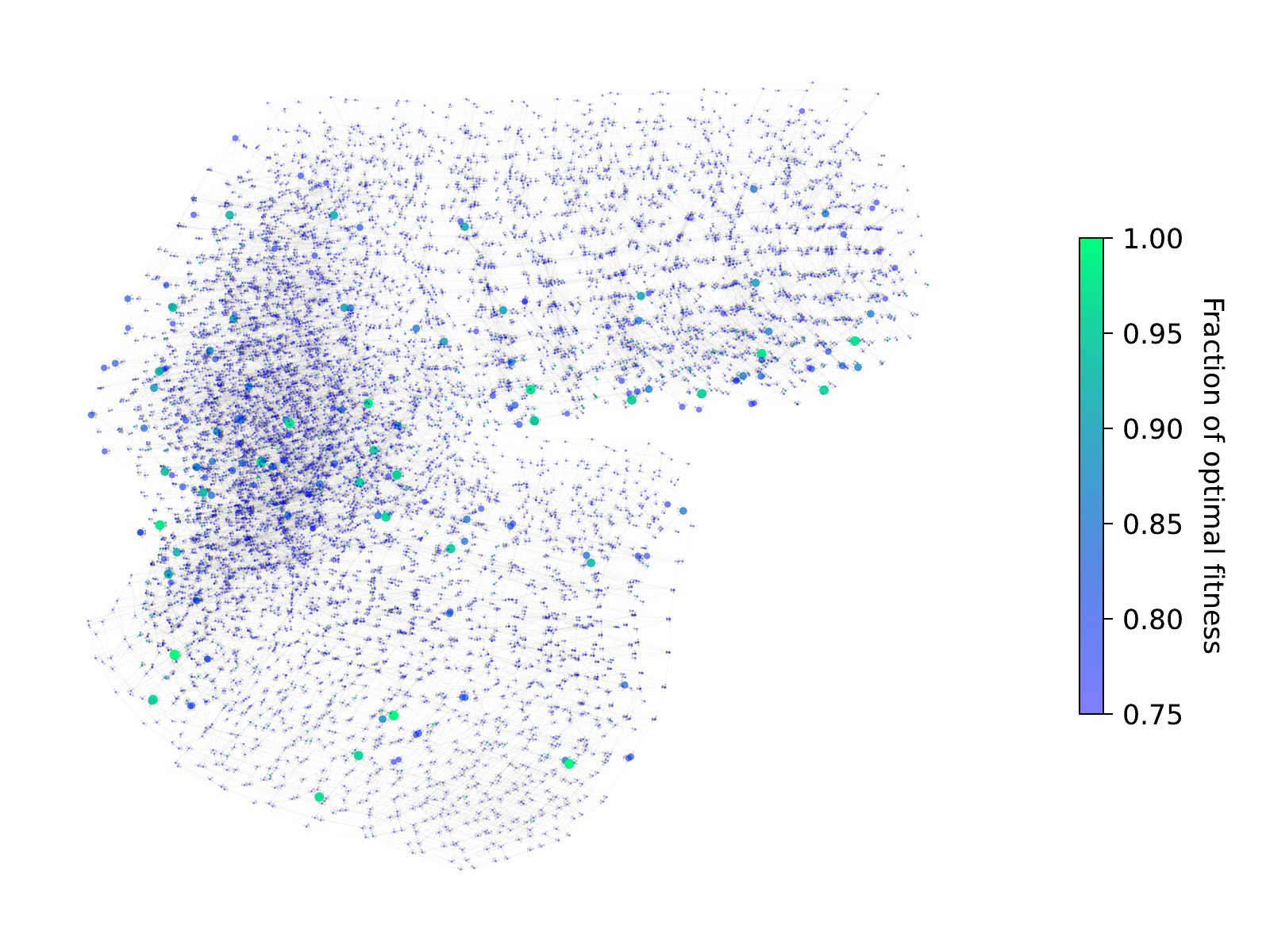}
    \vspace{-0.5cm}
    \caption{\scriptsize Fitness flow graphs of PnPoly kernel search spaces of the (left) NVidia Titan RTX, and (right) NVidia GTX 1080Ti. Each node is a point in the search space. There is a directed edge between neighbouring points from higher to lower fitness. Points are coloured within a fitness range of $+25\%$ with respect to the global minimal fitness (global minimum in green), i.e., each point is coloured by its fraction of optimal fitness, and points with a fraction below 0.75 are given the same colour. Local minima are represented as larger nodes.}
    \vspace{-0.5cm}
    \label{fig:ffg_MI50_conv}
\end{figure*}

\begin{figure}
    \centering
    \newcommand{\wdth}{0.485}
    \includegraphics[width=\wdth\textwidth]{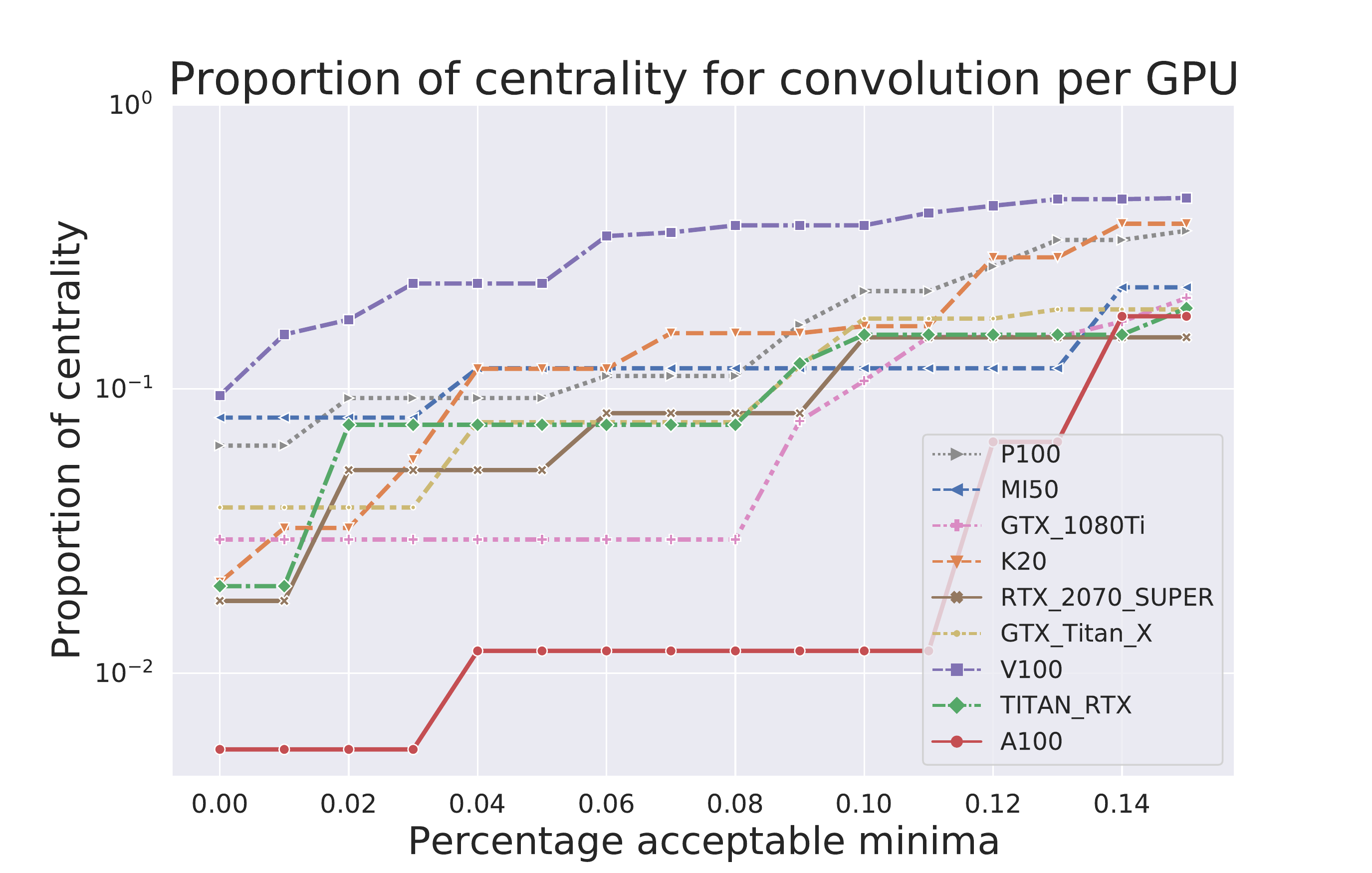}
    \includegraphics[width=\wdth\textwidth]{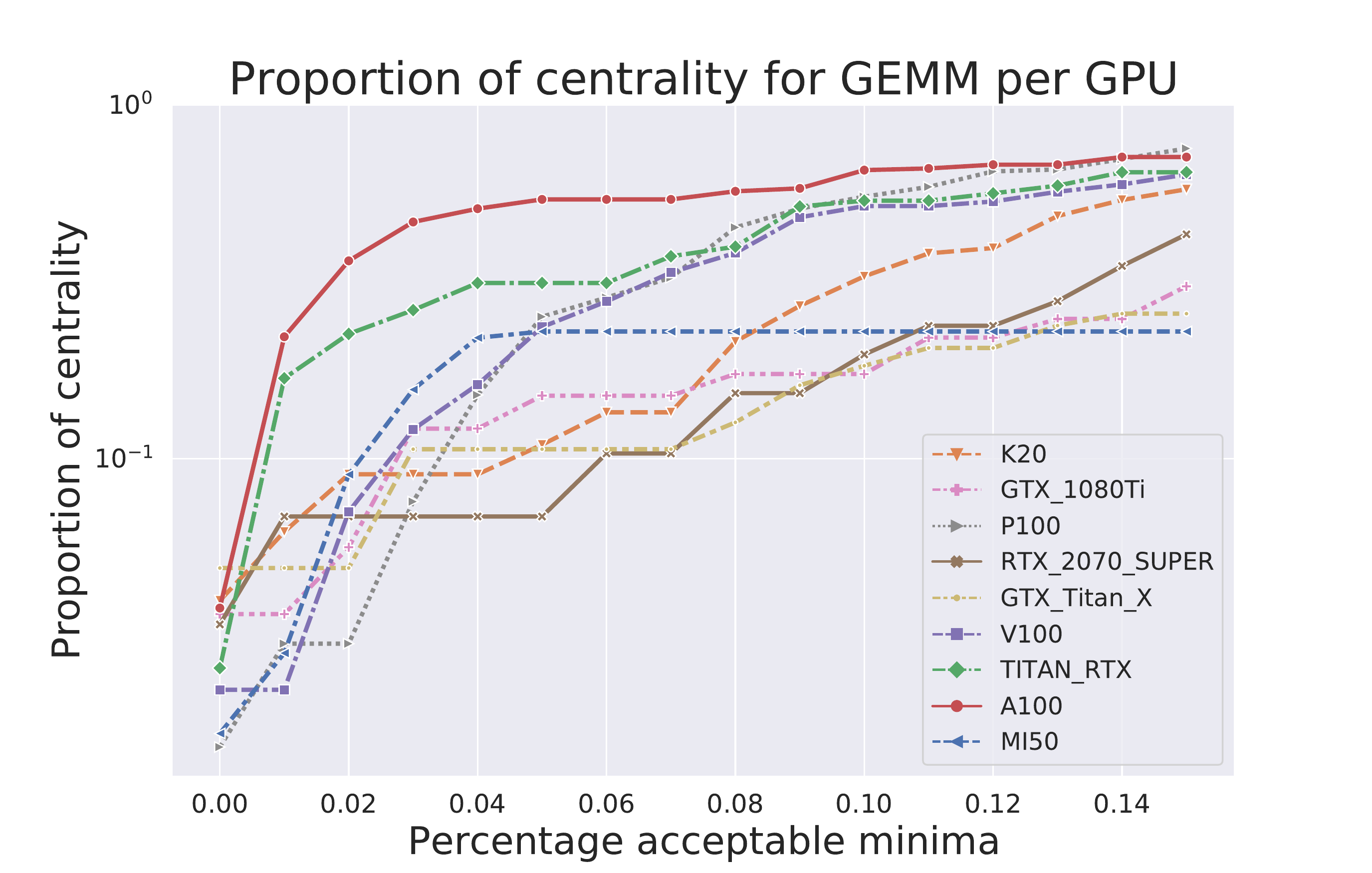}
    \includegraphics[width=\wdth\textwidth]{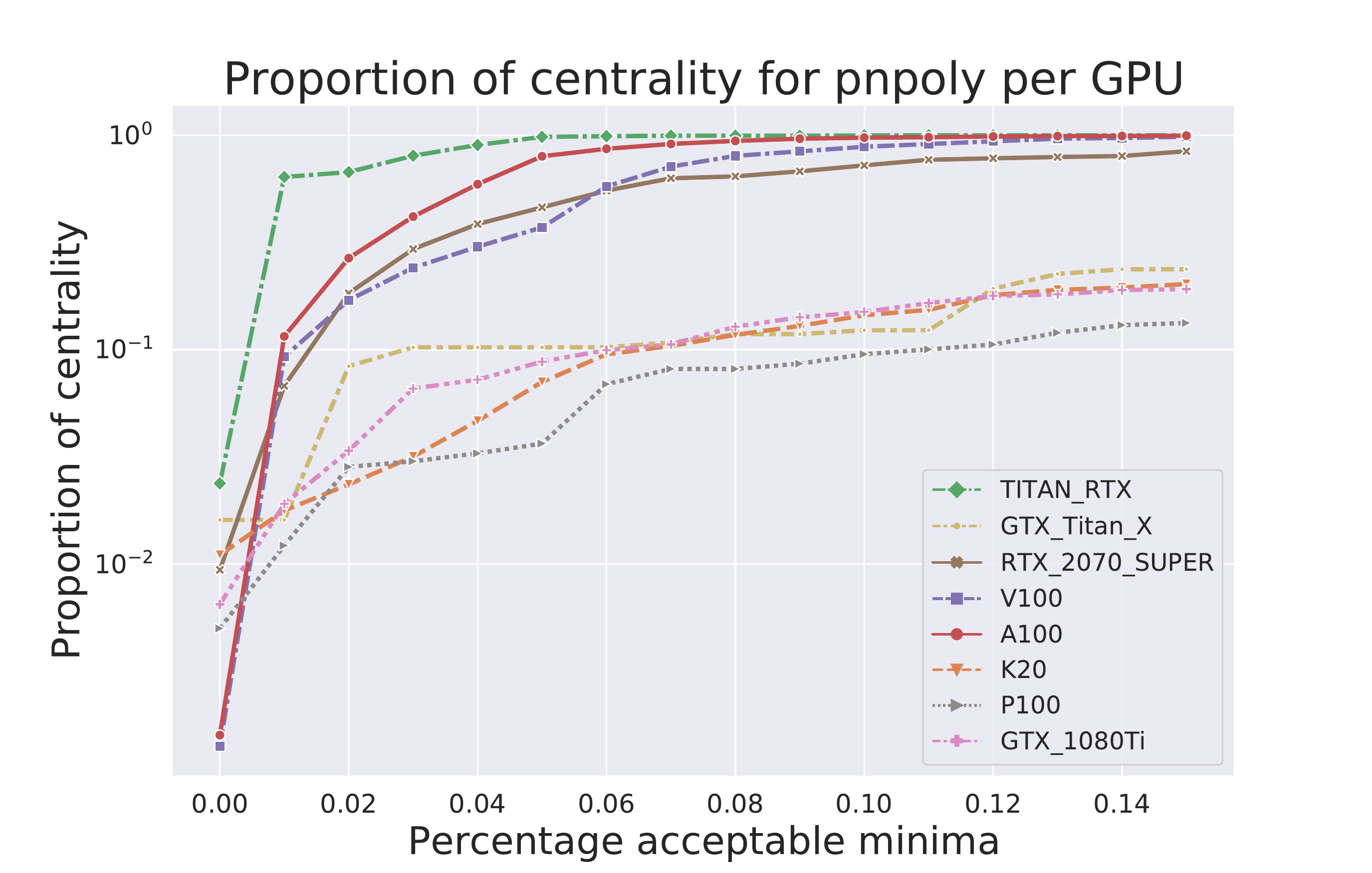}
    \caption{\scriptsize Proportion of centrality for fitness flow graphs for each GPU. The proportion of centrality is computed by taking the sum of PageRank centrality for local minima within $p$\% of the optimal fitness, divided by the total PageRank centrality of all local minima. From top to bottom are convolution, GEMM, and PnPoly kernel.}
    \label{fig:plotcentr}
\end{figure}

\subsection{Naive metric: fraction of optimal fitness of local minima}
\textbf{Method:} As an example of a simple metric that intuitively could explain the results, we consider the fraction of optimal fitness of local minima. For a minimum $x_i$, and optimal fitness $f_{opt}$, we can consider the \emph{fraction of optimal fitness} of the minimum $f_{opt} / f(x_i)$. In this case, we divide the global minimal runtime by the runtime of the minima.

\textbf{Results:} In Figure \ref{fig:boxstrip_frac_minima_conv_pergpu} we show a scatter plot of the fraction of optimal fitness for the local minima for the convolution kernel (per GPU model). According to this distribution, the V100 and A100 GPUs have the closest to optimal median fitness for local minima. This means that an algorithm that randomly explores local minima with equal probability will obtain the closest to optimal runtime for these kernels. 

\textbf{Analysis:} To empirically check how difficult the GPU kernels are to tune, we can plot the fraction of optimal fitness that optimization algorithms managed to achieve for certain budgets. If $f_j$ is the lowest fitness found for a single run for some budget $p$, a point in the plot is the average over 50 runs computed as $\tilde{f}_p:=(1/50)\cdot\sum_{j=1}^{50}(f_{opt} /f_{j,p})$. In Figures \ref{fig:dsa_pergpu} and \ref{fig:greedyils_pergpu} we plot $\tilde{f}_p$ for dual annealing and FirstILS.
We chose dual annealing and FirstILS as they represent the strongest algorithms for low, and medium budgets respectively. 

We see that for convolution on the A100 GPU both algorithms returned solutions which were furthest away from the optimum, while for the V100 both optimizers return close to optimal solutions for few function evaluations. These observations are opposite to what would be expected on the basis of Figure \ref{fig:boxstrip_frac_minima_conv_pergpu}. Hence, the distribution of fitness of the local optima does not properly explain tuning difficulty.

\subsection{Refined approach: Fitness flow graphs and PageRank}
\label{sec:critgraphs}
\textbf{Method:} A more refined metric to quantify GPU tuning difficulty may be to compute how likely local search algorithms terminate in local minima. For this purpose, we introduce the \emph{fitness flow graph} (FFG), which contains all points in the search space, and creates a directed edge to a neighbouring point if the neighbour has lower fitness. This means that a random walk across the FFG mimics the behaviour of a randomized first-improvement local search algorithm. The expected proportion of arrivals of each minimum then gives a metric for weighting reachability of each minimum. We show two example FFGs in Figure \ref{fig:ffg_MI50_conv}.

To compute the likelihood of arrival per local minima, we compute the PageRank \emph{node centrality}, which was originally used to determine the relevance of a webpage \cite{brin1998anatomy, page1999pagerank}. Let $A_G$ be the adjacency matrix of a directed graph $G$, rescaled such that each column adds up to 1. Essentially, this means that for every node, the column is a probability vector of visiting adjacent nodes with equal likelihood. The PageRank values are then the values of the dominant right eigenvector of $A_G$. For an FFG, this means that the PageRank value of a local minimum is the probability of arriving in that minimum after a long random walk through the graph.

As a measure of difficulty we consider how likely a certain subset of ``suitably good'' local minima are to be visited by a local search algorithm relative to the rest.
Suppose that $f_{opt}$ is the optimal fitness, and let $L(X)$ be the set of local minima of $X$. Given a proportion $p$, we take the set of nodes $L_p(X)$ consisting of local minima with fitness less than $(1+p)f_{opt}$ (for minimization problems, otherwise $(1-p)f_{opt}$). For a centrality function $c_G$, we define the $p$-\emph{proportion of centrality}
\begin{equation}
    C_p(G,X)=\frac{\sum_{x\in L_p(X)} c_G(x)}{\sum_{x\in L(X)} c_G(x)}.
\end{equation}

\textbf{Results:} The proportion of centrality for strong local minima for each FFG is shown in Figure~\ref{fig:plotcentr}. We calculate the proportion of centrality for different acceptance percentages with respect to the global minimum of $p=0, 1, 2, \ldots, 15\%$.

Revisiting the A100 and V100 convolution kernel comparison, we see that the proportion of centrality matches the experimental observations for dual annealing and FirstILS. Figure \ref{fig:plotcentr} shows that the NVidia V100 has the most central local minima, whereas the A100 has the least central local minima. For the GEMM and PnPoly kernels, Figures \ref{fig:dsa_pergpu} and \ref{fig:greedyils_pergpu} align with the expectations based on the proportion of centrality. For example, the group of PnPoly kernels with lowest proportion of centrality (P100, GTX Titan X, K20, GTX 1080Ti) are indeed the hardest to tune for both algorithms.

One exception is the K20 GEMM kernel, where proportion of centrality does not entirely reflect the perceived difficulty for dual annealing. This suggests that the proportion of centrality may correlate with GPU tuning difficulty better for certain optimization algorithms. This is to be expected since the PageRank centrality on the FFG in expectation mimics the performance of randomized first-improvement local search. Algorithms that are substantially different than first-improvement local search will therefore also correlate less with the expected difficulty on the basis of proportion of centrality.



\vspace{1.0em}
\textbf{Analysis:} Overall, the experimental results suggest that the proportion of centrality is a suitable metric for estimating tuning difficulty for GPU kernels. By using FFGs and the PageRank algorithm, we are able to observe kernel differences that were otherwise unknown. For example, both the A100 and V100 convolution kernels have few outlier minima with a close to optimal fitness. In fact, the existence of only a few kernel configurations that lead to large increases in performance is a general property of certain GPU kernels \cite{kerneltuner}. Crucially however, the likelihood of local search algorithms arriving in such minima differs greatly between the A100 and V100. The proportion of centrality of an FFG gives us a tool to quantify this likelihood. However, further research is necessary to quantitatively determine how well our proposed metric correlates with GPU tuning difficulty.


As a final remark on kernel differences, the experimental results shows that the difficulty of tuning a particular kernel can greatly differ from one GPU to the next, and that these changes do not appear to be correlated with release time of the models. The A100 is the most recent GPU in our set, while the K20 is the oldest. For GEMM and PnPoly, we can say that it has become easier to tune these kernels with more recent GPUs, but the convolution kernel has become more difficult to tune, except on the V100.

\section{Conclusion}
\label{sec:conclusion}
In this paper, we have investigated which optimization algorithms produce the fastest GPU kernel configurations across different tuning-time ranges. To do so, we analyzed 26 GPU kernel spaces for 9 GPUs. We computed sets of optimal hyperparameters for GPU tuning for each optimization algorithm. From among the tested algorithms in this set of experiments, we conclude that dual annealing performs best as GPU kernel tuner when a limited amount of function evaluations is desirable. When more evaluations are possible, first-improvement local searchers such as FirstILS proved the best GPU kernel tuners. Using these algorithms, we are convinced that GPU programmers can reliably auto-tune GPU kernels to close to optimal runtime while requiring relatively few re-compilations of the code. Furthermore, we conclude that treating GPU tuning as a deterministic optimization problem is preferred over treating the runtime as a stochastic variable.

We showed that the basic metric of fraction of optimality of local minima is not suitable for explaining the results observed in the experimental benchmarks. To make steps towards a metric for tuning difficulty, we introduced the concept of fitness flow graphs, and proportion of centrality. Our results suggest that the proportion of centrality can be used to quantify tuning difficulty. For future work, in cases where exhaustive exploration is infeasible, perhaps a procedure to dynamically update the proportion of centrality of FFGs can be used. Such dynamic estimates of tuning difficulty could be used for automatic algorithm selection within frameworks such as Kernel Tuner. Furthermore, the pagerank centrality of strong local minima within FFGs can be used to investigate why certain minima are unlikely to be visited, for example because neighbouring configurations fail to compile. Lastly, in this work we fully computed 26 kernel spaces, and made these publicly available. We aim to extend this to a benchmark dataset for evolutionary computation algorithms.

\section*{Acknowledgements}

This work has made use of the experimental systems on the Dutch national e-infrastructure with the support of the SURF Cooperative. The CORTEX project has received funding from the Dutch Research Council (NWO) in the framework of the NWA-ORC Call (file number NWA.1160.18.316). 
This work is also financially supported by the Netherlands Organization for Scientific Research (NWO), project number 639.073.506.





%

{\small
\bibliographystyle{ieee_fullname}
\bibliography{library}
}

\begin{table*}[hbt!]
    \centering
    \footnotesize
    \begin{tabular}{|l|ccc|}
        \hline
       \textbf{Kernel} & parameter to tune & list of values & number of possible values\\
       \hline
        Convolution     & \verb|block_size_x|  & $\{1, 2, 4, 8, 16, 32, 48, 64, 80, 96, 112, 128\}$ & 12 \\
          (except MI50) & \verb|block_size_y|  & $\{1, 2, 4, 8, 16, 32\}$ & 6 \\
         & \verb|tile_size_x|  & $\{1, 2, 3, 4, 5, 6, 7, 8\}$ & 8 \\
         & \verb|tile_size_y|  & $\{1, 2, 3, 4, 5, 6, 7, 8\}$ & 8 \\
         & \verb|use_padding|  & $\{0,1\}$ & 2 \\
         & \verb|read_only|  & $\{0, 1\}$ & 2 \\
        \hline
        Convolution (MI50)    & \verb|block_size_x|  & $\{16, 32, 48, 64, 80, 96, 112, 128\}$ & 8 \\
         & \verb|block_size_y|  & $\{1, 2, 4, 8, 16, 32\}$ & 6 \\
         & \verb|tile_size_x|  & $\{1, 2, 4\}$ & 3 \\
         & \verb|tile_size_y|  & $\{1, 2, 4\}$ & 3 \\
         & \verb|use_padding|  & $\{0,1\}$ & 2 \\
        \hline
        \hline
        GEMM & \verb|MWG|  & $\{16, 32, 64, 128\}$ & 4 \\
         & \verb|NWG|  & $\{16, 32, 64, 128\}$ & 4 \\
         & \verb|MDIMC|  & $\{8, 16, 32\}$ & 3 \\
         & \verb|NDIMC|  & $\{8, 16, 32\}$ & 3 \\
         & \verb|MDIMA|  & $\{8, 16, 32\}$ & 3 \\
         & \verb|NDIMB|  & $\{8, 16, 32\}$ & 3 \\
         & \verb|VWM|  & $\{1, 2, 4, 8\}$ & 4 \\
         & \verb|VWN|  & $\{1, 2, 4, 8\}$ & 4 \\
         & \verb|SA|  & $\{0, 1\}$ & 2 \\
         & \verb|SB|  & $\{0, 1\}$ & 2 \\
        \hline
        \hline
        Point-in-polygon  & \verb|block_size_x|  & $\{32, 64, 96, 128, 160, 192, 224, 256, 288, 320, 352, $ & 31 \\
         & & $384, 416, 448, 480, 512, 544, 576, 608, 640, 672,$ &\\
         & & $704, 736, 768, 800, 832, 864, 896, 928, 960, 992\}$ &\\
         & \verb|tile_size|  & $\{1, 2, 4, 6, 8, 10, 12, 14, 16, 18, 20\}$ & 11 \\
         & \verb|between_method| & $\{0, 1, 2, 3\}$ & 4 \\
         & \verb|use_precomputed_slopes| & $\{0, 1\}$ & 2 \\
         & \verb|use_method| & $\{0, 1, 2\}$ & 3 \\
        \hline
    \end{tabular}
    \caption{\small Tunable parameters per kernel, and list of possible values for each parameter.}
    \label{tab:tuneparams}
\end{table*}

\newpage\phantom{new}
\newpage\phantom{new}   

\section{Appendix}
\label{app:appendix}
\subsection{Appendix: Tunable parameters per GPU kernel}
\label{app:tuneparams}
In Table \ref{tab:tuneparams} we show the tunable parameters per kernel, and the values each parameter could take. For the convolution kernel, the MI50 GPU (the only AMD model) required a different problem setup due to hardware constraints.

\subsection{Appendix: Alternative splits for competition heatmaps}
\label{app:compheatmaps}
\begin{figure*}
    \centering
    \newcommand{\wdth}{0.46}
    \newcommand{\bspc}{0.0}
    \newcommand{\hspc}{-0.2}
    \newcommand{\vspc}{-0.91}
    \captionsetup[subfigure]{labelformat=empty}
    \subfloat{\makebox[\bspc\textwidth][c]{
    \subfloat{\includegraphics[width=\wdth\textwidth]{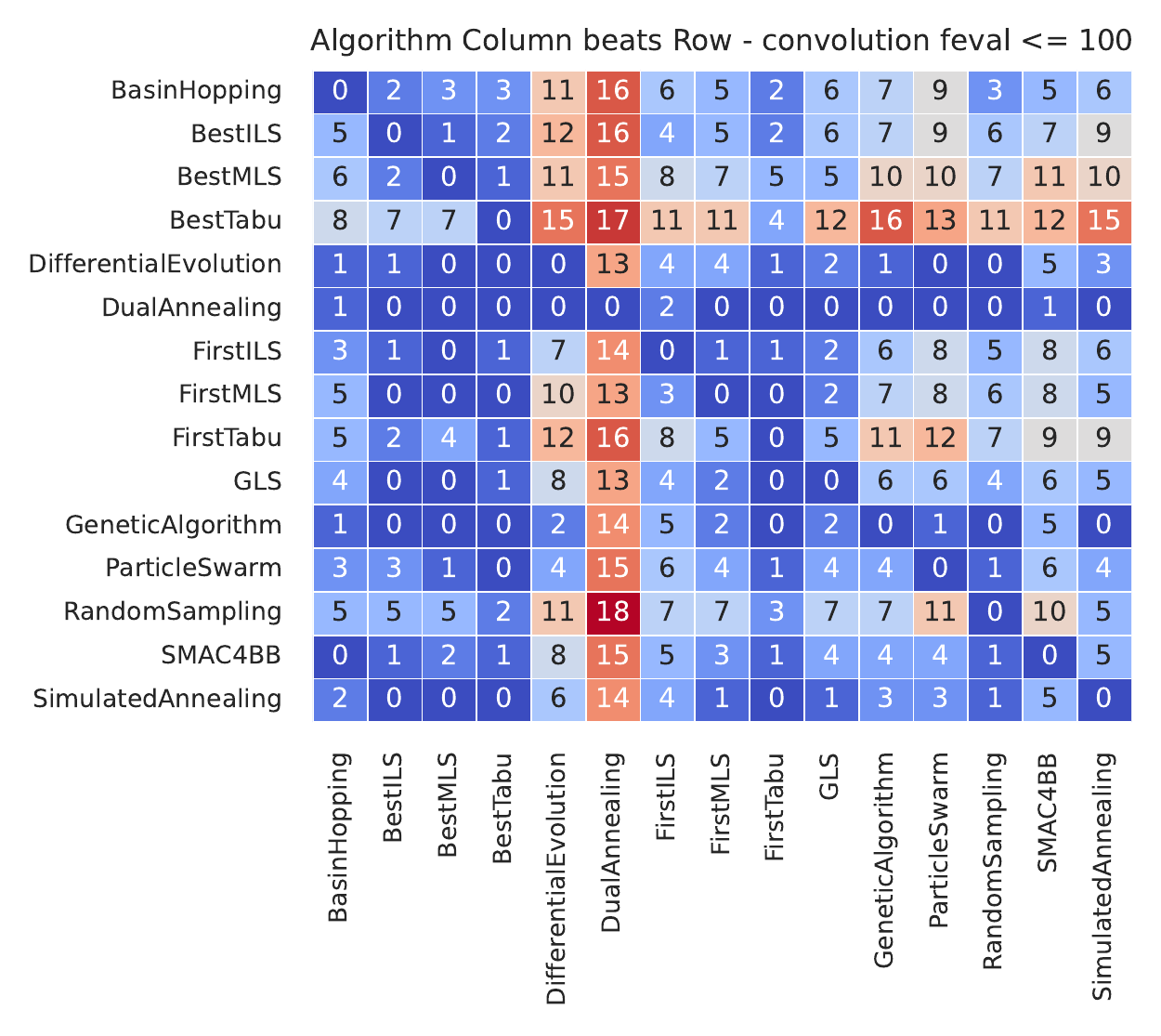}}
    \hspace{\hspc cm}
    \subfloat{\includegraphics[width=\wdth\textwidth]{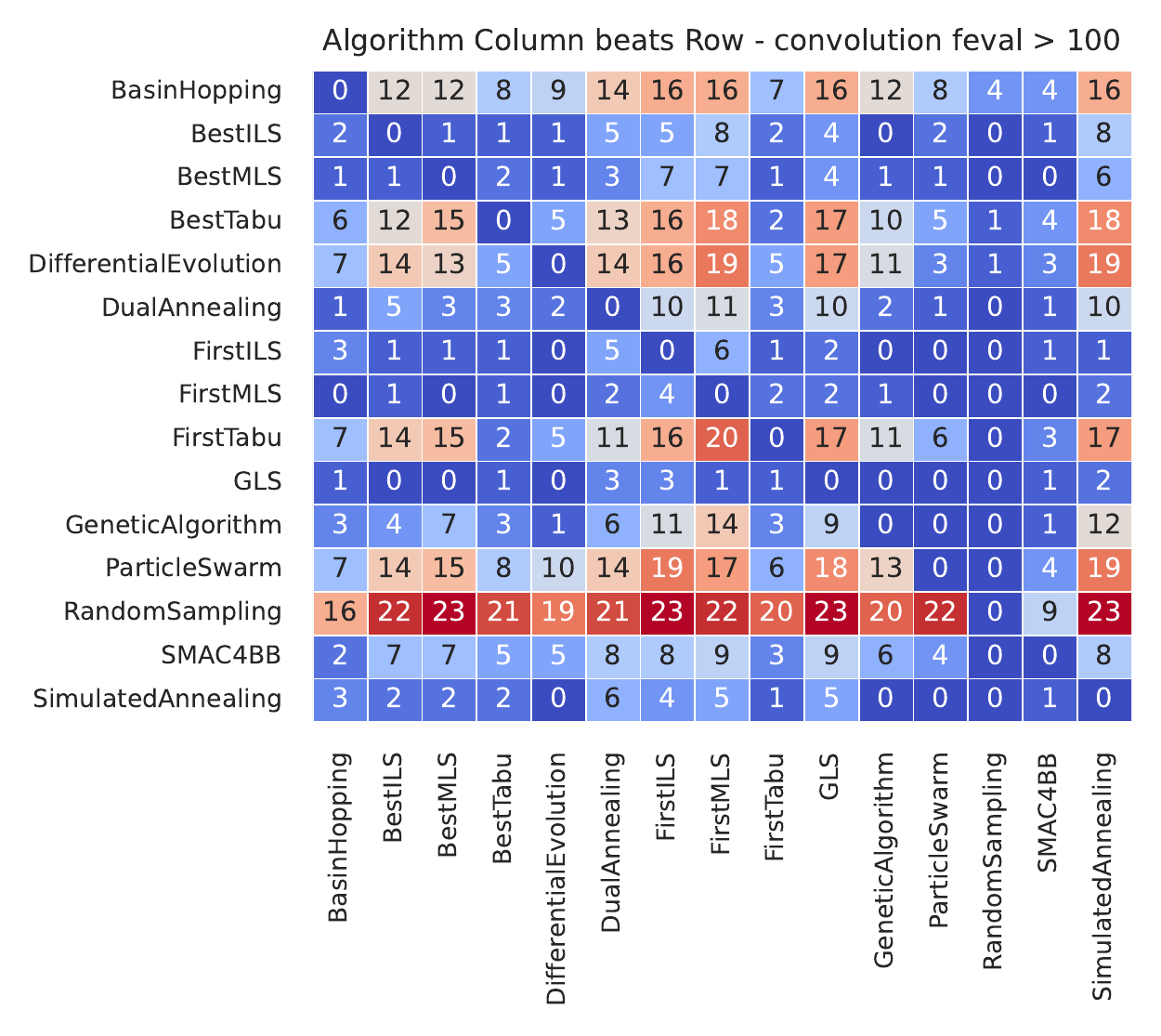}}
    }}\\
    \vspace{\vspc cm}
    \subfloat{\makebox[\bspc\textwidth][c]{
    \subfloat{\includegraphics[width=\wdth\textwidth]{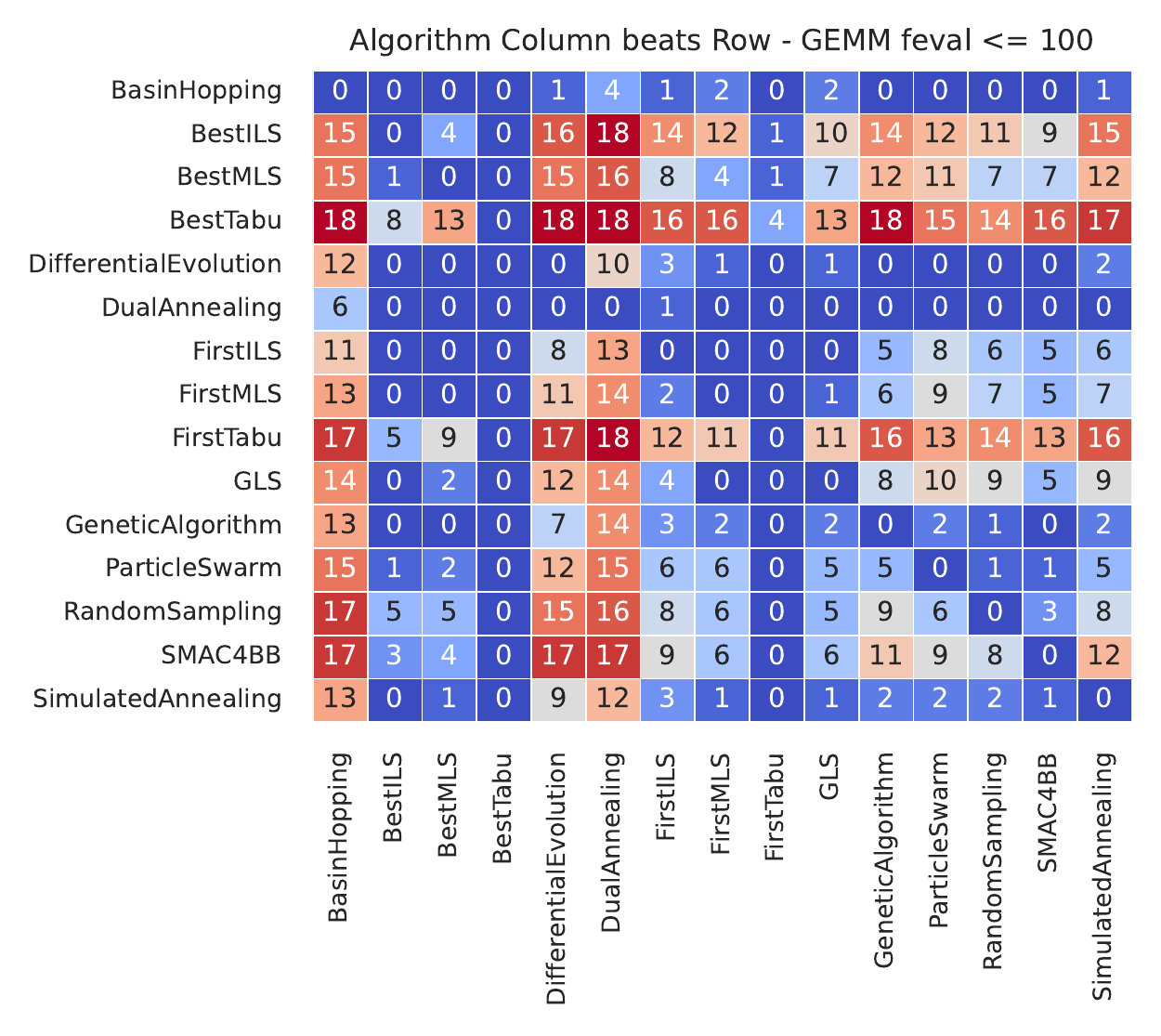}}
    \hspace{\hspc cm}
    \subfloat{\includegraphics[width=\wdth\textwidth]{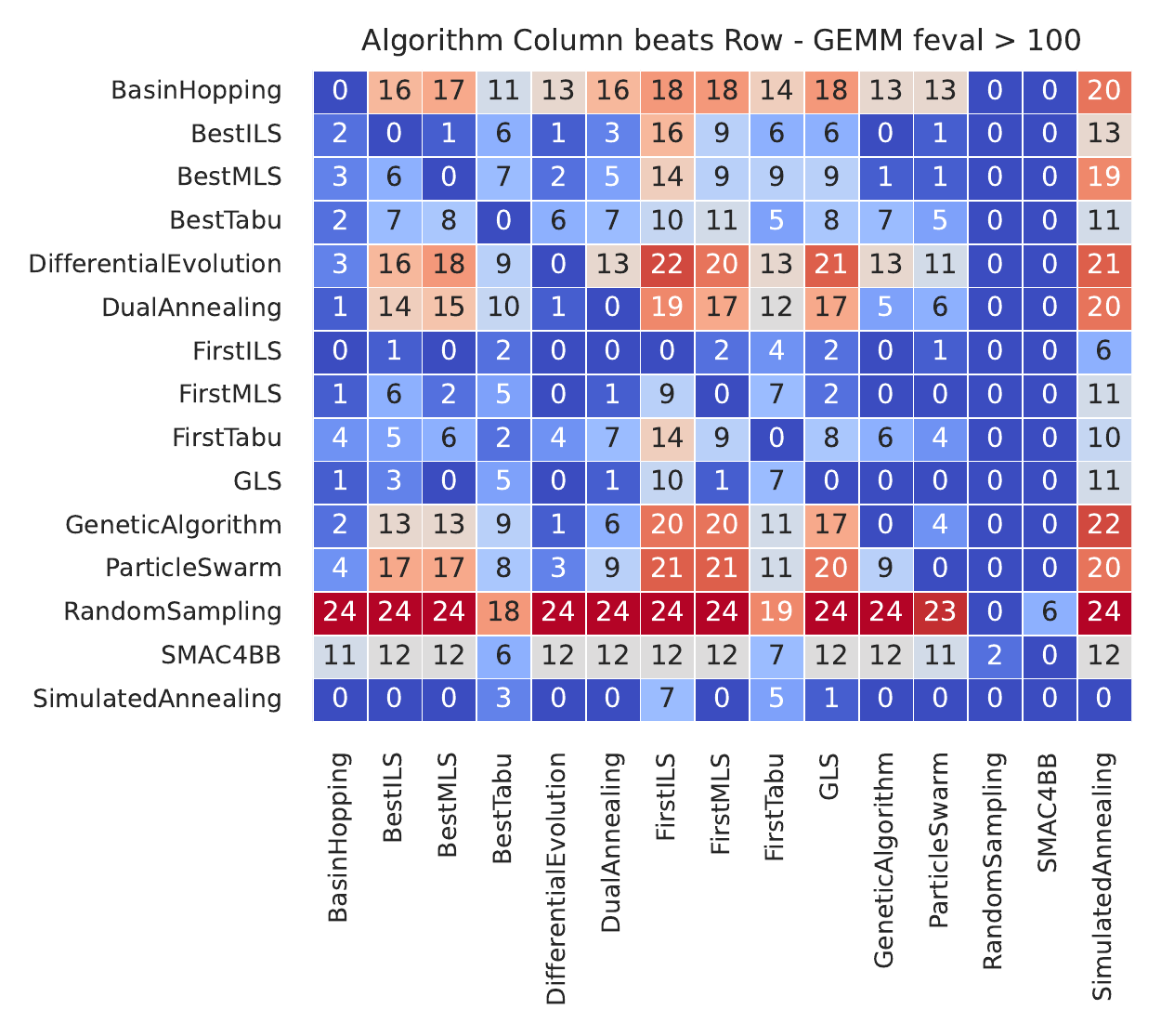}}
    }}\\
    \vspace{\vspc cm}
    \subfloat{\makebox[\bspc\textwidth][c]{
    \subfloat{\includegraphics[width=\wdth\textwidth]{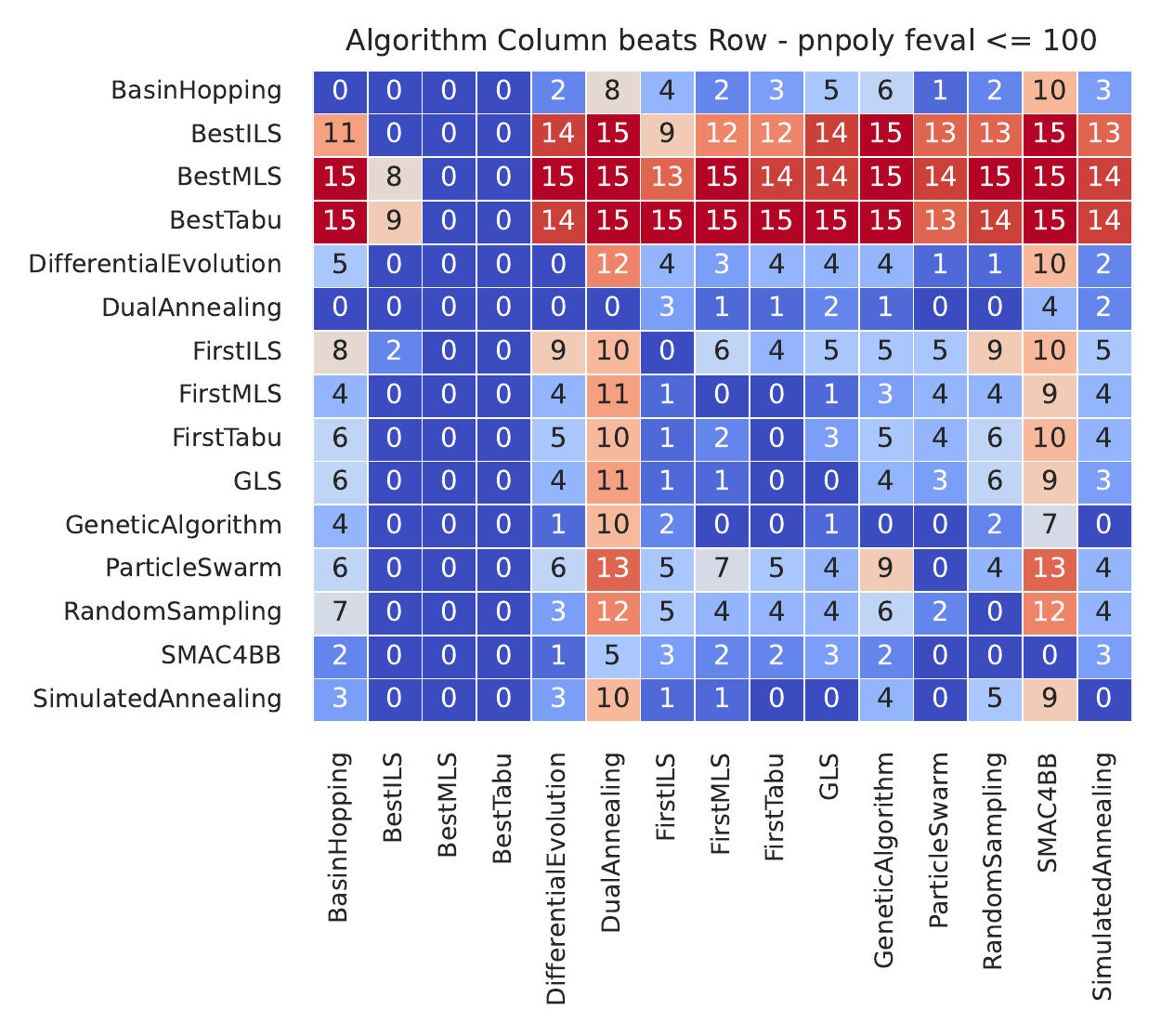}}
    \hspace{\hspc cm}
    \subfloat{\includegraphics[width=\wdth\textwidth]{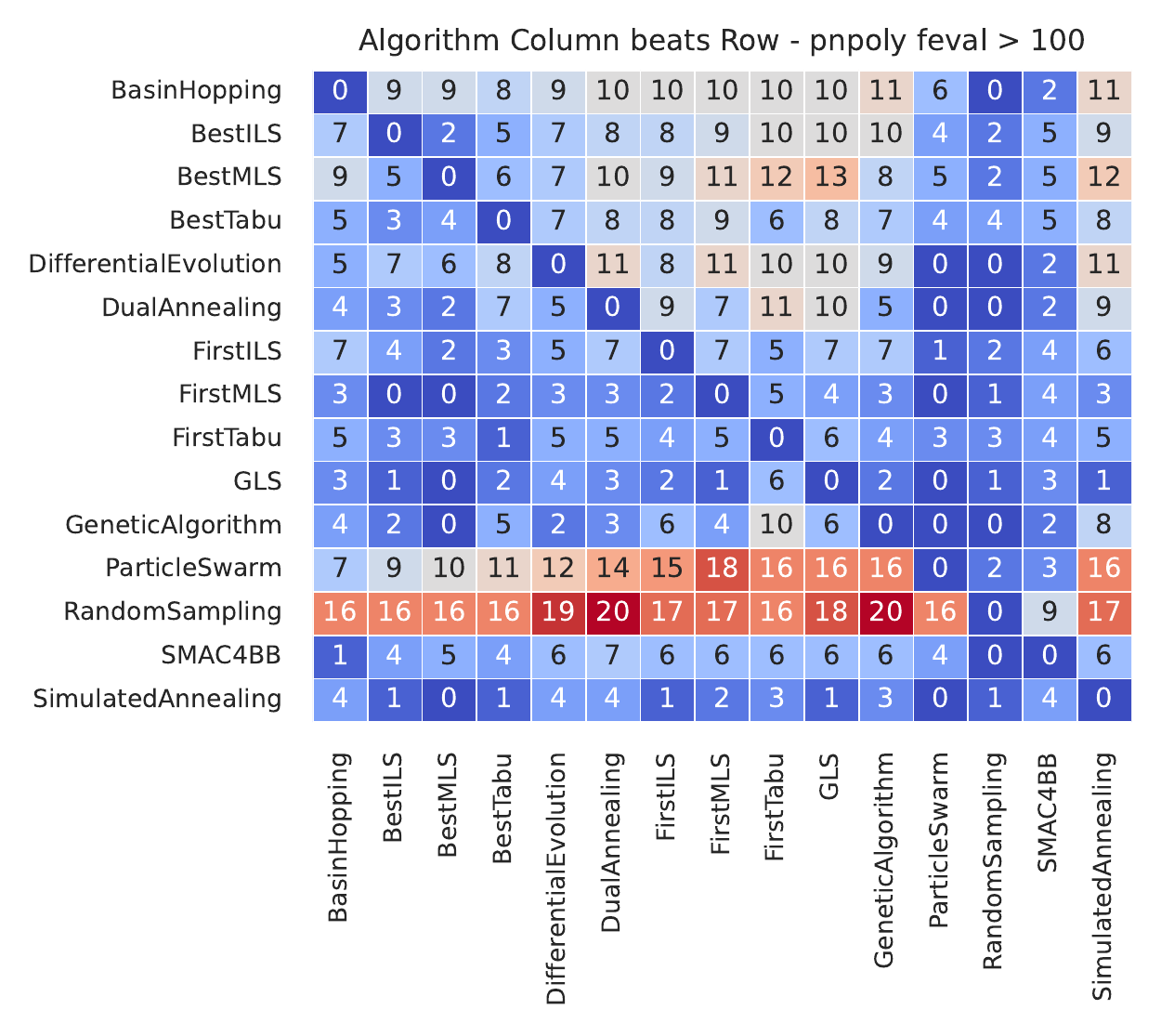}}
    }}
    \vspace{-0.15 cm}
    \caption{\scriptsize Heatmaps counting the occurrences when the column algorithm found statistically better solutions than the row algorithm for the (top) convolution, (middle) GEMM, and (botttom) PnPoly kernels. An occurrence is counted when 50 runs for a budget are statistically significantly better according to a two-sample independent t-test ($\alpha=0.05$). (Left): Heatmap for low $\leq100$ budgets, i.e., 25, 50, and 100. (Right): Heatmap for mid and high $>100$ budgets, i.e., 200, 400, 800, 1600. Algorithms with low values (blue) in their rows were not often beaten for those budgets, and algorithms with high values in their column (red) often beat other algorithms.}
    \label{fig:heatmapconv100}
\end{figure*}

\begin{figure*}
    \centering
    \newcommand{\wdth}{0.46}
    \newcommand{\bspc}{0.0}
    \newcommand{\hspc}{-0.2}
    \newcommand{\vspc}{-0.91}
    \captionsetup[subfigure]{labelformat=empty}
    \subfloat{\makebox[\bspc\textwidth][c]{
    \subfloat{\includegraphics[width=\wdth\textwidth]{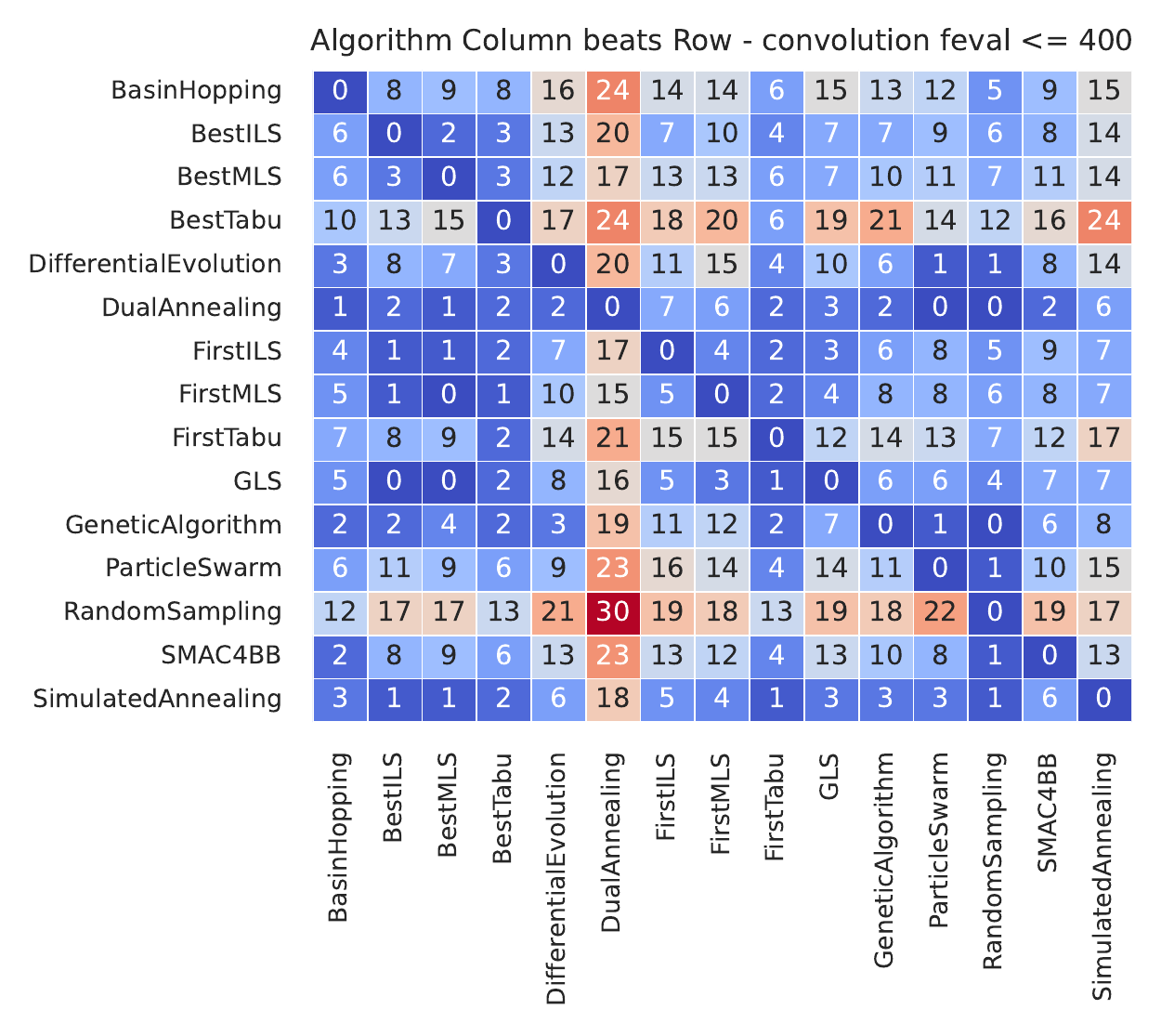}}
    \hspace{\hspc cm}
    \subfloat{\includegraphics[width=\wdth\textwidth]{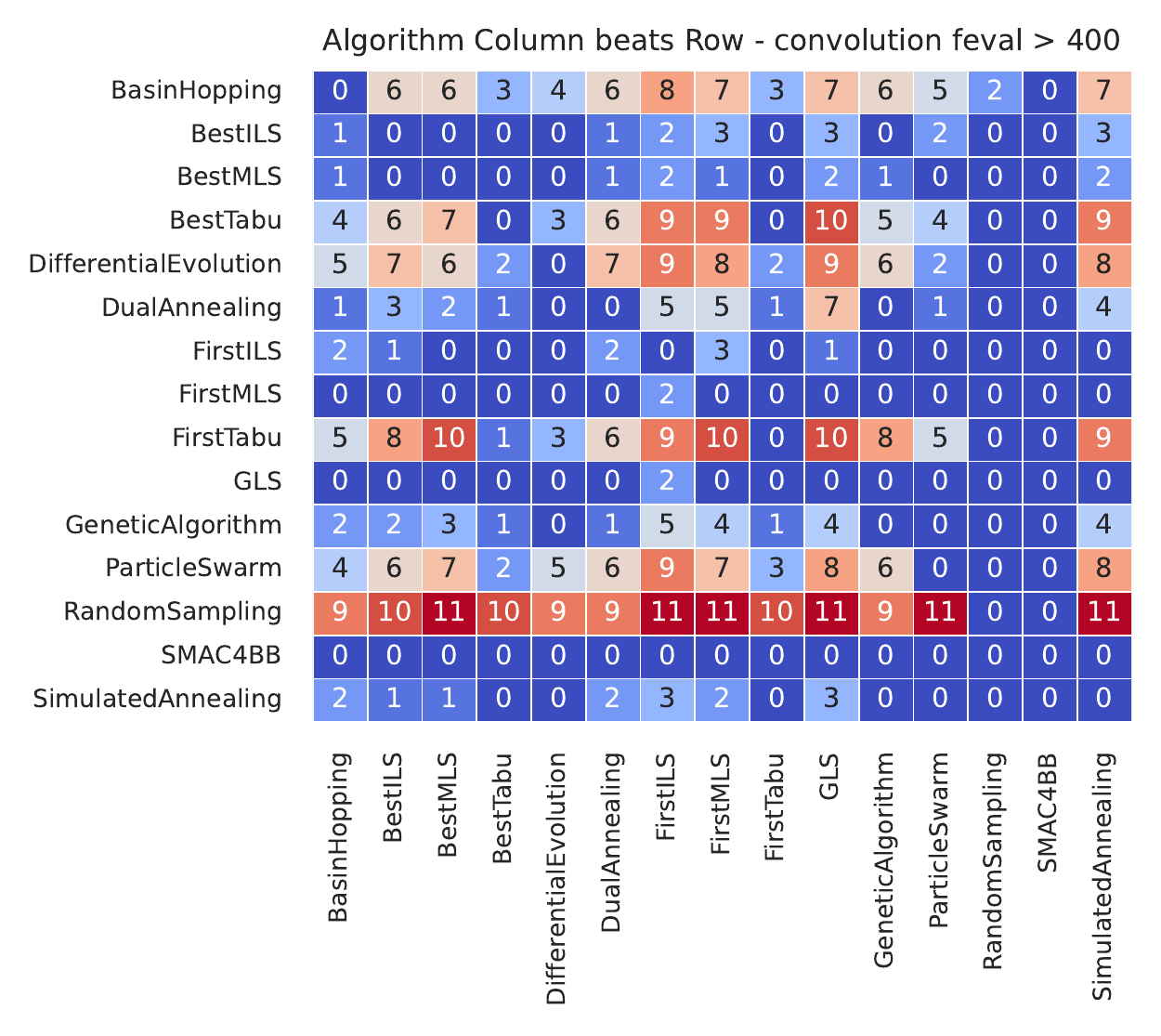}}
    }}\\
    \vspace{\vspc cm}
    \subfloat{\makebox[\bspc\textwidth][c]{
    \subfloat{\includegraphics[width=\wdth\textwidth]{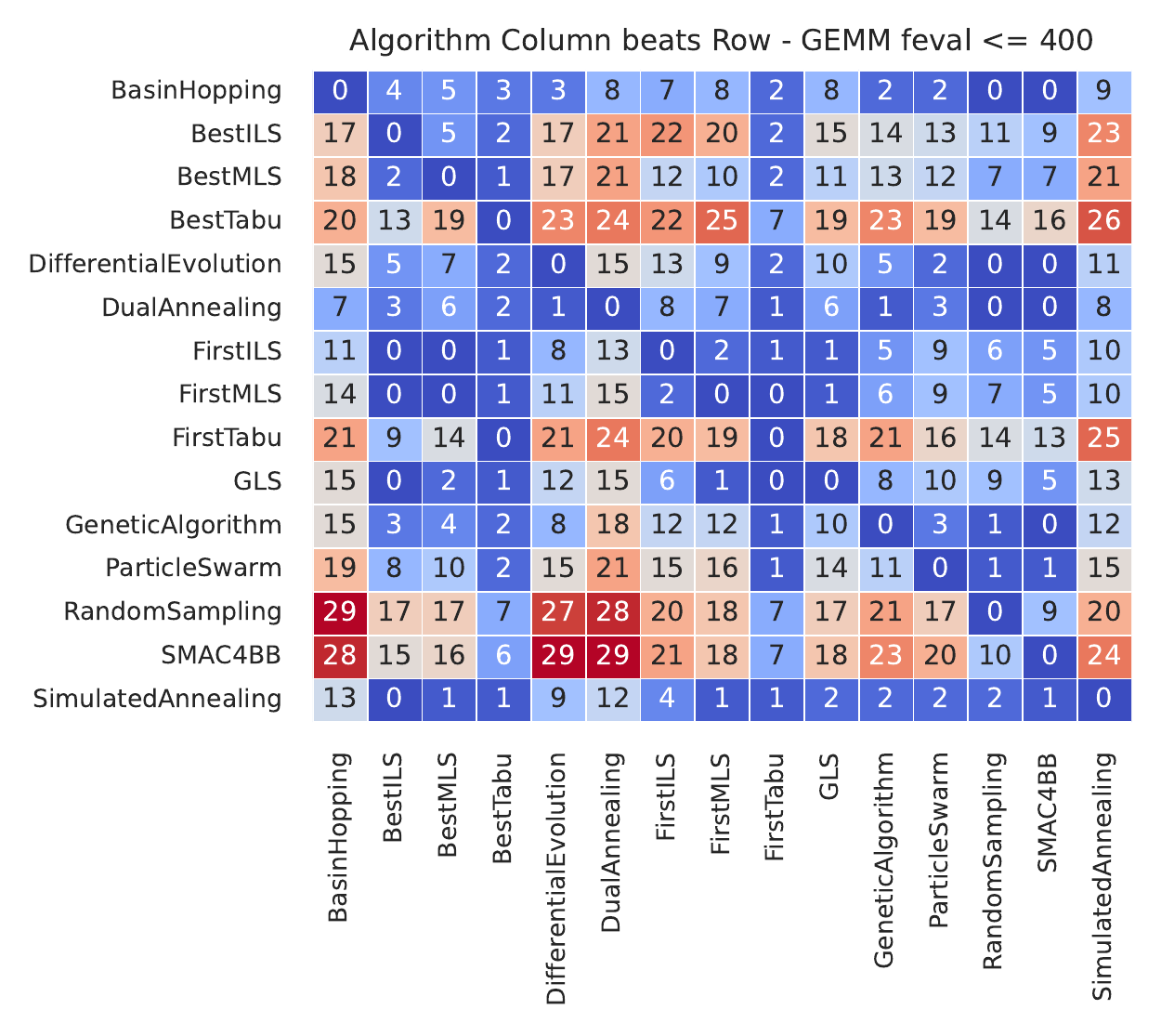}}
    \hspace{\hspc cm}
    \subfloat{\includegraphics[width=\wdth\textwidth]{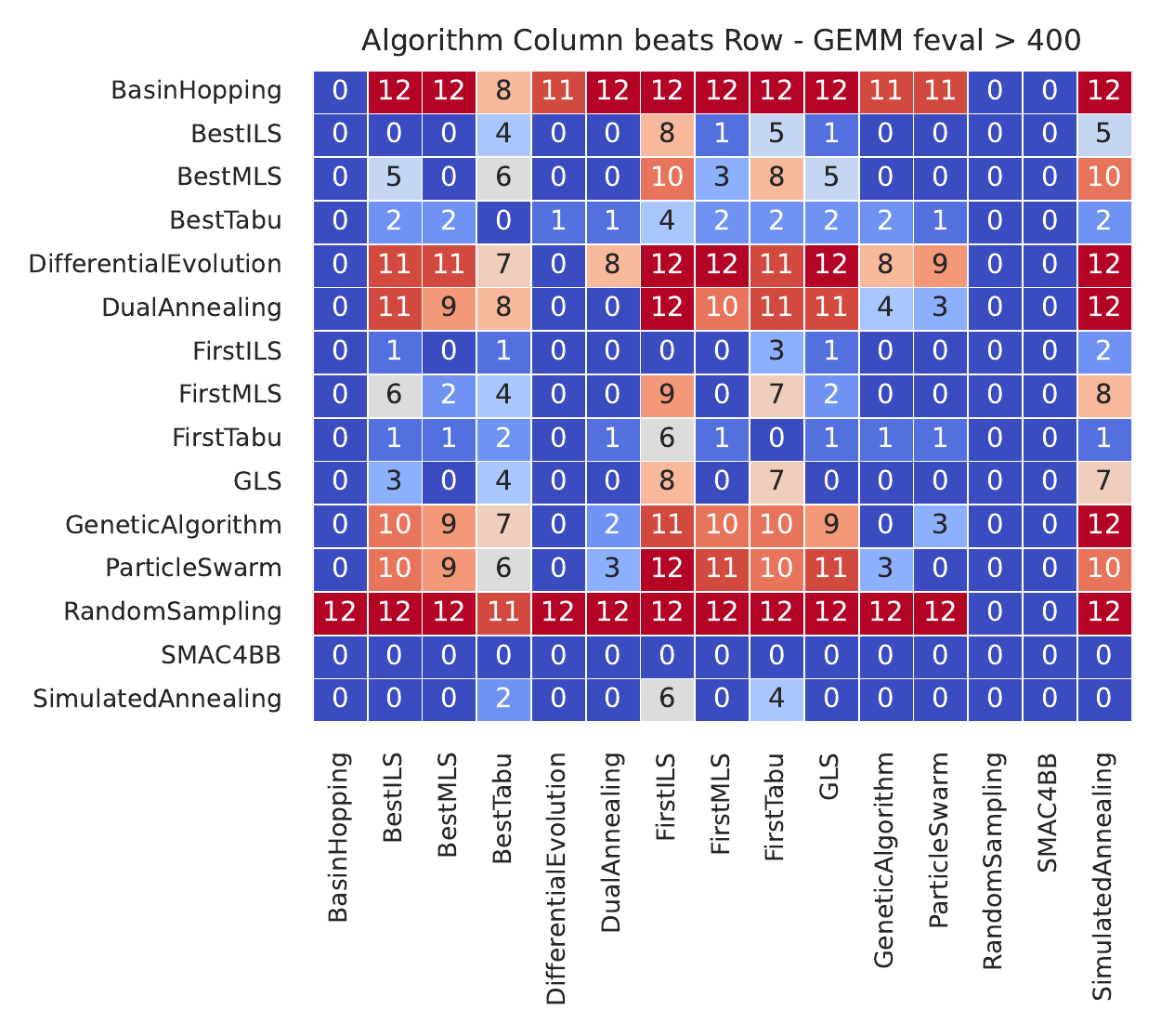}}
    }}\\
    \vspace{\vspc cm}
    \subfloat{\makebox[\bspc\textwidth][c]{
    \subfloat{\includegraphics[width=\wdth\textwidth]{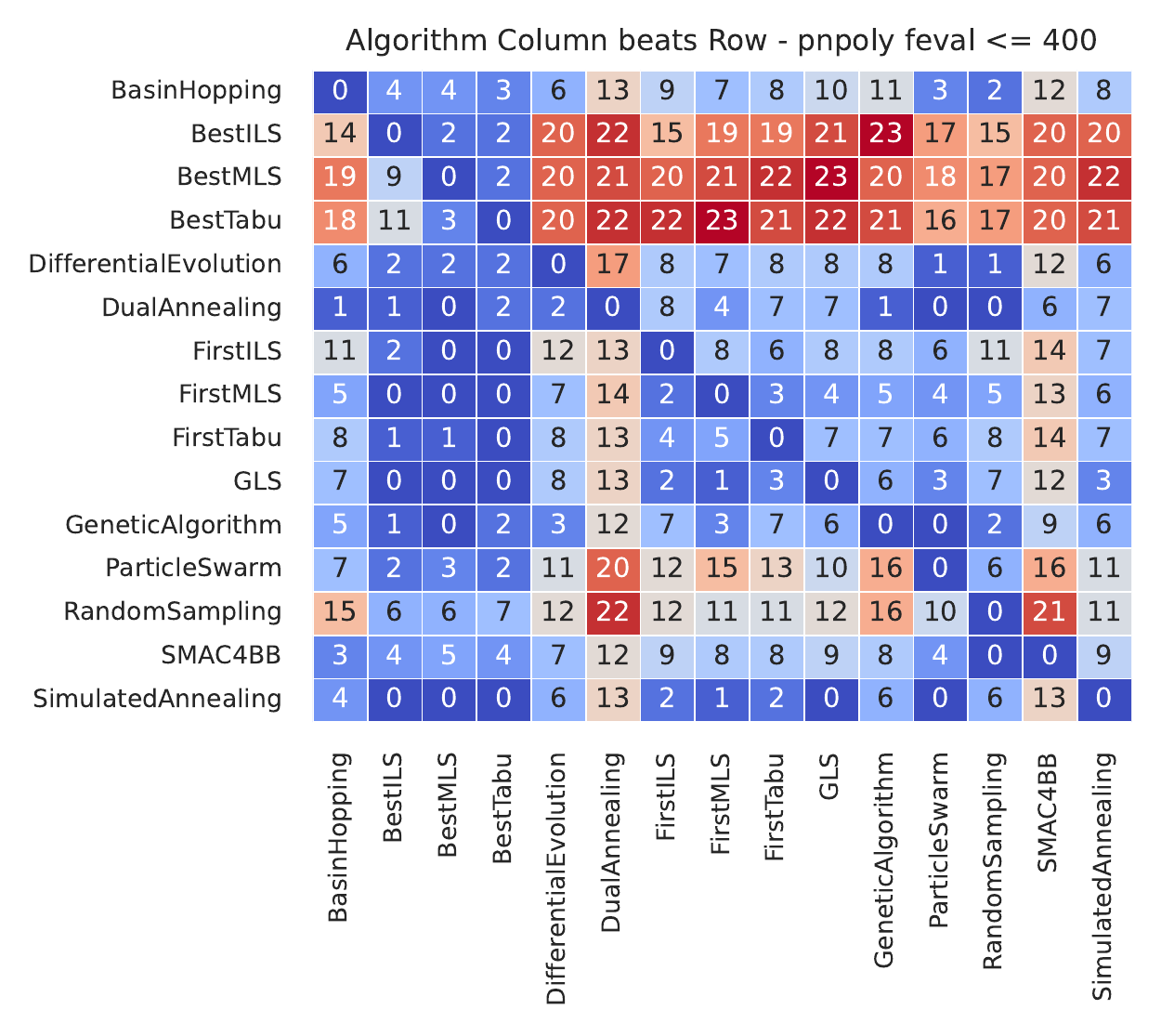}}
    \hspace{\hspc cm}
    \subfloat{\includegraphics[width=\wdth\textwidth]{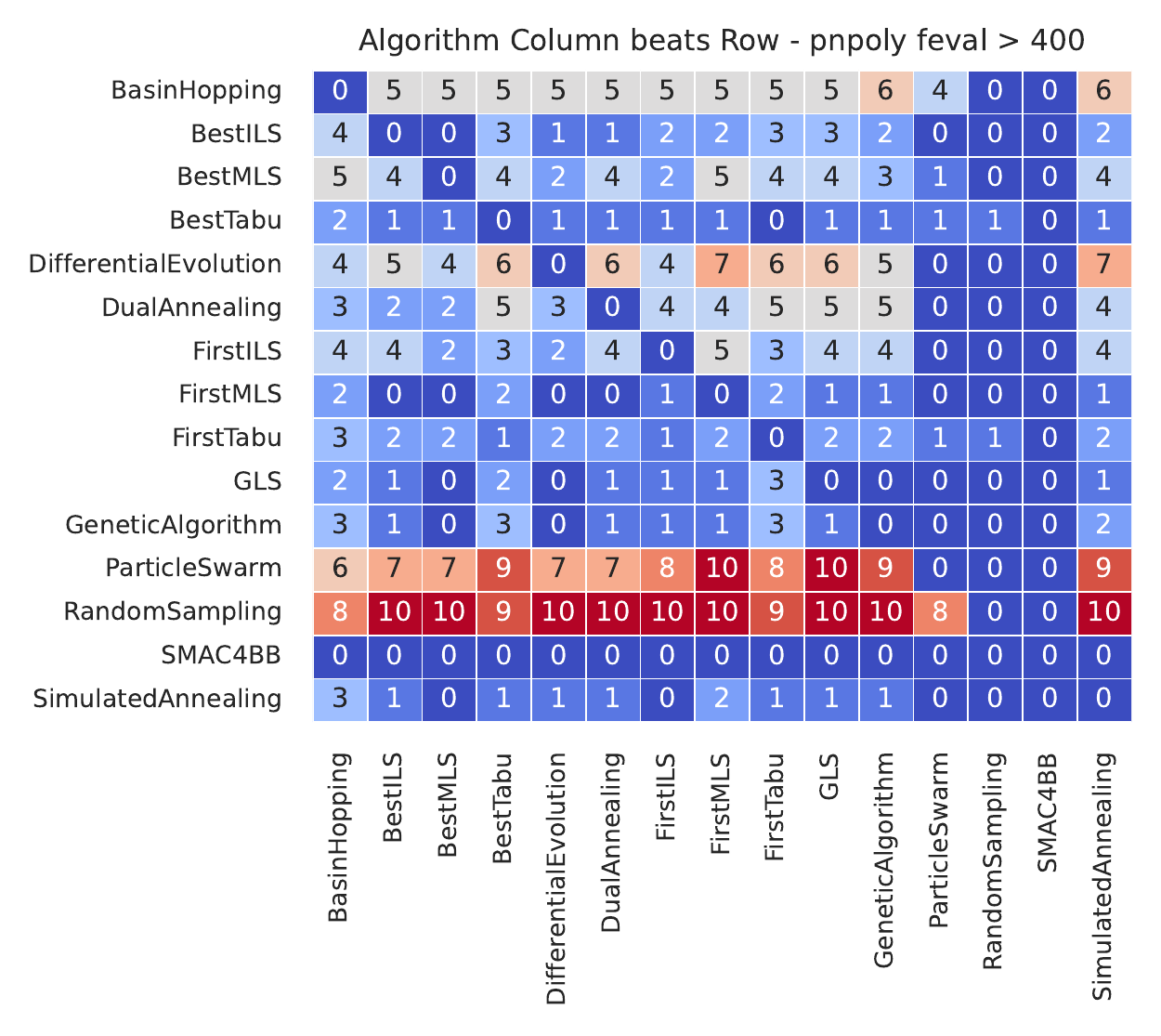}}
    }}
    \vspace{-0.15 cm}
    \caption{\scriptsize Heatmaps counting the occurrences when the column algorithm found statistically better solutions than the row algorithm for the (top) convolution, (middle) GEMM, and (botttom) PnPoly kernels. An occurrence is counted when 50 runs for a budget are statistically significantly better according to a two-sample independent t-test ($\alpha=0.05$). (Left): Heatmap for low $\leq400$ budgets, i.e., 25, 50, 100, 200, and 400. (Right): Heatmap for mid and high $>400$ budgets, i.e., 800, 1600. Algorithms with low values (blue) in their rows were not often beaten for those budgets, and algorithms with high values in their column (red) often beat other algorithms.}
    \label{fig:heatmapconv400}
\end{figure*}

In Figures \ref{fig:heatmapconv100} and \ref{fig:heatmapconv400} we show the algorithm competition heatmaps such as in Figure \ref{fig:heatmapconv} but when split at 100 and 400 budgets instead of 200.

\subsection{Appendix: Per kernel graphs of experimental results}
\label{app:separate_algos}
\begin{figure*}
    \centering
    \newcommand{\wdth}{0.99}
    \includegraphics[width=\wdth\textwidth]{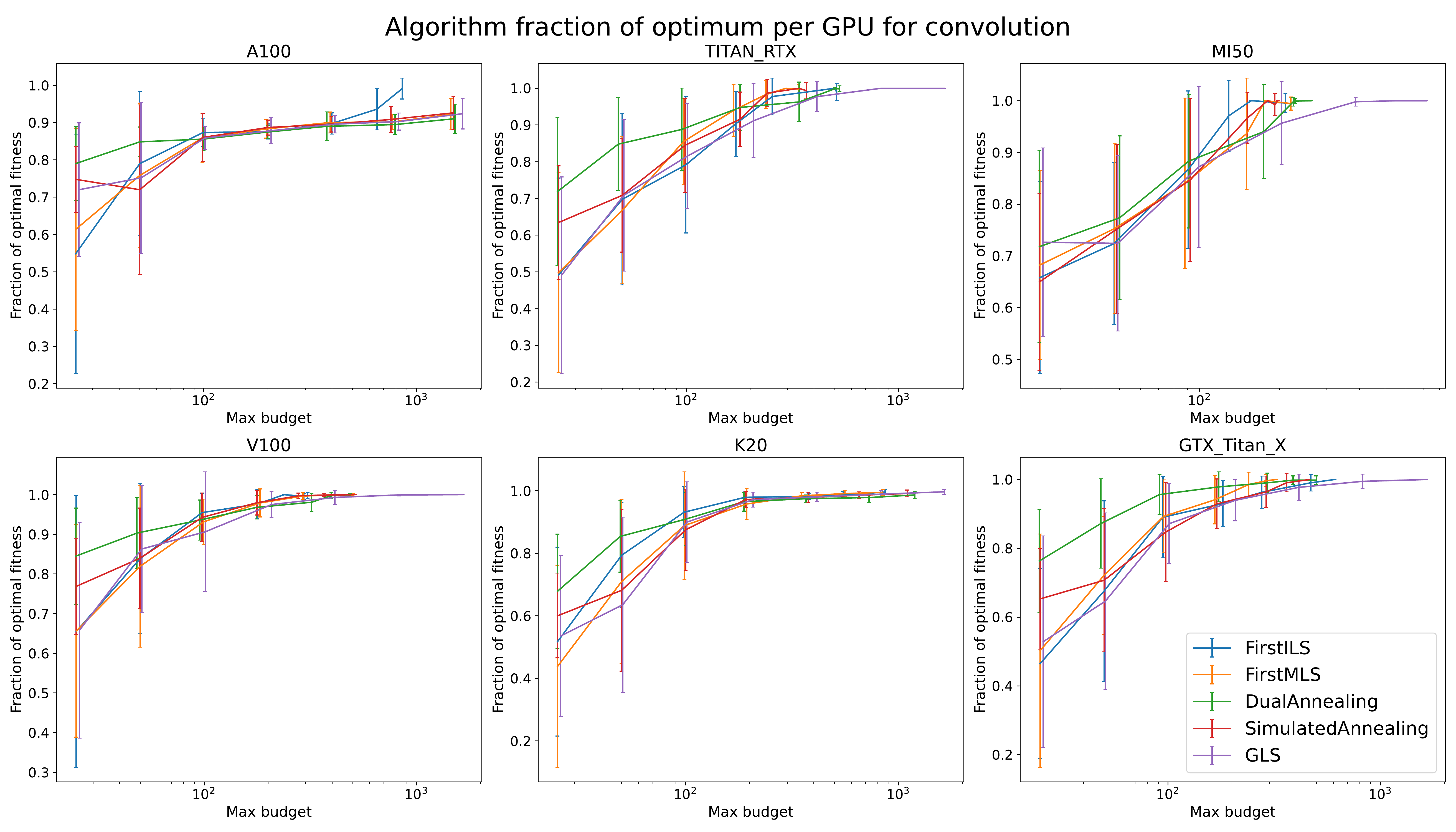}
    \caption{\scriptsize \textbf{Convolution:} Fraction of optimal runtime per GPU for FirstILS, FirstMLS, dual annealing, simulated annealing, and GLS over 50 runs. Each point is the mean fraction of optimal runtime found ($y$-axis) for mean budget used (logarithmic $x$-axis), with error bars indicating the standard deviation in fraction of optimum.}
    \label{fig:separate_sub1_conv}
\end{figure*}

\begin{figure*}
    \centering
    \newcommand{\wdth}{0.99}
    \includegraphics[width=\wdth\textwidth]{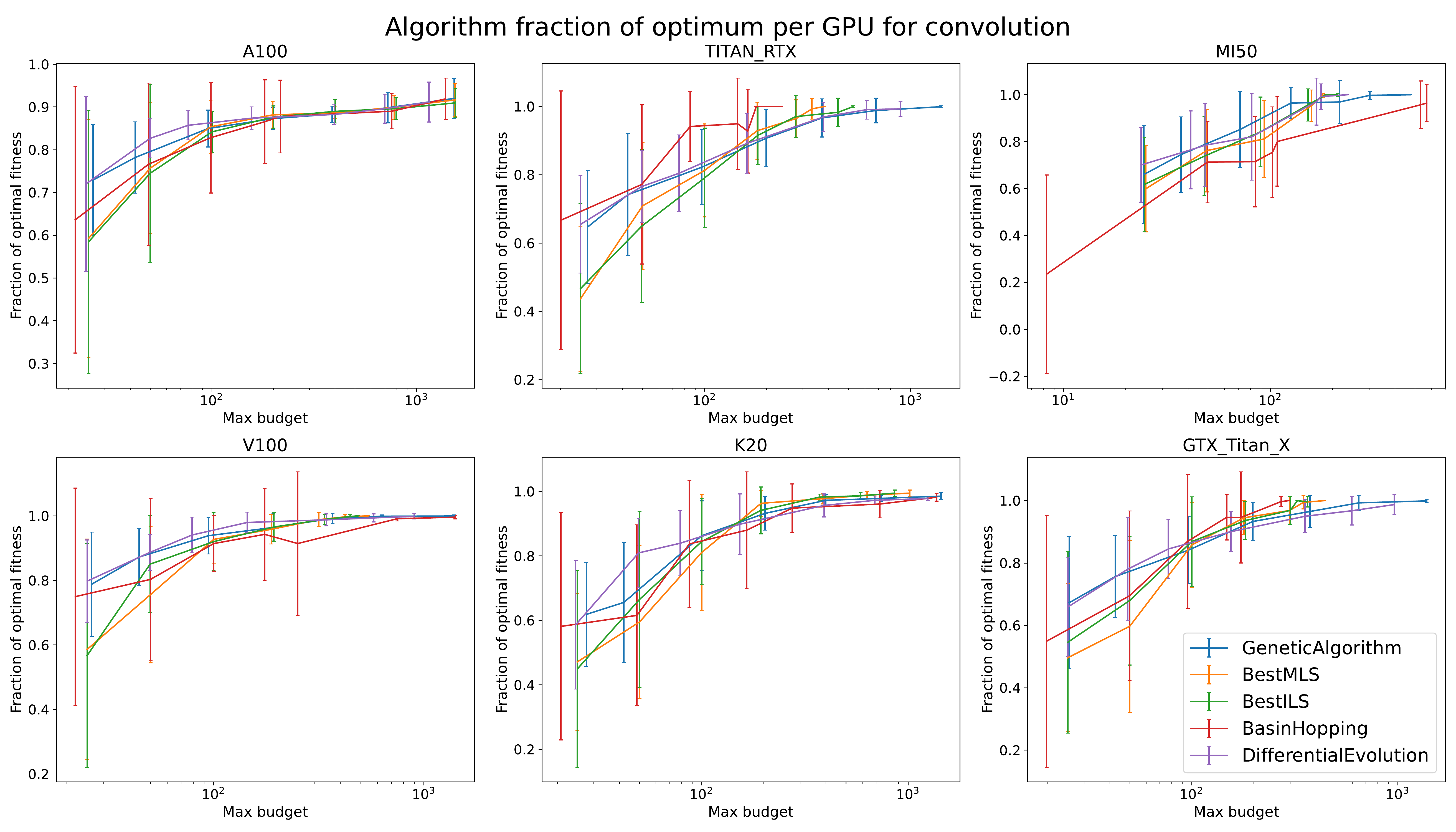}
    \caption{\scriptsize \textbf{Convolution:} Fraction of optimal runtime per GPU for GA, BestMLS, BestILS, basin hopping, and differential evolution over 50 runs. Each point is the mean fraction of optimal runtime found ($y$-axis) for mean budget used (logarithmic $x$-axis), with error bars indicating the standard deviation in fraction of optimum.}
    \label{fig:separate_sub2_conv}
\end{figure*}

\begin{figure*}
    \centering
    \newcommand{\wdth}{0.99}
    \includegraphics[width=\wdth\textwidth]{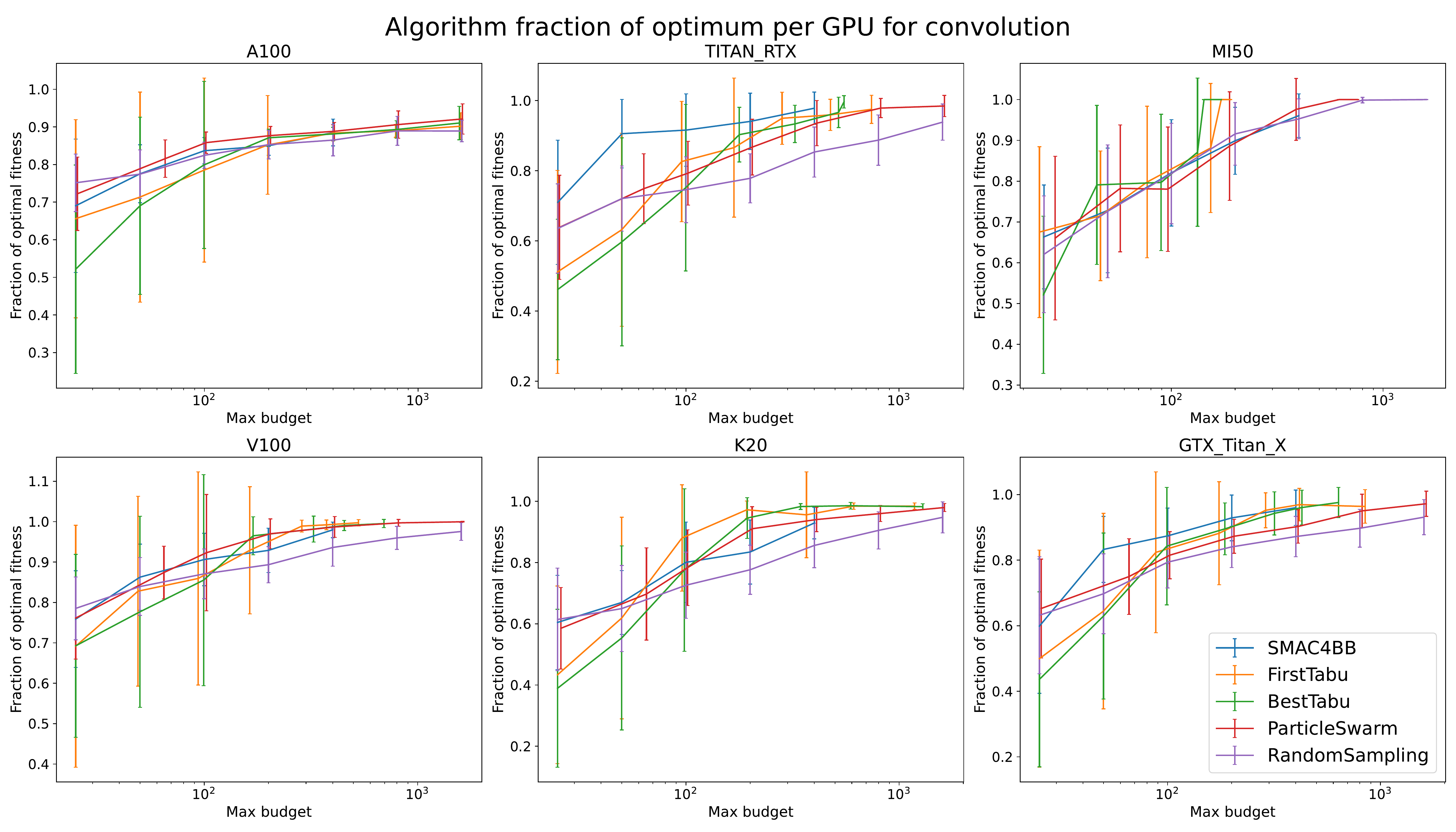}
    \caption{\scriptsize \textbf{Convolution:} Fraction of optimal runtime per GPU for SMAC, FirstTabu, BestTabu, PSO, and random sampling over 50 runs. Each point is the mean fraction of optimal runtime found ($y$-axis) for mean budget used (logarithmic $x$-axis), with error bars indicating the standard deviation in fraction of optimum.}
    \label{fig:separate_sub3_conv}
\end{figure*}

\begin{figure*}
    \centering
    \newcommand{\wdth}{0.99}
    \includegraphics[width=\wdth\textwidth]{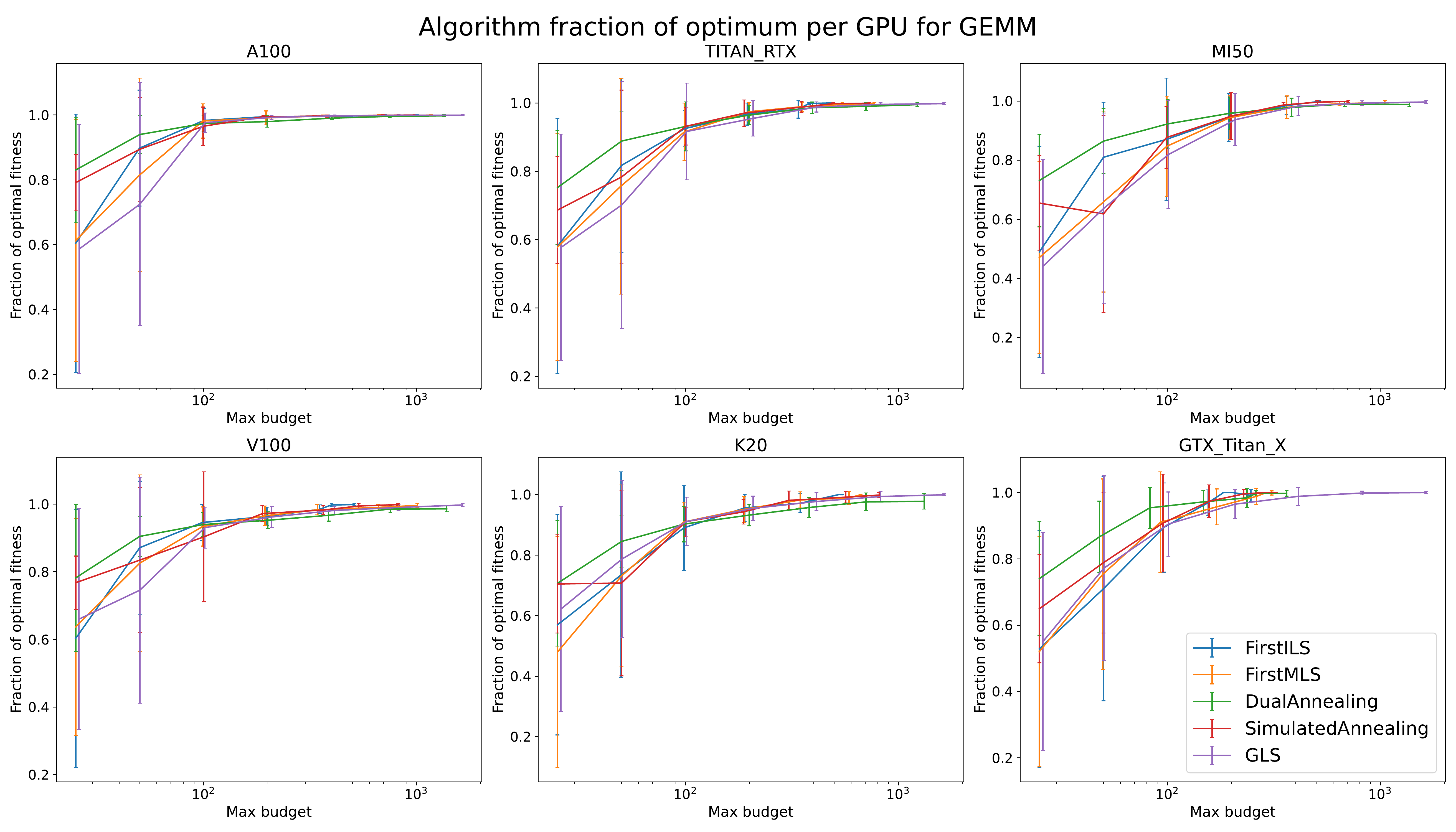}
    \caption{\scriptsize \textbf{GEMM:} Fraction of optimal runtime per GPU for FirstILS, FirstMLS, dual annealing, simulated annealing, and GLS over 50 runs. Each point is the mean fraction of optimal runtime found ($y$-axis) for mean budget used (logarithmic $x$-axis), with error bars indicating the standard deviation in fraction of optimum.}
    \label{fig:separate_sub1_GEMM}
\end{figure*}

\begin{figure*}
    \centering
    \newcommand{\wdth}{0.99}
    \includegraphics[width=\wdth\textwidth]{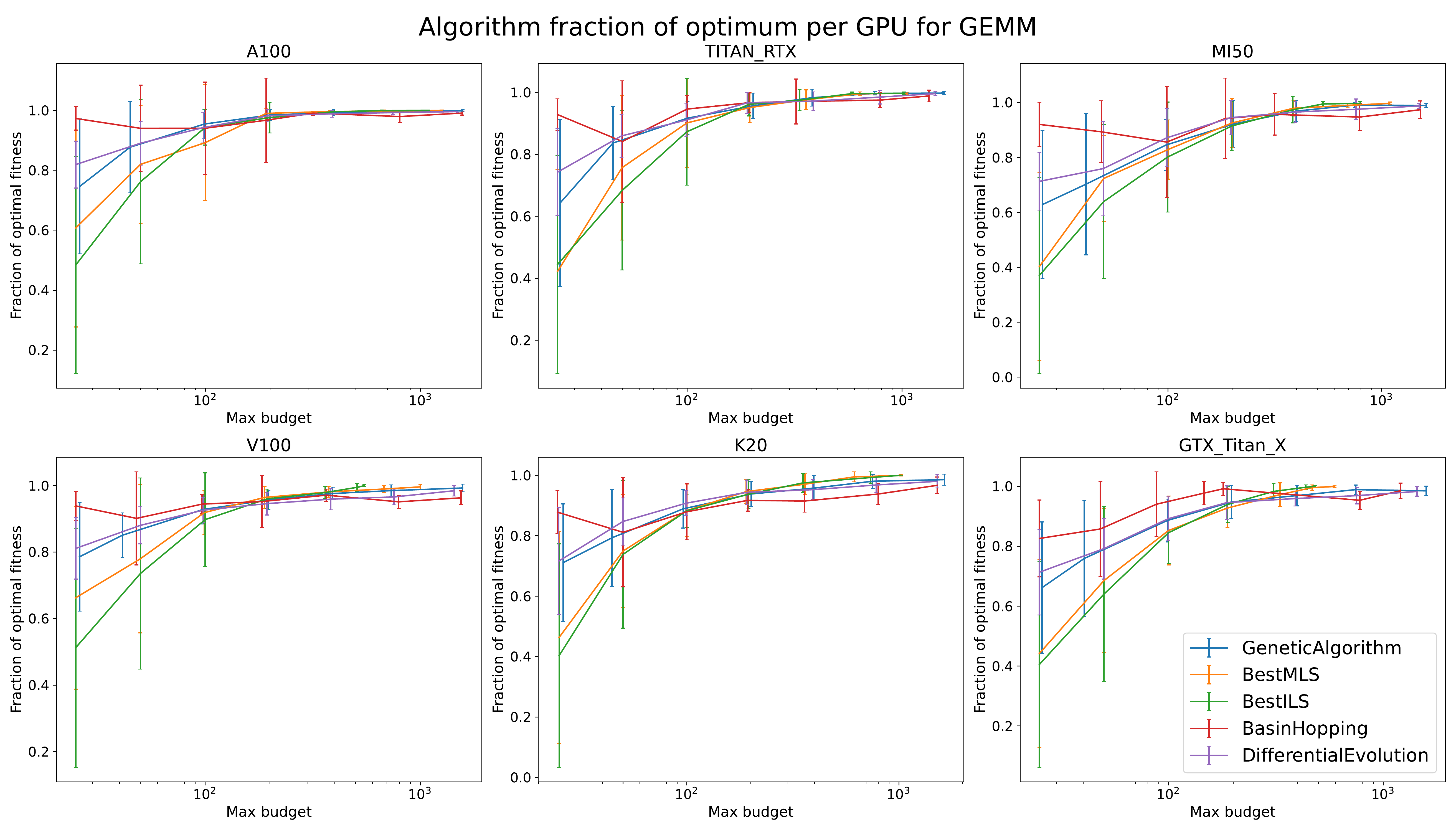}
    \caption{\scriptsize \textbf{GEMM:} Fraction of optimal runtime per GPU for GA, BestMLS, BestILS, basin hopping, and differential evolution over 50 runs. Each point is the mean fraction of optimal runtime found ($y$-axis) for mean budget used (logarithmic $x$-axis), with error bars indicating the standard deviation in fraction of optimum.}
    \label{fig:separate_sub2_GEMM}
\end{figure*}

\begin{figure*}
    \centering
    \newcommand{\wdth}{0.99}
    \includegraphics[width=\wdth\textwidth]{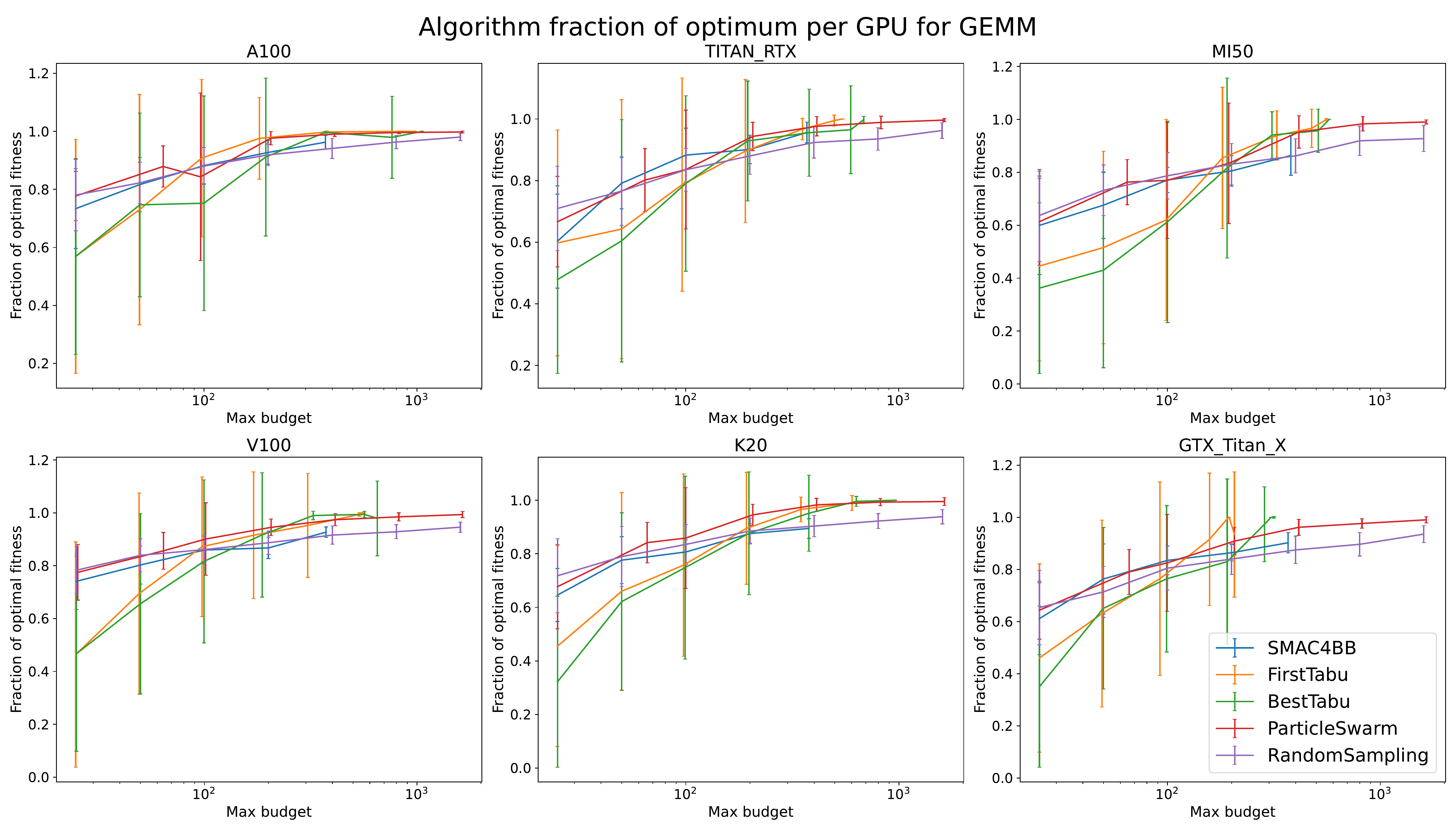}
    \caption{\scriptsize \textbf{GEMM:} Fraction of optimal runtime per GPU for SMAC, FirstTabu, BestTabu, PSO, and random sampling over 50 runs. Each point is the mean fraction of optimal runtime found ($y$-axis) for mean budget used (logarithmic $x$-axis), with error bars indicating the standard deviation in fraction of optimum.}
    \label{fig:separate_sub3_GEMM}
\end{figure*}

\begin{figure*}
    \centering
    \newcommand{\wdth}{0.99}
    \includegraphics[width=\wdth\textwidth]{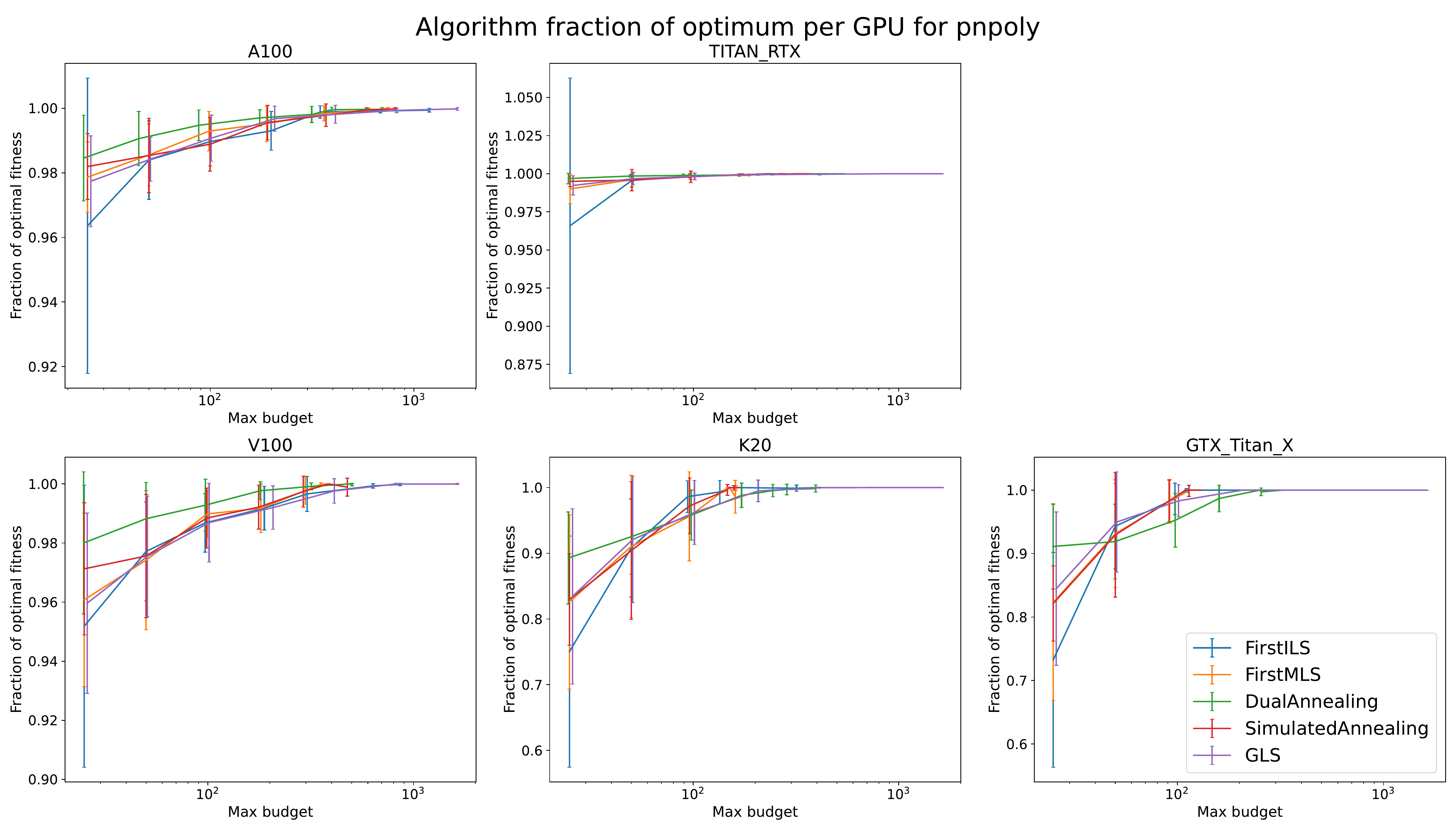}
    \caption{\scriptsize \textbf{Point-in-polygon:} Fraction of optimal runtime per GPU for FirstILS, FirstMLS, dual annealing, simulated annealing, and GLS over 50 runs. Each point is the mean fraction of optimal runtime found ($y$-axis) for mean budget used (logarithmic $x$-axis), with error bars indicating the standard deviation in fraction of optimum. The point-in-polygon kernel was not implemented for the MI50 GPU.}
    \label{fig:separate_sub1_pnpoly}
\end{figure*}

\begin{figure*}
    \centering
    \newcommand{\wdth}{0.99}
    \includegraphics[width=\wdth\textwidth]{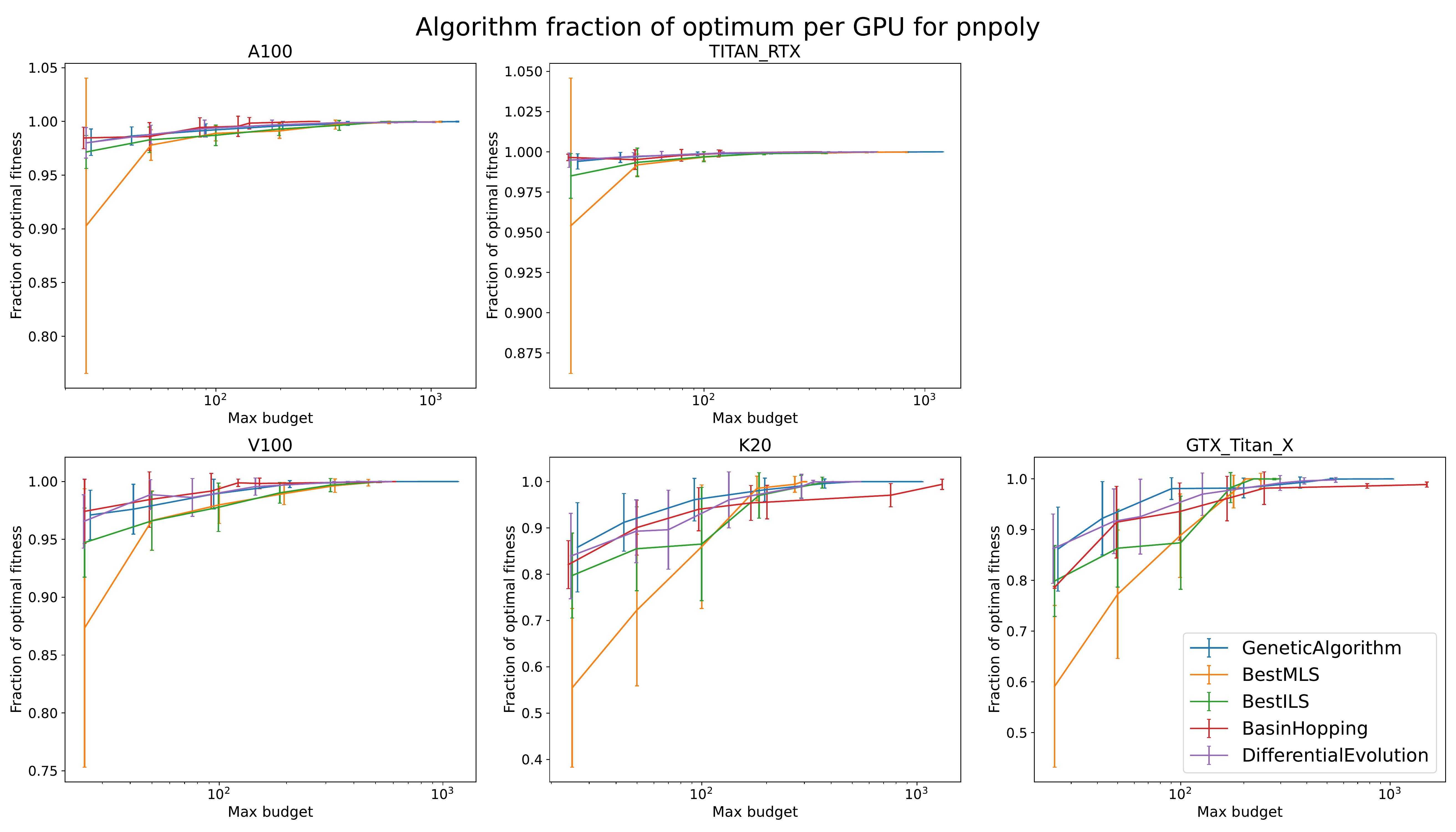}
    \caption{\scriptsize \textbf{Point-in-polygon:} Fraction of optimal runtime per GPU for GA, BestMLS, BestILS, basin hopping, and differential evolution over 50 runs. Each point is the mean fraction of optimal runtime found ($y$-axis) for mean budget used (logarithmic $x$-axis), with error bars indicating the standard deviation in fraction of optimum. The point-in-polygon kernel was not implemented for the MI50 GPU.}
    \label{fig:separate_sub2_pnpoly}
\end{figure*}

\begin{figure*}
    \centering
    \newcommand{\wdth}{0.99}
    \includegraphics[width=\wdth\textwidth]{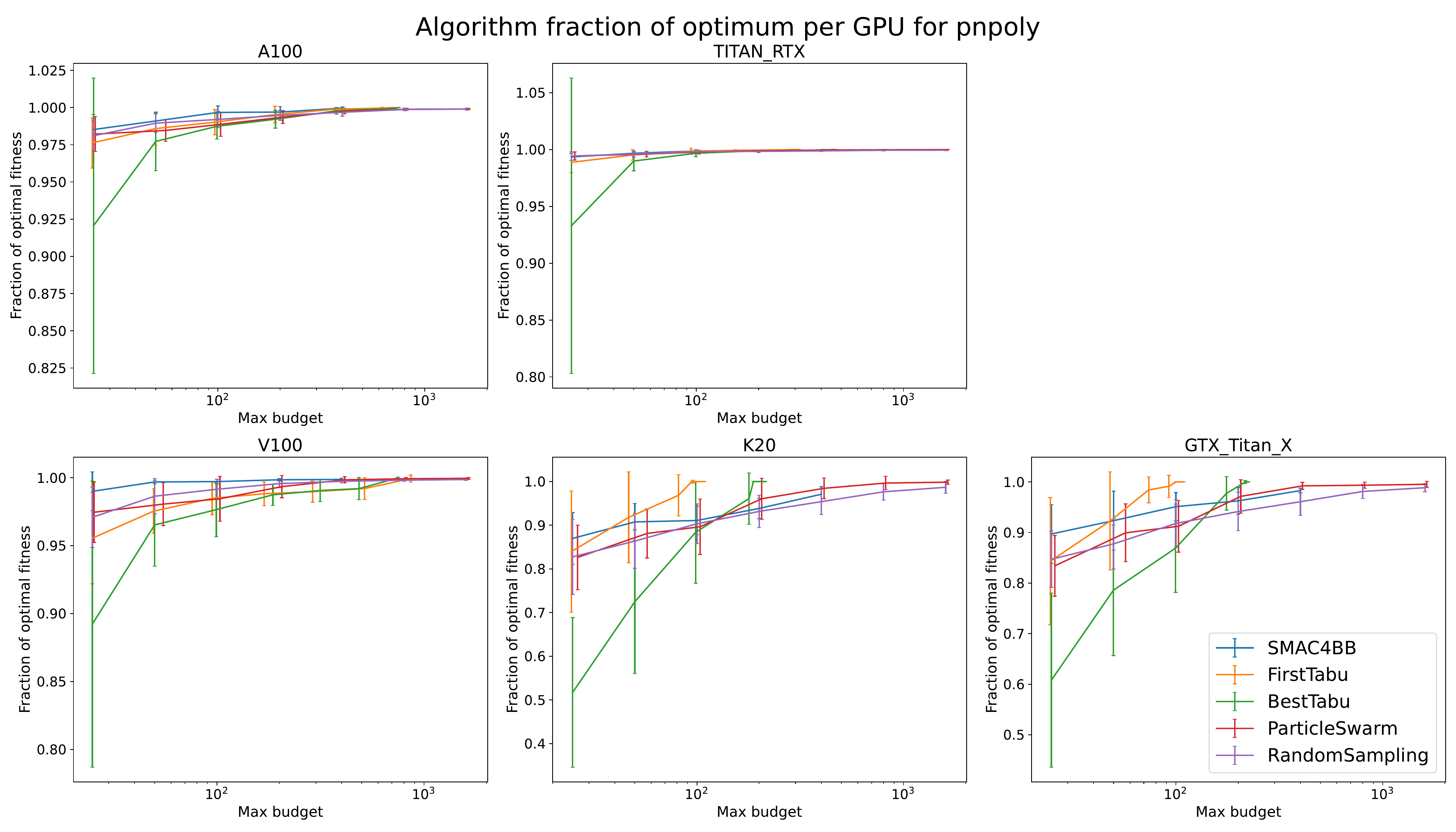}
    \caption{\scriptsize \textbf{Point-in-polygon:} Fraction of optimal runtime per GPU for SMAC, FirstTabu, BestTabu, PSO, and random sampling over 50 runs. Each point is the mean fraction of optimal runtime found ($y$-axis) for mean budget used (logarithmic $x$-axis), with error bars indicating the standard deviation in fraction of optimum. The point-in-polygon kernel was not implemented for the MI50 GPU.}
    \label{fig:separate_sub3_pnpoly}
\end{figure*}

In Figures \ref{fig:separate_sub1_conv} to \ref{fig:separate_sub3_pnpoly} we show plots of algorithm performance in terms of fraction of optimal fitness found for certain budget used (per GPU).

\clearpage

%








\end{document}